\documentclass[11pt,a4paper]{article}
\usepackage{amscd,amsmath,amssymb,amsfonts,xspace,mathrsfs,mathtools,amsthm}
\usepackage{xr-hyper}
\usepackage{jheppub}
\usepackage[dvipsnames]{xcolor}
\usepackage[utf8]{inputenc}
\usepackage{bm}
\usepackage{bbm}
\usepackage{graphicx}
\usepackage[color,matrix,arrow]{xy}
\usepackage{stmaryrd}
\SetSymbolFont{stmry}{bold}{U}{stmry}{m}{n}
\usepackage[shortlabels]{enumitem}
\usepackage{array,amsmath,booktabs}
\usepackage{slashed}
\usepackage{longtable}
\usepackage{tensor}
\usepackage{caption}
\usepackage{subcaption}
\usepackage{enumitem}
\usepackage[textsize=tiny]{todonotes}
\usepackage{lineno}
\usepackage{diagbox}
\usepackage{pdflscape}
\usepackage{silence}
\usepackage{booktabs, cellspace, hhline}
\newcommand{\sss}[1]{\medskip \noindent  \textbf{#1}}
\usepackage{booktabs, cellspace, hhline}
\newcommand{\RN}[1]{%
  \textup{\uppercase\expandafter{\romannumeral#1}}%
}
\newtheorem{conjecture}{Conjecture}
\newtheorem{theorem}{Theorem}

\newtheorem{proposition}{Proposition}

\makeatletter
\def\th@plain{%
  \thm@notefont{}%
  \itshape
  \addtolength\thm@preskip\parskip
  \addtolength\thm@postskip\parskip
}
\makeatother

\newcommand{\eg}{\textit{e.g.}}

\newcommand{\SW}{{\rm SW}}
\newcommand{\ws}{Weierstra{\ss}}

\newcommand{\vd}{\text{vd}\,}

\newcommand{\PM}{P^{\rm M}}

\newcommand{\tauD}{\tau_{\rm D}} 
\newcommand{\qD}{q_{\rm D}} 
\newcommand{\SU}{{\rm SU}} 
\newcommand{\SO}{{\rm SO}} 
\newcommand{\ie}{\emph{i.e.}} 
\renewcommand{\t}{\widetilde }

\newcommand{\br}[1]{\llbracket #1\rrbracket}

\newcommand{\mad}{m_{\text{AD}}}

\newcommand{\spinc}{\text{Spin}^c} 
\newcommand{\rk}{\text{rk}}

\newcommand{\slz}{\text{SL}(2,\mathbb Z)}

\newcommand{\nstar}{\CN=2^*}

\newcommand{\jt}{\vartheta}
\newcommand{\tn}{\tau_0} 

\newcommand{\chih}{\chi_{\rm h}}

\newcommand{\sw}{\mathcal{SW}}

\newcommand{\bea}{\begin{equation} \begin{aligned}}
\newcommand{\eea}{\end{aligned} \end{equation}}
\newcommand{\beasmall}{\begin{equation}\footnotesize\begin{aligned}}
\newcommand{\eeasmall}{\end{aligned}\footnotesize\end{equation}}
\newcommand{\be}{\begin{equation}} 
\newcommand{\ee}{\end{equation}} 
\newcommand{\bes}{\begin{equation*}}
\newcommand{\ees}{\end{equation*}}

\newcommand{\im}{i}

\newcommand{\tC}{\mathtt C}
\newcommand{\tD}{\mathtt D}
\newcommand{\tG}{\mathtt G}
\newcommand{\tH}{\mathtt H}
\newcommand{\tu}{\mathtt u}
\newcommand{\tm}{\mathtt m}
\newcommand{\tx}{\mathtt x}
\newcommand{\tv}{\mathtt v}
\newcommand{\point}{{\mathrm{pt}}}
\newcommand{\mud}{\mu_{\mathrm D}}
\newcommand{\segre}{\mathscr S}
\newcommand{\segregen}{\mathcal G}
\newcommand{\universal}{\mathscr U}
\newcommand{\decoup}{\mathscr D}
\newcommand{\Mi}{\CM_k^{\rm i}}
\newcommand{\Mm}{\CM_k^{\rm m}}

\newcommand{\CE}{\mathcal{E}}  
\newcommand{\CF}{\mathcal{F}}

\newcommand{\CI}{\mathcal{I}}
\newcommand{\CJ}{\mathcal{J}}
\newcommand{\CK}{\mathcal{K}}
\newcommand{\CL}{\mathcal{L}} 
\newcommand{\CM}{\mathcal{M}}  
\newcommand{\CN}{\mathcal{N}}
\newcommand{\CO}{\mathcal{O}} 

\newcommand{\CQ}{\mathcal{Q}}
\newcommand{\CR}{\mathcal{R}}
\newcommand{\CS}{\mathcal{S}}

\newcommand{\BQ}{\mathbb{Q}}

\newcommand{\bfm}{{\boldsymbol m}}

\newcommand{\bfx}{{\boldsymbol x}}

\newcommand{\bfk}{{\boldsymbol k}}
\newcommand{\bfz}{{\boldsymbol z}}

\newcommand{\bfmu}{{\mu}}

\title{Universal Functions for Topological Correlators}

\abstract{
We consider correlation functions of topologically twisted, $\mathcal{N}=2$ supersymmetric Yang-Mills theory with gauge group ${\rm SU}(2)$ and $N_f\leq 3$ massive hypermultiplets in the fundamental representation. For a smooth, compact, oriented four-manifold $X$ with $b_2^+>1$, the correlation functions are expressed in terms of a finite set of universal functions. The mass dependence of these functions encodes intersection numbers of the moduli space of instantons. We determine closed expressions for the universal functions by combining techniques of the Seiberg--Witten geometry, $u$-plane integral and the blowup formula. If $X$ is specialised to a complex algebraic surface $S$, the correlation functions can be identified with generating functions of Segre invariants for moduli spaces of sheaves on $S$. We verify that our results agree with the results by Göttsche and Kool for these generating functions. 
\\

\noindent\today
}

\author{Elias Furrer${}^\flat$, Jan Manschot${}^\sharp$\\
\vspace{5pt}
{${}^\flat\ $ \it NHETC and Department of Physics and Astronomy, Rutgers University,  \\ \it \quad 126 Frelinghuysen Rd., Piscataway NJ 08855, USA \\
${}^\sharp\ $ \it School of Mathematics, Trinity College, Dublin 2, Ireland\\
${}^\sharp\ $ \it Hamilton Mathematical Institute, Trinity College, Dublin 2, Ireland
\vspace{10pt} 
}}

\preprint{}
\begin{document}
\maketitle
\baselineskip=18pt

\newpage
\section{Introduction}
Path integrals of topologically twisted quantum field theories can be evaluated for many theories. Depending on the gauge group and matter content, the path integral evaluates to a generating function of topological invariants for solution spaces of a set of differential equations, such as the instanton or monopole equations~\cite{Witten:1988ze, Witten:1994cg, Vafa:1994tf, Cordes:1994fc, Moore:1997pc, LoNeSha, Marino:1998bm, Labastida:1998sk, Laba05, moore2017, Moore:2017cmm, Dedushenko:2017tdw, Manschot:2021qqe, Aspman:2022sfj,  Aspman:2023ate, Manschot:2023rdh, Cushing:2023rha}. In this paper, we consider $\CN=2$ supersymmetric Yang--Mills theory with gauge group $\SU(2)$ and $N_f \leq 3$ massive fundamental hypermultiplets with mass $m_j,j=1,\dots ,N_f$ (SQCD) on a smooth, compact four-manifold $X$. 

Recall that for $N_f=0$, the partition function $Z$ reduces to a generating function of invariants of moduli spaces of anti-self-dual instantons. This function $Z$ decomposes into a contribution  from the Coulomb branch $Z_u$ and contributions $Z_{{\rm SW},\pm}$ with delta function support on the two strong coupling singularities $u=\pm \Lambda_0^2$, known as the Seiberg--Witten (SW) contributions~\cite{Witten:1994cg, Moore:1997pc},
\be \label{Zi Zm intro}
Z=Z_u+Z_{{\rm SW},+}+Z_{{\rm SW},-}~.
\ee
While $Z_u$ is only non-vanishing for $b_2^+\leq 1$, it is possible to deduce from $Z_u$ also the contributions $Z_{{\rm SW},\pm}$ for generic $b_2^+$ using wall-crossing and topological invariance~\cite{Moore:1997pc}.
Besides the Seiberg--Witten invariants of $X$, $Z_{{\rm SW},\pm}$ only depend on a set of universal couplings and basic topological data such as the Euler number and signature of $X$~\cite{Witten:1994cg, Moore:1997pc}. For $b_2^+>1$ and with the inclusion of observables, Eq. (\ref{Zi Zm intro}) is the physical counterpart of the celebrated mathematical result that the generating function of Donaldson invariants of instanton moduli spaces is determined in terms of Seiberg--Witten invariants~\cite{Kronheimer1995, Feehan_2015, Gottsche:2010ig}.

For $N_f>0$, the partition function $Z(\bfm)$ depends on the masses $\bfm=(m_1,\dots,m_{N_f})$ and reduces to a generating function of topological invariants of moduli spaces of non-Abelian multi-monopole equations, since these are the $\CQ$-fixed equations for topologically twisted Yang-Mills theory with hypermultiplets~\cite{Labastida:1995zj, Hyun:1995hz, LoNeSha}. The non-Abelian monopole equations also provide a setup to understand the equivalence of Donaldson and Seiberg--Witten invariants through a bordism of moduli spaces~\cite{pidstrigach1995localisation, Feehan:1997gj}. The partition function $Z(\bfm)$ decomposes into a contribution from the Coulomb branch $Z_u$ and contributions from the $2+N_f$ strong coupling singularities,
\be 
\label{eq:Zm}
Z(\bfm)=Z_u(\bfm)+\sum_{j\in \{ \pm, 1,\dots, N_f\}} Z_{{\rm SW},j}(\bfm)~.
\ee 
When all masses $m_j$ are large, the two singularities labelled by $\pm$ are the singularities which become the familiar monopole and dyon singularities of the pure $\SU(2)$ theory in the infinite mass limit, while if the masses are also equal, the $N_f$ singularities labelled by $j=1, \dots, N_f$ merge to a single singularity. Similarly to $N_f=0$, it is possible to deduce the contributions $Z_{{\rm SW},j}$ using the wall-crossing formula in terms of a finite set of universal functions depending on the mass  $\bfm$~\cite{Moore:1997pc,Aspman:2023ate}. This approach has also been applied to various other twisted theories~\cite{Marino:1998bm, Manschot:2021qqe, Kim:2025fpz}.

The purpose of this paper is to derive closed form expressions for these universal functions from the physical path integral for a general four-manifold and gauge group $\SU(2)$. To this end, we consider all $N_f$ masses to be equal, $m_j=m$. This limit is smooth for $Z_{{\rm SW},+}+Z_{{\rm SW},-}$, and the coefficients in an expansion in $m^{-1}$ are expected to match with certain intersection numbers of the moduli space of instantons $\CM^{\rm i}$ mentioned above.

To derive the universal functions, we use that these functions correspond to the strong coupling limit, $g\to\infty$, of the effective coupling of various functions on the Coulomb branch, such as to the couplings of the theory to curvature, background fluxes etc. These couplings appear in the Seiberg--Witten solution for the Coulomb branch~\cite{Seiberg:1994rs, Seiberg:1994aj}, as well as in the $u$-plane integral~\cite{Moore:1997pc, Aspman:2022sfj} and the blowup formula~\cite{Fintushel1996, Edelstein:2000aj, Gottsche:2010ig}. Analogously to the use of the blowup formula for $N_f=1$ in Ref.~\cite{Gottsche:2010ig}, we derive closed-form expressions for couplings in the $N_f=2,3$ theories as functions of the effective coupling constant $\tau$.
Due to the electric-magnetic duality of the field theory, these couplings are related by modular forms, which allows to derive the universal functions as algebraic functions of the mass. 

To understand the coefficients of the mass expansion as intersection numbers of $\CM^{\rm i}$, we are motivated by the Donaldson--Uhlenbeck--Yau theorem~\cite{Donaldson1985, Uhlenbeck1986} to compare our results with universal functions derived in the context of moduli spaces of semistable sheaves on an algebraic surface with $b_2^+>1$. Many techniques are available for the determination of universal functions in this setting~\cite{Ellingsrud:1999iv, Mochizuki2009}. Building on previous work~\cite{Feehan_2015, feehan2019superconformal,feehan2019so, Marian:2015,marian2025segre, Marian2021, voisin2017segre, gottsche2022, Oberdieck_2022}, this has resulted in the determination of a complete set of universal functions~\cite{Gottsche:2020ass,Gttsche2024}. With a proper identification of the algebraic and physical parameters, we find exact agreement between the algebro-geometric universal functions and the physical universal functions derived in this paper. 

One intriguing aspect is that the set of physical couplings is smaller than the set of functions on the algebraic side, which implies several constraints for the algebro-geometric functions. Moreover, the agreement with the algebraic side provides support for the decomposition of $Z(\bfm)$  into contributions from different components of the fixed point locus of the moduli space of non-Abelian monopoles with respect to the action of the flavour symmetry group of the hypermultiplets. 
That is to say, 1) the instanton component, where the monopole fields vanish, and 2) the reducible or monopole component where the connection is reducible and some of the monopole fields are non-vanishing. Since $Z^{\rm i}$ is shown to be the contribution from the instanton component, the other terms, $Z^{\rm m}(\bfm)=\sum_{j=1}^{N_f} Z_{{\rm SW},j}(\bfm)$, are naturally identified as due to the reducible component. Such a decomposition of the partition function with respect to components of the fixed point locus is established for Vafa--Witten theory~\cite{Vafa:1994tf, Dijkgraaf:1997ce, Tanaka:2017jom} and $\CN=2^*$ Yang--Mills theory~\cite{Manschot:2021qqe}.

The main results of the paper are collected in a Proposition~\ref{prop:Segre=ZSW} and a physical Theorem~\ref{thm: DZ=G}. In order to give an overview of the results, let us start with the mathematical construction, which is best understood in the case where $S$ is an algebraic surface. The family of invariants relevant to our setup are the \emph{virtual Segre numbers}, which are intersection numbers on the moduli space $M$ of semistable sheaves on $S$. Given a $K$-theory class $\alpha$ on $S$, and an algebraic homology class $x\in H_\bullet(S)$ on $S$ (\eg~a point or surface class), they are defined as intersection numbers
\bea
\segre_\alpha=\int_M c(\alpha_M)\mud(x)~.
\eea
Here, $\mud$ is the Donaldson map, and $\alpha_M$ is a tautological class on the moduli space associated to $\alpha$, with total Chern class $c(\alpha_M)$.
Special cases of the moduli space $M$ are the Hilbert scheme of $n$ points for gauge group ${\mathrm U}(1)$, as well as the instanton moduli space in the algebro-geometric setting for higher rank gauge bundles. 
Specialising to gauge group $\SU(2)$, or rank two sheaves on $S$, the Segre numbers can be collected in a generating function, 
\bea\label{gen segre intro}
\segregen_{\segre}=\sum_{c_2}z^{\frac12\vd M}\segre_\alpha~,
\eea
where we sum over the second Chern class of sheaves on $S$, and $\vd M$ the virtual dimension of $M$. 
We denote by 
\bea\label{universal intro}
\universal_\alpha=V_s^{c_2(\alpha)}W_s^{c_1(\alpha)^2}\times\dots~
\eea
a product of in total 12 universal functions $V_s$, $W_s$, etc. They are all power series in $z^{\frac12}$, with $z$ some formal parameter. The universal series $\universal_\alpha$ depends on $\alpha$ only through its rank $s$ and Chern classes $c_i(\alpha)$.
Let $\segregen_{\universal}$ be the projection of $\universal_\alpha$ to its even or odd part, depending on details of the surface $S$ and the moduli space $M$. By a conjecture of Göttsche and Kool~\cite{Gottsche:2020ass}, which we display in Conjecture~\ref{conj_GK}, all Segre numbers of $S$ are determined using the product of universal functions:
\bea\label{conj1 intro}
\segregen_{\segre}=\segregen_{\universal}~.
\eea

To compare these expressions with topological correlators for SQCD, we express the $K$-theory classes $\alpha$ in terms of the bundles involved in the topological twist. This allows us to recast the monopole and dyon contributions $Z_{\SW,\pm}$ in a form that parallels the universal series $\universal_\alpha$. More precisely, in Proposition~\ref{prop:Segre=ZSW} we rigorously show that the contribution $Z_{\SW,-}$ from the monopole singularity (including a decoupling factor $\decoup$) agrees with half of the universal series~\eqref{universal intro},
\bea\label{prop 1 intro}
\decoup Z_{\SW,-}=\tfrac12 \universal_\alpha~.
\eea
This identification amounts for each number $N_f$ of flavours to expressing all Coulomb branch couplings at the monopole singularity in terms of universal series. The equality~\eqref{prop 1 intro} holds as a $z$-series, where we identify $z= \frac{\Lambda_0^2}{2m^2}$, with $\Lambda_0$ the scale of the pure $\SU(2)$ theory.

The contribution $Z_{\SW,+}$ from the dyon singularity results in a similar series $\universal_\alpha(-z)$, where some coefficients have the opposite sign. In the absence of the $u$-plane integral $Z_u$, the sum over the two Seiberg--Witten contributions, $Z_{\SW}^{\rm i}=Z_{\SW,+}+Z_{\SW,-}$, gives the full instanton partition function $Z^{\rm i}$ \eqref{Zi Zm intro}, and agrees with the projection $\segregen_\universal$ as before. This is the main result of our paper, which we make precise in Theorem~\ref{thm: DZ=G}:
\bea\label{D ZSW i = G_U}
\decoup Z_{\SW}^{\rm i}=\segregen_\universal~.
\eea
When $X$ is algebraic, by~\eqref{conj1 intro} the instanton component to the topological partition function agrees with the generating function~\eqref{gen segre intro} of virtual Segre numbers,
\bea
\decoup Z_{\SW}^{\rm i}=\segregen_\segre~.
\eea

As in Donaldson theory, the richness of the theory comes from the inclusion of homology classes $x\in H_\bullet(X)$, in our case points and surfaces. In physics, these correspond to observables in the topological QFT. All the above equations should be understood as containing the point and surface observables. Indeed, the main challenge is to identify and compute the various couplings and contact terms for the observables.

The outline of this paper is as follows. Section \ref{sec:TopSQCD} reviews aspects of Seiberg--Witten geometry, topological SQCD and its partition function, and includes an in-depth study of the blowup formula.
Section \ref{sec:universal functions from PF} reviews the generating functions in the algebraic geometric context and the universal functions involved. The main technical result is Proposition~\ref{prop:Segre=ZSW}, which develops the identification between quantities on the physical and mathematical side. It enables the formulation of Theorem~\ref{thm: DZ=G}, which is the main conceptual result of the paper. 
Section~\ref{sec:deriving_alg_functions} derives the universal functions on the physical side, which forms the bulk of the proof of Proposition~\ref{prop:Segre=ZSW}. The paper concludes with a discussion section~\ref{sec:discussion} with several directions for future work. The three appendices review modular and elliptic functions~\ref{app:modularforms}, Seiberg--Witten geometry and couplings~\ref{app:largemassSW}, and include a symbol list~\ref{app:symbol list}.

\acknowledgments We would like to thank Johannes Aspman, Cyril Closset, Lothar Göttsche, Martijn Kool, Ideal Majtara, Greg Moore, Vivek Saxena and Ranveer Kumar Singh for discussions and correspondence. We especially thank Johannes Aspman, Greg Moore, Vivek Saxena and Ranveer Kumar Singh for comments on a draft of the paper.
The work of E.F. was supported by the EPSRC grant ``Local Mirror Symmetry and Five-dimensional Field Theory'' and the US Department of Energy under grant DE-SC0010008.

\section{Topological SQCD}
\label{sec:TopSQCD}

In this section, we review the Coulomb branch geometry relevant to the formulation of topological correlators, the topological twist and the computation of the contribution from the $u$-plane to topological correlators. 

\sss{Notation.} While we try to keep as much notation from previous literature as possible, we have to introduce various new symbols for the rather detailed universal results of Section~\ref{sec:universal functions from PF}. For functions on the Coulomb branch (or more specifically of the low-energy effective coupling $\tau$), we use ordinary Latin letters, such as $G_{N_f}(\tau)$. For the values of such functions at the strong coupling singularities, we use the $\mathtt{Typewriter}$ font, such as $\tG_{N_f}$. We compiled a list of the most important quantities in Appendix~\ref{app:symbol list}.

\subsection{Brief review of Coulomb branch geometry}
\label{sec:CB geometry}
In this subsection, we describe relevant aspects of the Coulomb branch geometries of SQCD with $N_f\leq 3$ massive fundamental hypermultiplets. We do this in a unified way in terms of their Seiberg--Witten solutions.\footnote{See~\cite{Klemm:1997gg} or~\cite[Section 2]{Aspman:2022sfj} for a detailed introduction to SW geometry.}

\sss{SW geometry.} The low-energy effective dynamics of SQCD is famously encoded in the Seiberg--Witten curve, which we list in Appendix~\ref{app:largemassSW} for $N_f=1,2,3$. The SW curve is fibered over the Coulomb branch (or $u$-plane), and becomes singular at $2+N_f$ points, where hypermultiplets become massless. This generalises the solution for pure ($N_f=0$)  $\CN=2$ super-Yang--Mills theory. In that case, the Coulomb branch has two strong coupling singularities, $\tu^-=-\Lambda_0^2$ and $\tu^+=+\Lambda_0^2$. At $\tu^-$, the monopole becomes massless, while at $\tu^+$ a dyon becomes massless.\footnote{In most of the previous literature this assignment is reversed. We choose this convention since it is the most natural when generalising to SQCD.}

The position of the singularities on the $u$-plane depends in a complicated way on the masses $\bfm=(m_1,\dots, m_{N_f})$ of the $N_f$ hypermultiplets. In this article, we consider the case where all masses $m_i=m$ are equal, and $m$ is large. Since the large mass limit decouples hypermultiplets from the theory, for $m\to\infty$ we should then recover the original $N_f=0$ theory. Thus, when the (equal) mass $m$ of the hypermultiplets is large, this singles out two singularities $\tu_{N_f}^\pm$ which flow to the monopole and dyon singularities $\tu_{N_f}^\pm\to \pm\Lambda_0^2$ of the pure $\SU(2)$ $\CN=2$ theory. 
The equal mass case is characterised by the Kodaira singular configuration of the Seiberg--Witten curve
\begin{equation}
    (I_{4-N_f}^{\infty,*},I_1,I_1,I_{N_f})~.
\end{equation}
Here, $I_{4-N_f}^{\infty,*}$ corresponds to the weak coupling region, and the two $I_1$'s become the monopole and dyon singularities for $m\to \infty$. The remaining singularities for all $N_f=1,2,3$ are organised by the flavour symmetry into a single  $I_{N_f}$ singularity $\tu_{N_f}^*$.

\begin{figure}[ht]\centering
	\includegraphics[scale=1.2]{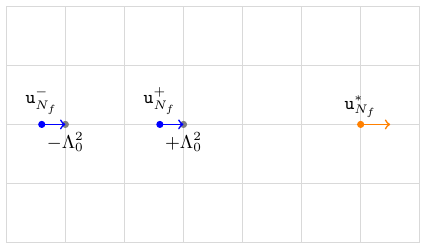}
    \caption{The large mass singular structure on the Coulomb branch of $\SU(2)$ SQCD with $N_f=1,2,3$ hypermultiplets is as follows. For $m\gg \Lambda_0$ and positive, the two singularities $\tu_{N_f}^\pm$ approach the two $N_f=0$ singularities $\pm\Lambda_0^2$ from below, while the third singularity $\tu_{N_f}^*$ moves to $+\infty$. In the limit $m\to\infty$, this recovers the pure $N_f=0$ $u$-plane.}
	\label{fig:equalmass}
\end{figure}

Since in the strict $m\to\infty$ limit we obtain the $N_f=0$ theory, we express the geometry in terms of the scale $\Lambda_0$ for $N_f=0$. It is generated in the decoupling limit as a double scaling relation
$m^{N_f}\Lambda_{N_f}^{4-N_f}=\Lambda_0^4$~\cite{Seiberg:1994rs}. Then for $N_f=1,2,3$, we have the large $m$ series
\bea\label{upm}
\tu_{N_f}^\pm &=\pm \Lambda_0^2-\frac{N_f\Lambda_0^4}{16m^2}+\CO(m^{-4})~, \\
\tu_{N_f}^*&=m^2+\frac{\Lambda_0^4}{8m^2}+\CO(m^{-4})~,
\eea
which we depict in Fig.~\ref{fig:equalmass}.

The two $N_f=0$ singularities $\tu_{N_f}^\pm$ are `mirrored' for large $m$. While $\tu_{N_f}^\pm$ have dimension 2, the quantities $\tu_{N_f}^\pm/m^2$ are dimensionless and admit a Taylor series in $z\coloneqq \frac{\Lambda_0^2}{2m^2}$. Let us denote the Taylor series for $-\tu_{N_f}^-/m^2$ by $T_{N_f}(z)$. The second singularity is then obtained as 
\begin{equation}\label{u/m2}
-\frac{\tu_{N_f}^+}{m^2}=T_{N_f}(-z)~.
\end{equation}
This is because the discriminant $\Delta_{N_f}$ of the SW curve depends only on $z^2$ and is thus invariant under $z\mapsto -z$, and as a consequence the two roots $\tu_{N_f}^\pm$ are interchanged. It can be understood as the emergent $R$-symmetry of the pure $\SU(2)$ theory, which acts as a $\mathbb Z_2$ symmetry on the $u$-plane~\cite{Seiberg:1994rs,Seiberg:1994aj}.

This article focuses on determining contributions of the vacua $\tu_{N_f}^\pm$ to the topological partition function on four-manifolds. As will be explained in detail in Section~\ref{sec:universal functions from PF}, due to the symmetry~\eqref{u/m2}, it is sufficient to determine the contribution from one singularity only. For the remainder of this section, we focus on the singularity $\tu_{N_f}^-$, where we drop the label in the following and denote it as $\tu_{N_f}$. 

\sss{Coulomb branch couplings.}
The calculation of topological correlation functions requires various Coulomb branch functions. While some of these functions are obtained directly from the Seiberg--Witten curve, this is not immediately the case for other couplings, such as mass derivatives of the prepotential. 

Let us first discuss the more standard Coulomb branch functions: the period $\frac{da}{du}$ and the physical discriminant $\Delta_{N_f}$.
These two functions enter the $u$-plane integral as effective gravitational couplings $A^\chi B^\sigma$, where $A$ depends only on $\frac{da}{du}$, while $B$ is given by $\Delta_{N_f}^{\frac18}$.
The period $\frac{da}{du}$ is obtained simply from the SW curve by~\cite{Brandhuber:1996ng}:
\bea\label{dadu def tau}
\frac{da}{du}=\frac16\sqrt{\frac{g_2}{g_3}\frac{E_6}{E_4}}~,
\eea
where $g_2$ and $g_3$ are the \ws{} invariants of the Seiberg--Witten curve, as given in Appendix~\ref{app:curves}, and $E_k$ are the Eisenstein series, which we define in~\eqref{eisenstein}. The physical discriminant~\cite{Shapere:2008zf} is for generic masses defined as the monic polynomial $\Delta_{N_f}(u)=\prod_{j=1}^{N_f+2}(u-\tu_j^*)$, where $\tu_j^*$ label the $2+N_f$ singularities. Using the choice of \ws{} invariants of Appendix~\ref{app:curves}, it is given by 
\bea
\Delta_{N_f}=(-1)^{N_f}\Lambda_{N_f}^{2(N_f-4)}(g_2^3-27g_3^2)~,
\eea
which up to prefactors is the mathematical discriminant of the SW curve.
An important relation between the period $\frac{da}{du}$ and the discriminant $\Delta_{N_f}$ is~\cite{Aspman:2021vhs}:
\bea\label{dadu Delta}
\eta^{24}=2^6(-1)^{N_f}\Lambda_{N_f}^{2(4-N_f)}\left(\frac{da}{du}\right)^{12}\Delta_{N_f}~,
\eea
with $\eta$ the Dedekind eta function~\eqref{dedekind eta}.

The remaining CB functions required for the formulation of topological correlators are most easily expressed as derivatives of the prepotential $\CF$. Since the latter is obtained by a standard calculation, we refer to Appendix~\ref{app:largemassSW} for an exhaustive discussion.

The prepotential of $\SU(2)$ QCD with massive hypermultiplets is a function of the masses $\bfm$, the coordinate $a$ and the dynamical scale $\Lambda_{N_f}$. While $\partial^2 \CF/\partial a^2$ gives the effective gauge coupling $\tau$ of the theory and describes the non-perturbative effective dynamics on flat spacetime, the other second derivatives find an interesting application only in the topological theory. 

Let us introduce the couplings $v_j$ and $w_{jk}$ with $j,k\in 1,\dots, N_f$~\cite{AlvarezGaume:1997fg, AlvarezGaume:1997ek, Marino:1998ru,Edelstein:1999xk,Manschot:2021qqe,Aspman:2022sfj,Aspman:2023ate},
\be\label{vj wij}
v_j=\sqrt{2}\frac{\partial^2 \CF}{\partial a \partial m_j}~, \qquad w_{jk}=2\frac{\partial^2 \CF}{\partial m_j\partial m_k}~.
\ee
They appear in the topological path integral in exponentiated form:
\begin{equation}\label{C D couplings}
    D_j=e^{-\pi i v_j}~, \qquad C_{jk}=e^{-\pi i w_{jk}}~.
\end{equation}
We furthermore require all the second derivatives of the prepotential $\CF$ with respect to $m_i$ and $\Lambda_{N_f}$. These derivatives provide the order parameter $u$, along with a contact term $G_{N_f}$ for the surface observable $\bfx$, and the couplings $H_{N_f,j}$ of background fields $\bfk_j$ to the surface observable $\bfx$~\cite{Moore:1997pc, LoNeSha,Mari_o_1999,Manschot:2021qqe},
\bea\label{contactterm}
u&=\frac{4\pi i}{4-N_f} \Lambda_{N_f} \frac{\partial \CF}{\partial \Lambda_{N_f}}~, \\
G_{N_f}&=\frac{4\pi i}{(4-N_f)^2\Lambda_{N_f}^2}(\Lambda_{N_f}\partial_{\Lambda_{N_f}})^2 \CF~, \\
H_{N_f,j}&=\frac{4\sqrt2\pi}{4-N_f}\frac{\partial^2 \CF}{\partial \Lambda_{N_f}\partial m_j}~,
\eea
The contact term for the surface observable may be readily evaluated as~\cite{Moore:1997pc, LoNeSha, mm1998,Mari_o_1999} 
\bea\label{contactterm explicit}
G_{N_f}=-\frac{1}{24\Lambda_{N_f}^2}E_2\left(\frac{du}{da}\right)^{2}+\frac{1}{3\,\Lambda_{N_f}^2}\left(u+\frac{\Lambda_3^2}{64}\delta_{N_f,3}\right)~.
\eea

We discuss the calculation of these couplings in depth in Appendix~\ref{app:largemassSW}.  Several other important Coulomb branch functions can be extracted either simply from the SW curve, or generally through the prepotential~\cite{Aspman:2021vhs}.

\sss{Couplings at the singularities.}
Naturally, these derivatives and couplings are complicated functions of $a$, the masses $\bfm$ and the scale $\Lambda_{N_f}$. For the calculation of partition functions, we are interested in the asymptotic behaviour of the Coulomb branch of SQCD with $N_f$ equal mass $m$ hypermultiplets, in the regime where $m$ is large. Moreover, under certain circumstances, only their value at the strong coupling singularities is required.

The monopole (dyon) singularity $\tu_{N_f}^\pm$ is approached on the $u$-plane as $\tau\to 0$ ($\tau \to 2$)~\cite{Aspman:2021vhs}. Let thus $\tauD$ be the local coupling constant, that is, the low-energy effective coupling for the dual photon vector multiplet. For the monopole singularity $\tu_{N_f}^-$, we have $\tau=-1/\tauD$, while for the dyon singularity  $\tu_{N_f}^+$, we have $\tau=2-1/\tauD$. As before, we let $\tu_{N_f}=\tu_{N_f}^-$ be the monopole singularity. The vector multiplet scalar for the dual photon then has an expansion
\bea\label{u_exp}
u_{\rm D}(\tauD)=\tu_{N_f}+\tm_{N_f} \qD+\CO(\qD^2)~,
\eea
which is defined as $u_{\rm D}(\tauD)=u(-1/\tauD)$, and $q_{\rm D}=e^{2\pi i\tauD}$.\footnote{In general, for a modular form of weight $k$ we define $f_{\rm D}(\tauD)=\tauD^{-k}f(-1/\tauD)$} The $\qD$-series has integer exponents due to the fact that $\tu_{N_f}$ is an $I_1$ singularity. The first coefficient~$\tm_{N_f}$ is of main importance for the calculation later. By expanding the $\CJ$-invariant of the SW curve~\eqref{eq:curves} around a singularity $\tu_{N_f}$, the coefficient $\tm_{N_f}$ can be computed as 
\begin{equation}\label{mu def}    \tm_{N_f}=12^3(-1)^{N_f}\Lambda_{N_f}^{2(N_f-4)}\frac{g_2(\tu_{N_f})^3}{\Delta_{N_f}'(\tu_{N_f})}~,
\end{equation} 
where $\Delta_{N_f}'$ is the derivative of $\Delta_{N_f}$ with respect to $u$. 

Other important quantities are singular at strong coupling, and we are interested in the finite piece after removing the modular weight factor. The period $\frac{da}{du}$ has weight $1$ and thus is singular at strong coupling. We are interested in the constant term (that is, the coefficient of $\qD^0$) in the $\qD$ series of $(\frac{da}{du})_{\rm D}(\tauD)=\tauD^{-1}\frac{da}{du}(-1/\tauD)$. 
Adding the label $N_f$ again, it follows from~\eqref{dadu def tau} that\footnote{The period $\frac{da}{du}$ is not single-valued on the Coulomb branch, as under a large $u$ monodromy it goes to $-\frac{da}{du}$. Hence, in principle, one should always work with $(\frac{da}{du})^2$. However, below we will relate universal functions to $\tx_{N_f}$, rather than $\tx_{N_f}^2$, and it is therefore crucial to fix this sign. We will argue that with the normalisation of the \ws{} invariants (given in Appendix~\ref{app:largemassSW}) and the phases we determine for the partition functions, this choice of square root is the correct one.}
\begin{equation}\label{dadu def}
    \tx_{N_f}\coloneqq (\tfrac{da}{du})_{\rm D}(\tauD)\Big|_{\qD^0}=\frac16\sqrt{\frac{g_2(\tu_{N_f})}{g_3(\tu_{N_f})}}~.
\end{equation}

Since the prepotential $\CF$ is necessarily fully symmetric in the masses $\bfm$, so are the mass derivatives once evaluated at $m_i=m_j=m$. Therefore, all the single-mass derivatives $v_j$~\eqref{vj wij} become identical, and the same holds for the contact terms $H_{N_f,j}$~\eqref{contactterm}. We denote the values of the former couplings at the monopole singularity (up to their weight factor) by $\tv_{N_f}$, while the contact terms become series $\tH_{N_f}$. 
For the second mass derivatives $w_{jk}$ (or $C_{ij}$), there are two cases.\footnote{The derivatives $F^{(n_1,n_2,\dots)}(m_1,m_2,\dots)$ are symmetric in the $n_1,n_2,\dots$ once evaluated at $m_i=m$. For $n_i=0,1,2$, there are two distinct classes of possibly different derivatives: There are the derivatives with $n_i=2$ for some $i$, and the mixed derivatives $n_i=n_j$ for some $i\neq j$.} This means that in the equal mass limit, there are only two functions: $C_{jj}$ for all $j$, and $C_{jk}$ for any $i\neq j$. As we prove below, all those functions become simply a function of $\Lambda_{N_f}/m$. More precisely, with $z=\frac{\Lambda_0^2}{2m^2}$,
we find that they admit Taylor series at $z=0$ in the variable $z^{\frac12}$. In order to distinguish the couplings as global functions (\eg~$D_j$ and $C_{jk}$) and their expansions at the monopole singularity for equal masses, we denote the latter by
\bea\label{couplings tau=0}
\tC_{N_f}&\coloneqq C_{jj}^{\rm D}(\tauD)\big|_{\qD^0}~, \\
\hat{\tC}_{N_f}&\coloneqq C_{jk}^{\rm D}(\tauD)\big|_{\qD^0}~, \\
\tD_{N_f}&\coloneqq e^{-\pi i v_j^{\rm D}(\tauD)}\big|_{\qD^0}~,  \\
\tG_{N_f}&\coloneqq  G_{N_f}^{\rm D}(\tauD)\big|_{\qD^0}~, \\
\tH_{N_f}&\coloneqq  H_{N_f,j}^{\rm D}(\tauD)\big|_{\qD^0}~, \\
\tv_{N_f}&\coloneqq v_j^{\rm D}(\tauD)\big|_{\qD^0}~.
\eea
for all $j\neq k=1,\dots, N_f$.\footnote{Note that here we label the series (\eg~$\mathtt D_{N_f}$) by $N_f$, while we have labels $j,k=1,\dots, N_f$ on the rhs. } The form of the large mass expansions of those functions are given in~\eqref{large m series modular forms}, and as a consequence, the dual functions are holomorphic $\qD$-series with integer exponents in a neighbourhood of $q_D=0$.\footnote{This is because $\jt_4(\tauD)\jt_3(\tauD)=\mathbb C[[\qD]]$, while $\jt_2(\tauD)\jt_3(\tauD)=\mathbb C((\qD^{\frac 18}))$.} Taking the coefficient of $\qD^0$ is therefore identical to taking the limit $\tauD\to i\infty$. 
The series $\tC_{N_f}$, $\hat \tC_{N_f}$, $\tD_{N_f}$, $\tH_{N_f}$, $\tH_{N_f}$ together with $\tm_{N_f}$, $\tx_{N_f}$ and $\tu_{N_f}$ form the set of 8 functions required for the contribution of the singularity $\tu_{N_f}$ to the equal mass partition function. 

The contact term $\tG_{N_f}$ is straightforward to determine from~\eqref{contactterm explicit},
\bea\label{contact term series}
\tG_{N_f}&=-\frac{1}{24\Lambda_{N_f}^2\tx_{N_f}^2}+\frac{1}{3\,\Lambda_{N_f}^2}\left(\tu_{N_f}+\frac{\Lambda_3^2}{64}\delta_{N_f,3}\right)~,
\eea
which thus depends only on $\tx_{N_f}$ and $\tu_{N_f}$. The couplings $\tC_{N_f}$, $\hat \tC_{N_f}$ and $\tH_{N_f}$ do not have such a simple expression in terms of the curve, and the determination is one of the key results below.

\subsection{Topological twist}
\label{top twist}

Before we review the topological twist of $\CN=2$ SQCD, let us set some notation.  We are closely following the more recent discussion~\cite{Aspman:2022sfj, Aspman:2023ate}, and refer to~\cite{Hyun:1995hz,Hyun:1995mb,Labastida:1995zj,Moore:1997pc,LoNeSha, Labastida:1997rg,Labastida:1996tz} for an overview.

\sss{Compact four-manifolds.}
We let $X$ be a smooth, oriented, compact and simply-connected four-manifold. We assume throughout that $X$ has no boundary. Basic topological data of $X$ are its Betti numbers $b_i=b_i(X)$, Euler number $\chi$ and signature $\sigma$. 
Let $L$ be the embedding of $H^2(X,\mathbb Z)$ in $H^2(X,\mathbb Z)\otimes \mathbb R$, modding out torsion. The intersection form on $H^2(X,\mathbb Z)$ provides a  bilinear form $B:L\otimes \mathbb \mathbb R\times L\otimes \mathbb R\to\mathbb R$ that pairs degree two cocycles, and we abbreviate $\bfk_1\bfk_2\coloneqq B(\bfk_1,\bfk_2)$. It furthermore gives rise to a quadratic form $\bfk^2\coloneqq B(\bfk,\bfk)$.

For the topological correlators to be non-zero, we require that the moduli space of instantons has even real dimension. Since $X$ is assumed simply-connected, this requires that $b_2^+$ is odd. 
Moreover, one can deduce that $H^2(X,\mathbb{Z})$ contains an element $c$ such that $c^2=2\chi+3\sigma$~\cite[p. 377]{scorpan:2005}. The existence of this element implies that $X$ admits an almost complex structure whose first Chern class is $c$. As a result, $X$ is an almost complex four-manifold, and its tangent bundle is a complex manifold~\cite{Donaldson90}.

To this holomorphic tangent bundle, we can then associate a Chern character. Let $T_X^*$ be the cotangent bundle, which is a rank 2 complex vector bundle. The canonical bundle $K_X=\det(T_X^*)$ has a well-defined first Chern class, which is the canonical class $K=c_1(K_X)=-c_1(T_X^*)$, and satisfies $K^2=2\chi+3\sigma$. Finally, we define the holomorphic Euler characteristic  $\chih=\frac14(\chi+\sigma)\geq 1$, which is an integer for almost complex four-manifolds~\cite{Witten:1994cg}.
To facilitate the comparison with the mathematical literature, we use $(\chih,K^2)$ rather than $(\chi,\sigma)$, for which we use the inverse relations $\chi=-K^2+12\chih$ and $\sigma=K^2-8\chih$.

Moreover, we let $E\to X$ be a principal $\text{SO}(3)$-bundle with connection. The second Stiefel-Whitney class $w_2(E)\in H^2(X,\mathbb{Z}_2)$ measures the obstruction to lift $E$ to an $\SU(2)$ bundle, which exists globally only if $w_2(E)=0$. 
We denote a lift of $w_2(E)$ to $L$ by $\bar w_2(E)\in L$, and define the 't Hooft flux $\bfmu=\bar w_2(E)/2\in L/2$. The instanton number of the principal bundle is defined as $k=-\frac14\int_X p_1(E)$  and satisfies $k\in -\bfmu^2 + \mathbb{Z}$, where $p_1$ is the first Pontryagin class.

\sss{Topological twist.}
To formulate the massive $\CN=2$ SQCD on a compact four-manifold $X$, we perform a refined Donaldson--Witten twist. Aside from the usual $\text{SU}(2)$ R-symmetry bundle, for each hypermultiplet we require a principal bundle $\CL_j$ with connection for the flavour symmetries~\cite{Manschot:2021qqe,Aspman:2022sfj,Moore:2024vsd}.

In the Donaldson--Witten twist, the R-symmetry bundle is isomorphic to the chiral spin bundle $S^+$. The fields combine to sections of this non-trivial R-symmetry bundle. Effectively, the fields transform under the diagonal group of $\SU(2)_+\times \SU(2)_R$. 
While the vector multiplet fields combine to differential forms, the
hypermultiplet bosons become spinors, that is, sections $M^j$ of the spin bundle $S^+$, while the fermions are sections of $S^+$ and $S^-$. Thus, the twisted hypermultiplets can a priori only be formulated on four-manifolds which are spin, \ie~$w_2(X)=0$~\cite{Hyun:1995mb, Moore:1997pc}. 
However, if the hypermultiplets are charged under a gauge field, the product of these bundles with $S^\pm$ may be a Spin$^c$ bundle, $W^+$ or $W^-$~\cite{Hyun:1995mb,Laba05, Manschot:2021qqe}. The topologically twisted hypermultiplets are then well-defined on non-spin four-manifolds $X$ whenever $\bar w_2(X)=\bar w_2(E) \mod 2L$~\cite{Moore:1997pc}. Thus, generally not all 't Hooft fluxes $\bfmu$ are admissible. 

To consider more general 't Hooft fluxes, we can couple the $j$'th hypermultiplet to a line bundle $\CL_j$. For this, let $\CL_E^{\frac12}$ be the line bundle whose sections are components of the fundamental representation of $\SU(2)$.
For $\CE_j=\CL_E\otimes \CL_j$, the requirement that $S^\pm\otimes \CE_j^{\pm1/2}$ is globally well-defined is that $c_1(\CL_j)\in \bar w_2(X)+\bar w_2(E) +2L$ for each $j$. This is consistent with the above constraints for $c_1(\CL_j)=0$.
In the following, we will abbreviate $\bfk_j\coloneqq \frac12 c_1(\CL_j)$. Then the topological twist is well-defined if~\cite{Aspman:2022sfj}
\be \label{constraint_c1CL}
\bfk_j\equiv \frac K2-\bfmu\mod L~,
\ee 
for each $j=1,\dots,N_f$. This can be derived using the general approach of the transfer of structure groups between the twisted and untwisted theories \cite{Manschot:2021qqe, Moore:2024vsd}. The combination of the fluxes $\bfk_j$ is an example of a ``generalized Spin$^c$ structure", introduced in~\cite{Moore:2024vsd}.

\sss{$\mathcal Q$-fixed equations.}
For our case of interest, 
the $\CQ$-fixed equations are the non-Abelian monopole equations for gauge group SU(2) with $N_f$ matter fields in the fundamental representation. They read
\bea
\label{Qfixedeqs}
0&=\left(F^{a}_{\dot \alpha \dot \beta}\right)^++\frac{i}{2} \sum_{j=1}^{N_f}\bar M^j_{(\dot \alpha} T^a M^j_{\dot \beta)}~,\\
0&=\slashed DM^j~,
\eea
where $T^a$ are the generators of the Lie algebra in the fundamental representation SU(2)~\cite{pidstrigach1995localisation, Labastida:1995zj, LoNeSha, Feehan:1997gj}. 

We denote the moduli space of solutions to these equations by $\CM^{\CQ,N_f}_{k,\mu}(X)$, where we suppress the dependence on the fluxes $\bfk_j$, and occasionally drop other dependences. It is known that $\CM$ can become non-compact for vanishing masses~\cite{bryan1996,Moore:1997dj, Dedushenko:2017tdw}. This is improved upon turning on masses and localising with respect to the $\text{U}(1)^{N_f}$ flavour symmetry,  $M^j\to e^{i\,\varphi_j} M^j$, which leave the $\CQ$-fixed equations \eqref{Qfixedeqs} invariant.
There are two components~\cite{pidstrigach1995localisation}:
\begin{itemize}
    \item the instanton component $\Mi$, with $F^+=0$ and $M^j=0$, $j=1,\dots,N_f$. Since the hypermultiplet fields vanish, this component is associated with the Coulomb branch.
\item the monopole component $\Mm$, for which a U$(1)$ subgroup of the flavour group acts as pure gauge. Here, the connection is reducible, and a U(1) subgroup of the  $\SU(2)$ gauge group is preserved. For generic, non-equal masses, there are $N_f$ such components. For each component, all but one monopole field $M^j$ vanish.
As the monopole fields originate from the hypermultiplets, this component is associated with the Higgs branch~\cite{Moore:1997dj,Marino:1998uy,Dedushenko:2017tdw}.
\end{itemize}
In this article, we focus on the contributions to the path integral from the instanton component $\Mi$. We leave the analysis of the monopole component for future work~\cite{FMM:future}.

The $\CQ$-fixed equations (\ref{Qfixedeqs}) include a Dirac equation for each hypermultiplet $j=1,\dots,N_f$ in the fundamental representation. The corresponding index bundle $W_k^j$ defines an element of the K-group of $\Mi$. 
Its virtual rank $\rk\, W^j_k$ is the formal difference of two infinite dimensions. It is given by an index theorem and reads $\rk\, W^j_k=-k+\bfk_j^2-\frac \sigma4$, which is an integer due to~\eqref{constraint_c1CL}~\cite{Aspman:2023ate}. In fact, the full Chern character of the index bundle is given by the Atiyah--Singer index theorem~\cite{Atiyah:1970ws}. See also Ref.~\cite{Berline1992}.

The moduli space $\CM^{\CQ,N_f}_{k}$ for $N_f$ hypermultiplets corresponds to the vanishing locus of the obstructions for the existence of $N_f$ zero modes of the Dirac operator. As a result,  the virtual complex dimension of the moduli space $\CM^{\CQ,N_f}_{k}$ is that of the instanton moduli space plus the sum of ranks of the index bundles $W_k^j$,
$\vd \CM^{\CQ,N_f}_{k}=\vd\CM^{\CQ,0}_k+\sum_{j=1}^{N_f}\rk\, W_k^j $~\cite{Hyun:1995hz, bryan1996, Feehan:2001jc, Manschot:2021qqe}.
This gives
\begin{equation}
\label{vdimML}
	\vd \CM_{k}^{\CQ,N_f}=(4-N_f)k
    -3\chih-\tfrac{N_f}{4}\sigma+\sum_{j=1}^{N_f}\bfk_j^2~,
\end{equation}
This reduces for $N_f=0$ to the dimension of the instanton moduli space $\Mi$, with real dimension
\begin{equation}
\label{eq:dimMi}
\vd_{\!\mathbb{R}} \CM_{k}^{\rm i}=8k-\tfrac{3}{2}(\chi+\sigma),
\end{equation}
as expected.

\sss{Correlation functions.}
The correlation functions on $X$ in the theory with $N_f$ massive fundamental hypermultiplets reduce to a finite-dimensional integral over $\CM^{\CQ,N_f}_{k,\mu}$.
Localisation to the fixed point locus in $\CM^{\CQ,N_f}_{k}$ with respect to the $U(1)^{N_f}$ action gives a sum of contributions from the instanton component $\Mi$ and the monopole component $\Mm$ (see~\cite{Mathai:1986tc,Atiyah:1990tm, Vafa:1994tf,Cordes:1994fc, LoNeSha,Lossev:1997bz, Manschot:2021qqe,Aspman:2022sfj,Aspman:2023ate} for related work).

Let $\CO$ be a $\CQ$-closed observable in the UV theory, and denote by $\langle \CO\rangle$ its correlation function on $X$. Our interest is in the contribution $\langle \CO\rangle^{\rm i}$ to $\langle \CO\rangle$ from the instanton branch.
We use the equivariant Chern classes of the direct sum of index bundles for the hypermultiplets, 
\bea\label{W_k}
W_k=\bigoplus_{j=1}^{N_f} W_k^j~.
\eea
Abbreviating $c_{l,j}\coloneqq c_l(W_k^j)$, the instanton contribution then reads~\cite{LoNeSha}
\begin{equation}\label{O instanton}
	\begin{split}
	&\langle \CO\rangle^{\rm i}=\sum_k  \Lambda_{N_f}^{\vd \CM^{\CQ,N_f}_{k}} \int_{\Mi} \left(\prod_{j=1}^{N_f}  m_j^{-\rk\,  W^j_k}\sum_l \frac{c_{l,j}}{m_j^l}\right) \mud(\CO)~,
	\end{split}
\end{equation}
where $\mud(\CO)$ is the associated Donaldson class of $\CO$.

\sss{Decoupling limit.}
Increasing the mass of a hypermultiplet to infinity decouples it from the theory. For $N_f$ distinct masses, this decoupling limit is a double scaling limit, where $m_{N_f}\to \infty$, $\Lambda_{N_f}\to 0$, while their product $m_{N_f}\Lambda_{N_f}^{4-N_f}=\Lambda_{N_f-1}^{4-(N_f-1)}$ remains fixed and becomes the new scale $\Lambda_{N_f–1}$ of the decoupled theory.
The decoupling limit of the correlation functions~\eqref{O instanton} has been derived in~\cite{Aspman:2022sfj,Marino:1998uy}. It expresses the decoupling of a correlator $\langle \CO\rangle_{N_f}$ to $\langle \CO\rangle_{N_f-1}$, and is given explicitly in~\cite[(3.20)]{Aspman:2022sfj}. 
For a large mass $m_{N_f}$, the dominant contribution for $j=N_f$ in~\eqref{O instanton} is from $c_{0,N_f}=1$. 

For the situation of interest, we consider $N_f$  hypermultiplets with equal mass $m$, and want to decouple them simultaneously. This can be achieved by multiplying all decoupling factors for each of the $N_f$ hypermultiplets, and using $m_{N_f}\Lambda_{N_f}^{4-N_f}=\Lambda_{N_f-1}^{5-N_f}$ for each `intermediate' decoupling. With $m^{N_f}\Lambda_{N_f}^{4-N_f}=\Lambda_0^4$, this gives the correlation function in the $N_f=0$ theory,
\begin{equation}
    \lim_{m\to\infty}\decoup\, 
    \langle \CO\rangle_{N_f} =\langle \CO\rangle_{0}~,
\end{equation}
with decoupling factor
\begin{equation}\label{decoup equal mass}
    \decoup=\left(\frac{\Lambda_{N_f}}{m}\right)^{\frac{N_f}{4}(K^2-5\chih)-(\bfk^2)}~,
\end{equation}
where we set $(\bfk^2)=\bfk_1^2+\dots+\bfk_{N_f}^2$.

\sss{Large mass expansion.}
By including the decoupling factor $\decoup$, we can significantly simplify the instanton contribution to the partition function $Z^{\rm i}=\langle \mathbbm 1\rangle^{\rm i}$~\eqref{O instanton} for equal masses $\bfm=(m,\dots, m)$. It becomes
\bea
\decoup Z^{\rm i}(m)= \sum_k \left(\frac{\Lambda_0}{m}\right)^{\vd \Mi}\!\!\!\!\!\!\!\sum_{\sum_j l_j=\vd\Mi}\int_{\Mi}\prod_{j=1}^{N_f} c_{l_j,j}~,
\eea
where we emphasise the mass dependence.
The selection rule $l_1+\dots l_{N_f}=\vd\Mi$ matches the degree of the Chern classes with the dimension of the moduli space. This can be more elegantly written in terms of the direct sum $W_k$~\eqref{W_k}, since $c(W_k)=\prod_{j=1}^{N_f} c(W_k^j)$ contains all the combinations of the $c_{l_j,j}$:
\bea\label{DZSWi}
\decoup Z^{\rm i}(m)= \sum_k \left(\frac{\Lambda_0}{m}\right)^{\vd \Mi}\!\int_{\Mi}c(W_k)~.
\eea
In Section~\ref{sec:universal functions from PF}, we match this series with generating functions of virtual Segre numbers. 
The inclusion of point and surface classes follows analogously.

\subsection{The effective theory}
\label{sec: contr singularities}

The calculation of topological correlation functions is significantly easier in the infrared, where the gauge group $\SU(2)$ is broken to ${\rm U}(1)$, and the field theory becomes Abelian. For $\SU(2)$ SQCD with $N_f$ flavours, the effective theory coupled to $N_f$ background fluxes can be modelled as a theory with gauge group ${\rm SU}(2) \times {\rm U}(1)^{N_f}$, where the fields of the U(1) factors are frozen in a specific way~\cite{Nelson:1993nf,Manschot:2021qqe,Aspman:2022sfj}. To start, we use the topological Lagrangian of a general rank $N_f+1$ theory, which is worked out explicitly in~\cite{Marino:1998bm}. The masses of the hypermultiplets are then set as the expectation values of the frozen scalar fields of the U(1) vector multiplets, $m_j\cong \langle\phi_j\rangle$, for $j=1,\dots, N_f$. The curvature for the U(1) components is $[F_j]=4\pi \bfk_j$, where $\bfk_j=\frac12 c_1(\CL_j)$, with $\CL_j$ as above.

\sss{$\boldsymbol{u}$-plane integral and SW contributions.}
For any four-manifold $X$ as before, the topological partition function of SQCD with generic masses $\bfm=(m_1,\dots, m_{N_f})$ takes the form of a sum of a $u$-plane integral $Z_u$ and a Seiberg--Witten contribution $Z_\SW$~\cite{Moore:1997pc},
\be \label{Z_contr}
Z(\bfm)=Z_u(\bfm)+Z_{\SW}(\bfm)~.
\ee 
The partition function depends in general on three distinct collections of parameters, some of which we suppress in the notation: The masses $\bfm$, the metric $J$ (which is the generator of $H_2^+(X,\mathbb R)$ normalised to $J^2=1$) and a set of fluxes for the theory (such as a 't Hooft flux $\bfmu$ for the gauge bundle and background fluxes $\bfk_j$ for the flavour group).
The $u$-plane integral $Z_u(\bfm)$ vanishes for manifolds with $b_2^+>1$~\cite{Moore:1997pc}.\footnote{An in-depth study of the $u$-plane integral for SQCD can be found in~\cite{Aspman:2022sfj,Aspman:2023ate}.} For $b_2^+=1$, the SW contribution can be found from the $u$-plane integral $Z_u(\bfm)$ by wall-crossing as a function of the metric $J$.
The SW contribution consists of contributions from each of the $2+N_f$ singularities on the Coulomb branch,
\bea\label{ZSW def}
Z_{\SW}(\bfm)=\sum_{j\in \{\pm, 1,\dots,N_f\}} Z_{\SW,j}(\bfm)~.
\eea

As explained in Section~\ref{sec:CB geometry}, for large and equal masses $m$, the Coulomb branch contains two singularities $\tu_{N_f}^\pm$ which become the monopole and dyon singularities of the $N_f=0$ theory for $m\to \infty$, together with an $I_{N_f}$ singularity $\tu_{N_f}^*$ that scales as $m^2$ for $m\to\infty$. The SW contribution then becomes
\bea\label{ZSW pm}
Z_{\SW}(m)=\sum_{\pm} Z_{\SW,\pm}(m)+Z_{\SW,\tu^*_{N_f}}\!(m)~.
\eea
In the decoupling limit, the monopole fields are integrated out and we obtain the ordinary instanton moduli space. Since $\tu_{N_f}^\pm$ become the $N_f=0$ singularities, we can therefore associate these contributions to the instanton component $\Mi$, while the remaining contribution from $\tu_{N_f}^*$ is associated with the monopole component $\Mm$. 
Writing these contributions as $Z_{\SW}=Z_{\SW}^{\rm i}+Z_\SW^{\rm m}$, we then denote by
\bea\label{Z_inst}
Z_{\SW}^{\rm i}(m)=\sum_{\pm} Z_{\SW,\pm}(m)~
\eea
the instanton component of the SW partition function. For $b_2^+>1$, it is the full contribution from the instanton component, while generically the full partition function~\eqref{ZSW def} also receives a contribution from $\tu^*_{N_f}$. Let us review the calculation of the contributions $Z_{\SW,\pm}$.

\sss{The $u$-plane integral.}
The $u$-plane integral is the contribution from the continuous family of vacua to partition functions. We summarise the $u$-plane integral for SQCD with background fluxes~\cite{Moore:1997pc, Aspman:2022sfj, Aspman:2023ate}, which, using the notation of~\cite[Sec. 5.1]{Aspman:2022sfj}, takes the following form,\footnote{Ref.~\cite{Aspman:2022sfj} considered implicitly the special case that $\bfx$ is orthogonal to $\bfk_j$, such that the mixed contact terms $H_{N_f,j}$ are absent.}
\be \label{u plane integral}
\begin{split}
\Phi^J_{\bfmu,\{\bfk_j\}}(p,\bfx,\bfm, \Lambda_{N_f})&=\int_{\CF_{N_f}(\bfm)}d\tau\wedge d\bar \tau\,\nu(\tau,\{\bfk_j\})\,\Psi_{\bfmu}^J(\tau,\bar \tau,\bfz,\bar \bfz)\\
&\times e^{2pu/\Lambda_{N_f}^2+\sum_j B(\bfx,\bfk_j)H_{N_f,j}+\bfx^2 G_{N_f}}~.
\end{split}
\ee 
Here, $\bfk_j$ are the background fluxes satisfying~\eqref{constraint_c1CL}, the domain $\CF_{N_f}(\bfm)$ is the modular fundamental domain for the theory~\cite{Aspman:2021vhs}, and the $u$-plane measure reads,
\be\label{nu measure}
\nu(\tau,\{\bfk_j\})=\mathcal{K}_{N_f}\frac{da}{d\tau} A^\chi B^\sigma \prod_{i,j=1}^{N_f} C_{ij}^{\bfk_i\bfk_j}~.
\ee
with $\CK_{N_f}=(\pi \Lambda_{N_f})^{-1}$ a normalisation factor.
Moreover, $\Psi^J_\bfmu$ is a Siegel-Narain theta series with elliptic variables specified by
\be\label{elliptic variable u-plane}
\begin{split}
&\bfz=\frac{\bfx}{2\pi \Lambda_{N_f}}\frac{du}{da}+\sum_{j=1}^{N_f} v_j\bfk_j~,\\
&\bar \bfz=\sum_{j=1}^{N_f} \bar v_j\bfk_j~.
\end{split}
\ee
The effective gravitational couplings have a universal form in terms of $da/du$ and the discriminant $\Delta$ of the SW-curve~\cite{Witten:1995gf, Moore:1997pc},
\bea\label{A B couplings}
A&=2^{\frac14}\Lambda_{N_f}^{-\frac12}\left(\frac{du}{da}\right)^{\frac12}~, \\
B&=2e^{\frac{\pi i}{8}N_f}\Lambda_{N_f}^{-\frac14(N_f+2)}\Delta_{N_f}^{\frac18}~.
\eea
The $\Lambda_{N_f}$-dependent prefactor is consistent with the decoupling limit from $N_f\to N_f-1$ hypermultiplets~\cite{Marino:1998uy, Aspman:2022sfj}. These couplings can also be derived in the $\Omega$-background~\cite{Manschot:2019pog}.

To determine the Seiberg--Witten contributions from the $u$-plane integral, we introduce the monopole moduli space.

\sss{The SW equations.}
The Seiberg--Witten/monopole equations describe supersymmetric configurations at the monopole singularities $\tu^\pm$, which consist each of an abelian vector multiplet whose gauge field is a $\spinc$ connection with curvature $F$, coupled to a massless hypermultiplet with section $M$~\cite{Witten:1994cg}:
\bea
F_+&=(\bar MM)_+~, \\
\slashed DM&=0~.
\eea
The moduli space of solutions to the SW equations with given $\spinc$ structure $c$ has complex dimension $n(c)=\frac 18(c^2-K^2)$.
The integral over the SW moduli space may be evaluated as a residue~\cite{Moore:1997pc},
\be 
\label{SWcont}
Z_{\text{SW},j}(\bfm)=\sum_c \Lambda_{N_f}^{n(c)}\,\SW(c)\,  \mathop{\mathrm{Res}}_{a_j=0}\,\left[ \frac{e^{-S_{\SW,j}}}{a_j^{1+n(c)}}\right]~,
\ee 
with local coordinate $a_j$ at the singularity $\tu_j$, and the sum is over SW basic classes $c$, \ie~the elements of $H^2(X,\mathbb{Z})$ for which $\SW$ is non-vanishing. The action $S_{\SW,j}$ near a singularity $\tu_j$ can be determined explicitly from a wall-crossing argument~\cite{Moore:1997pc,Aspman:2023ate}.
A four-manifold $X$ is of SW simple type if $n(c)=0$ for all basic classes $c$, that is, $c^2=K^2$. The residue~\eqref{SWcont} for such manifolds is simply the leading coefficient in the series at $a_j=0$.
In the following, we will assume that $X$ is of SW simple type.

\sss{Contribution from singularities.}
For generic masses, the contribution from an $I_1$ singularity $\tu_j$ can be evaluated by explicitly determining the action $S_{\SW,j}$, and it reads for topological SQCD~\cite{Aspman:2023ate}:
\begin{equation}\label{SWuniversal}
Z_{\SW,j}= 2\prod_{i,j=1}^s C_{ij}^{\bfk_i\bfk_j}\left(-2i\Lambda_{N_f}^5\tm\,\tx^7\right)^{\chih}\left(\Lambda_{N_f}^{-1}\tx^{-1}\right)^{K^2}\sum_c(-1)^{\bfmu(c+K)}\SW(c)  \prod_{j=1}^s D_j^{\bfk_j c }~.
\end{equation}
The couplings $C_{ij}$ and $D_j$ are defined in~\eqref{C D couplings} and evaluated at the singular point $\tu_j$, as in~\eqref{couplings tau=0}. The quantities $\tm$ and $\tx$ given in~\eqref{mu def} and~\eqref{dadu def}, which can be defined for generic masses. For equal masses $\bfm=(m,\dots, m)$, the number of independent couplings is reduced. As discussed above~\eqref{couplings tau=0}, the remaining couplings become series in the mass $m$ only, and thus power series labelled by $N_f$. Omitting these labels for now, the contribution from the singularity $\tu_{N_f}=\tu_{N_f}^-$~\eqref{upm} simplifies to
\begin{equation}\label{SW all Nf}
Z_{\SW,-}=2 \tC^{(\bfk^2)}\hat{\tC}^{2\br{\bfk^2}}\left(-2i\Lambda_{N_f}^5\tm\,\tx^7\right)^{\chih}\left(\Lambda_{N_f}^{-1}\tx^{-1}\right)^{K^2}\sum_c(-1)^{\bfmu(c+K)}\SW(c) \tD^{c (\bfk)}~,
\end{equation}
where we defined $(\bfk^p)\coloneqq\bfk_1^p+\dots+\bfk_{N_f}^p$ for $p=1,2$, and $\br{\bfk^2}\coloneqq \sum_{i<j}^{N_f}\bfk_i\bfk_j$. In the absence of background fluxes, $\bfk_i=0$, this matches with previous results~\cite{Moore:1997pc,Marino:1998uy}. 
The factors of $\Lambda_{N_f}$ in the bases of $\chih$ and $K^2$ are important to guarantee $Z_\SW$ being dimensionless: with $\tm$ having dimension $2$, $\tx$ having dimension $-1$, $Z_\SW$ is dimensionless.

\sss{Decoupling factor.}
In order to find agreement with results on Segre numbers, we need to include the decoupling factor~\eqref{decoup equal mass} for equal masses. The inclusion of the decoupling factor $\decoup$ is natural from the physical perspective as well, since we are interested in the large mass regime, where the infinite mass limit is only well-defined after the inclusion of $\decoup$.
Using  $\frac{\Lambda_{N_f}}{m}=\left(\frac{\Lambda_0}{m}\right)^{\frac{4}{4-N_f}}$,~\eqref{SW all Nf} and~\eqref{decoup equal mass} combine rather nicely,
\begin{equation}
\begin{aligned}
    \decoup Z_{\SW,-}&= 2(\tfrac{m}{\Lambda_{N_f}}\tC)^{(\bfk^2)}\hat{\tC}^{2\br{\bfk^2}}\left(-2i\Lambda_{0}^5\tm\tx^7\right)^{\chih}\left(\Lambda_{0}^{-1}\tx^{-1}\right)^{K^2} \\ 
    &\quad \times \sum_c(-1)^{\bfmu(c+K)}\SW(c) \tD^{c (\bfk)}~.
\end{aligned}
\end{equation}

\sss{Observables.}
Finally, we include a set of observables. For four-manifolds $X$ with $b_1(X)=0$, the observables of the topological theory consist of points and surfaces. Adding the point observable amounts to inserting $\exp(2pu/\Lambda_{N_f}^2)$ into the partition function. The inclusion of the surface observable gives rise to contact terms in the infrared~\cite{Moore:1997pc,LoNeSha}. For $x\in H^2(X,\mathbb Q)$ and a fixed basic class $c$, the insertion into the path integral for generic masses is~\cite{Moore:1997pc, Manschot:2021qqe}
\begin{equation}\label{observables}
   \CO[p,x]= \exp\left(x^2 G_{N_f}+(\bfk)x H_{N_f} -\tfrac{i}{2\Lambda_{N_f}\tx} cx +2p\tfrac{u}{\Lambda_{N_f}^2}\right)~,
\end{equation}
where the contact terms $G_{N_f}$ and $H_{N_f}$ are defined in~\eqref{contactterm}. This can be derived from the $u$-plane answer~\eqref{u plane integral} using the wall-crossing argument explained above.
The generating function of correlation functions for point and surface observables for $N_f=1,2,3$ equal masses then reads:
\begin{equation}\label{D ZSW p x}
\begin{aligned}
    \decoup Z_{\SW,-}[p,x]&=2\, (\tfrac{m}{\Lambda_{N_f}}\tC_{N_f})^{(\bfk^2)}\hat{\tC}_{N_f}^{2\br{\bfk^2}}\left(-2i\Lambda_{0}^5\tm_{N_f}\tx_{N_f}^7\right)^{\chih}\left(\Lambda_{0}^{-1}\tx_{N_f}^{-1}\right)^{K^2} \\ 
&\quad \times e^{x^2 \tG_{N_f}+(\bfk)x \tH_{N_f}+2p\tu/\Lambda_{N_f}^2} \\
&\quad \times \sum_c(-1)^{\bfmu(c+K)}\SW(c) \tD^{c (\bfk)}e^{(2i\Lambda_{N_f}\tx)^{-1} cx}~.
\end{aligned}
\end{equation}
As explained in~\cite{Aspman:2023ate,Manschot:2021qqe}, the decoupling affects the normalisation of observables as well, which we will discuss below. Before we present the main result, we require one final ingredient: the blowup formula.

\subsection{Blowup formula}
\label{sec:blowup}
In this section, we relate the $u$-plane integral for a four-manifold $X$ to the $u$-plane integral of its blowup $\widehat X$. Combination with results on the moduli spaces of $\CQ$-fixed equations provides non-trivial results for the physical couplings. See~\cite{Fintushel1996, Gttsche2024,Nakajima:2003pg,Nakajima:2005fg,Nakajima2011,Kim:2019uqw,Bonelli:2024wha,Bonelli:2025juz,Bonelli:2025bmt} for a selection of important results on the blowup equation.

\sss{Notation.}
Before we review the blowup formula for $\CN=2$ QCD, which generalises the arguments of Moore--Witten~\cite{Moore:1997pc}, let us set some notation. We denote by $C'$ the exceptional divisor of the blowup, which has self-intersection $C'^2=-1$ and is orthogonal to all curves pulled back from $X$ by the blowdown map $\phi\colon \widehat X\to X$. We have the following relations for the basic topological quantities,
\be
\begin{split}
 b_2(\widehat X)&=b_2(X)+1,\\
\chi(\widehat X)&=\chi(X)+1,\\
\sigma(\widehat X)&=\sigma(X)-1,\\
K_{\widehat X}&=\phi^*K_X+C'.
\end{split}
\ee
We let $\widehat \bfk=\phi^*\bfk+\bfk'$ for elements $\widehat \bfk\in H^2(\widehat X)$, $\bfk\in H^2(X)$ and $\bfk'$ proportional to the Poincar\'e dual of $C'$. We choose a basis for $H^2(\widehat X)$ such that $\widehat \bfk$ and other quantities are represented as
\be\label{components exceptional divisor}
\begin{split}
\widehat \bfk=(\bfk, k')~, \quad
 \widehat \bfk_j=(\bfk_j, k'_j)~, \quad
\widehat \bfx=(\bfx,x')~, \quad
\widehat \bfmu=(\bfmu,\mu')~, \quad
 \widehat \bfz= (\bfz, z')~,
\end{split}
\ee
with $k', k'_j, \mu'\in \frac{1}{2}\mathbb{Z}$,  $x'\in \mathbb{R}$ and $z'\in \mathbb{C}$. In order to compare moduli spaces on $X$ and its blowup, we consider gauge bundles $E$ on $X$ and $\widehat E$ on $\widehat X$ that satisfy $\int_{X} c_2(E)=\int_{\widehat X} c_2(\widehat E)$,
such that the instanton numbers $\widehat k$ and $k$ are related as 
\be\label{instanton numbers}
\widehat k = k +\mu'^2~.
\ee

\sss{The blowup factor in the UV.}
In Donaldson theory, the blowup formula relates the Donaldson invariants for the blown-up manifold $\widehat X$ in terms of those for $X$, in the limit where the area of the exceptional divisor $C'$ is small~\cite{Fintushel1996}. 
Since the Donaldson invariants are given by correlation functions of the Donaldson--Witten TQFT, we wish to relate correlation functions of the TQFT on $\widehat X$ through those on $X$.

Since the magnitude of the exceptional divisor can be made arbitrarily small, for $N_f=0$ the correlation function on $\widehat X$ equals that of $X$ for $x'\to 0$ and in the absence of an 't Hooft flux through $C'$, that is, $\mu'=0$. Indeed, the dimensions of instanton moduli spaces for $\widehat X$ and $X$ are equal.

In the case with $N_f$ hypermultiplets, the moduli spaces $\CM^{\CQ,N_f}_{k,\mu}$ are  relevant. The difference in the dimensions of the non-Abelian monopole moduli spaces is
\be\label{vdim = vdim 1}
\vd \CM^{\CQ,N_f}_{k,\mu}(X)- \vd \CM^{\CQ,N_f}_{\widehat k,\widehat \mu}(\widehat X)=-(4-N_f)(\mu')^2-\frac{N_f}{4}+\sum_{j=1}^{N_f} (k_j')^2~.
\ee
Let us consider choices of $k_j'$ such that the difference (\ref{vdim = vdim 1}) vanishes. If $\mu'=0$ we require $k'\in\frac12+\mathbb Z$. From $\sum_jk_j'^2=\frac{N_f}{4}$ we conclude that $k_j'^2$ is bounded from above by $\frac14$ for each $j$, that is, $k_j'=\pm\frac12$ for all $j=1,\dots, N_f$.
When $\mu'=\frac12$ we have $k_j'\in\mathbb Z$ for all $j$. The virtual dimension is invariant if $\sum_j k_j'^2=1$, which is only possible with $k_i'=\pm1$ for one $i=1,\dots, N_f$, while $k_j'=0$ for all others $j\neq i$. We thus find that~\eqref{vdim = vdim 1} vanishes if:
\bea\label{constraints blowup equations}
\mu'=0&\colon \qquad k_j'=\pm \tfrac12~,\, \forall j=1, \dots, N_f~, \\
\mu'=\tfrac12&\colon \qquad k_i'=\pm 1 \text{ for one $i$, and } k_j'=0~, \, \forall j\neq i~.
\eea
Consider the blowup $\widehat{\mathbb P}^2$ of the complex projective plane $\mathbb P^2$, with $N_f$ generic masses, $\mu'=0$, and background fluxes $k_1'=\dots=k_{N_f}'=\frac12$.
An intricate analysis of the moduli spaces demonstrates that the blowup formula admits a Taylor series with a gap~\cite[Theorem 2.1]{Nakajima2011}\footnote{See also~\cite{Gottsche:2010ig}. For the pure $N_f=0$ case, it is $\CO(x'^4)$ instead of $\CO(x'^3)$. In this case, it is an even function, such that the $x'$ and $x'^3$ terms are absent. The quadratic term $x'^2$ is zero as a consequence of the semi-classical vanishing of the contact term~\cite{Mari_o_1999,Edelstein:2000aj}.}
\bea
\label{Vanishing theorem UV}
\left\langle e^{I(\widehat \bfx)}\right\rangle_{\widehat X,\widehat \mu}=\left\langle e^{I(\bfx)}\right\rangle_{X,\mu}+\CO(x'^3)~.
\eea 
This is a striking relation since it means that for all instanton numbers $\widehat k=k$, the leading three coefficients in the $x'$-expansion are fixed. 

To determine all the universal functions, we also consider the case $\mu'=\frac12$. Then the dimension of the instanton moduli space \eqref{eq:dimMi} differs by 2, ${\widehat \CM}^{\rm i}_k=\CM^{\rm i}_k+2$. At the same time, the ranks of the index bundles are related by $|{\rm rk}(\widehat W_k^j)|=|{\rm rk}(W_k^j)|+1$, which increases the degree of the top Chern class by two. From the intersection number of $\int_{\widehat \CM^{\rm i}_k} c(\widehat W_k^j) e^{I(\bfx)}$, we thus deduce for $\mu'=\frac12$,
\bea
\label{Vanishing theorem UV 2}
\left\langle e^{I(\widehat \bfx)}\right\rangle_{\widehat X,\widehat \mu}=c\,\left\langle e^{I(\bfx)}\right\rangle_{X,\mu}+\CO(x')~.
\eea 
with $c$ a numerical constant. With the constraints \eqref{Vanishing theorem UV} and \eqref{Vanishing theorem UV 2}, we will be able to explicitly determine all required Coulomb branch functions.

\sss{The blowup factor in the IR.} 
The IR provides a complementary perspective. 
By topological invariance,  correlators are metric-independent, and one may rescale the metric as $g\to t^2g$. Blowing up a point on $X$ produces an impurity, which is small by assumption that the area of $C'$ is small. As a consequence, it appears as a local $\CQ$-invariant observable, supported at a point~\cite{Moore:1997pc}. In the TQFT, any $\CQ$-invariant 0-form observable is a holomorphic function $\CR_{\mu'}(u,x',k_j')$ of $u=\frac{1}{16\pi^2}{\rm Tr }\, \phi^2$, with $\phi$ the adjoint-valued complex scalar. Conjecturally, we then have:
\bea\label{UV corr}
\left\langle e^{I(\bfx)+I(x')}\right\rangle_{\widehat X,\widehat \mu}=\left\langle e^{I(\bfx)}\,\CR_{\mu'}(u,x',\{k_j'\})\right\rangle_{X,\mu}~.
\eea 

We can indeed deduce $\CR_{\mu'}(u,x',k_j')$ from the $u$-plane integral for $b_2^+=1$. As discussed in~\cite{Moore:1997pc,LoNeSha}, the surface observable $I(\bfx)$ flows to $\widetilde I(\bfx)+\bfx^2 G_{N_f}$ in the IR, with $G_{N_f}$ the contact term. The observable $u$ becomes a coordinate on the Coulomb branch with the same name, and we can change variables to the effective coupling $\tau$. The correlation functions can be expressed as $u$-plane integrals~\eqref{u plane integral}, such that~\eqref{UV corr} becomes
\bea\label{blowup u-plane integrand}
\int_\CF \CI_{\widehat \mu}^{\widehat X}[\widehat \bfz]=\int_\CF \CI_{\mu}^{ X}[\bfz]\,\CR_{\mu'}(\tau, x',\{k_j'\})~,
\eea
where $\CI_{ \mu}^{ X}$ is the $u$-plane integrand~\eqref{u plane integral} for $X$, and similarly for $\widehat X$. This allows to determine the blowup factor
\bea\label{blowup def}
\CR_{\mu'}(\tau, x',\{k_j'\})=\frac{\CI_{\widehat \mu}^{\widehat X}[\widehat \bfz]}{\CI_{\mu}^{ X}[\bfz]}~.
\eea
In order to compute it, we need the ratio of the sum over fluxes $\Psi_{\mu}^J$ for $X$ and $\widehat X$, the latter we denote by $\widehat \Psi_{\widehat \mu}^{\widehat J}$.  If we choose $\widehat J=(J,0)$, $\widehat \Psi_{\widehat \bfmu}^{\widehat J}$ factorises as 
\be
\widehat \Psi_{\widehat \bfmu}^{\widehat J}(\tau,\bar \tau,\widehat \bfz, \widehat {\bar \bfz})=\left\{\begin{array}{rr} \Psi_{\bfmu}^J(\tau,\bar \tau,\bfz,\bar \bfz)\,\vartheta_4(\tau,z')~, & \qquad \mu'=0~, \\ 
\Psi_{\bfmu}^J(\tau,\bar \tau,\bfz,\bar \bfz)\,\vartheta_1(\tau,z')~,& \qquad \mu'=\frac{1}{2}~,\end{array}\right.
\ee
where the Jacobi theta functions $\jt_\ell$ are defined in Appendix~\ref{app:modularforms}.
We can then evaluate the blowup factor~\eqref{blowup def}: 
\be\label{blowup R}
\begin{split}
\CR_{\mu'}(\tau, x',\{k_j'\})&=iAB^{-1} \prod_{i,j=1}^{N_f} C_{ij}^{-k_i'k_j'}\times \exp\left(-x'\textstyle\sum_j k'_jH_{N_f,j}-x'^2\,G_{N_f}\right) \\
&\quad  \, \times \vartheta_\ell\!\left(\tau,\tfrac{x'}{2\pi \Lambda_{N_f}}\tfrac{du}{da}+\textstyle\sum_{j}k_j' v_j \right),
\end{split}
\ee
where $\ell=4$ if $\mu'=0$, and $\ell=1$ if $\mu'=\frac{1}{2}$. By including the background fluxes and arbitrary masses, this result generalises previous findings in the literature~\cite{Moore:1997pc,LoNeSha,Lossev:1997bz,Marino:1998bm,Takasaki:1998vm,Mari_o_1999,Edelstein:2000aj, Nakajima:2003pg,Nakajima2011, Kim:2019uqw, Gottsche:1996aoa}.\footnote{We can collect the terms in the exponent together with the theta function as follows: Let $\Pi=(x',k_1',\dots, k_{N_f}')$ and $L=(\Lambda_{N_f},m_1,\dots, m_{N_f})$ be two $N_f+1$ component vectors. The blowup factor~\eqref{blowup R} then reads, schematically, $\CR\sim iAB^{-1}\exp\left(\Pi^iL^j\frac{\partial^2\CF}{\partial L^i\partial L^j}\right)\jt_l\left(\tau,\Pi^j\frac{\partial^2\CF}{\partial a\partial L^j}\right)$.} We note also that the blowup function is (anti-)symmetric under the reflection of the fluxes $k_j'$ and the fugacity $x'$, 
\bea\label{CR symmetry}
\CR_{\mu'}(\tau, x',\{-k_j'\})=(-1)^{4\mu'^2}\CR_{\mu'}(\tau,-x',\{k_j'\})~.
\eea
Moreover, $\CR_{\mu'}$ is independent of $\bfk$, $\bfmu$ and $\bfx$. 

We explore next the constraints from the UV identity~\eqref{Vanishing theorem UV} for the IR, with the purpose of determining the unknown couplings $C_{ij}$, $H_{N_f,j}$ and $G_{N_f}$. In the equal mass case, these are three unknown functions for $N_f=1$ and four unknown functions for $N_f=2,3$. To determine them, we can specialise to a specific four-manifold and choices of $k_j'$ for which the moduli spaces are sufficiently well understood. We choose the blowup of the projective plane, $\phi: \mathbb{\widehat P}^2 \to \mathbb{P}^2$, for which we know the identity \eqref{Vanishing theorem UV} from the UV. 

For this, we study the Taylor series of~\eqref{blowup R} in $x'$ for these special choices of fluxes $k_j'$ and find non-trivial identities involving the unknown couplings, which we are able to solve for explicitly. In the following, we present a detailed derivation of these couplings, which we express in terms of elliptic and modular functions. The results are~\eqref{GNf HNf solutions} for $G_{N_f}$ and $H_{N_f}$, and~\eqref{C11 C12 solutions} for the $C_{ij}$. Because \eqref{Vanishing theorem UV} has to hold for all instanton numbers $\widehat k$, we deduce that 
\bea\label{Vanishing theorem}
\CR_0(\tau, x',\{\tfrac12\})=\begin{cases}
1+\CO(x'^4)~, &  \quad N_f=0~, \\
1+\CO(x'^3)~,  & \quad N_f=1,2,3~.
\end{cases}
\eea
Similarly, from \eqref{Vanishing theorem UV 2}, we deduce for $\mu'=\frac12$, $k'_i=\pm 1$ for one $i$, and $k'_j=0$ for $i\neq j$,
\bea\label{CR Taylor}
\CR_{\frac12}(\tau, x',k_j')=c+\CO(x')~,
\eea
for some complex number $c$, and any admissible fluxes $\mu'$ and $k_j'$.

\sss{Contact terms from blowup equation.}
Eq. \eqref{Vanishing theorem} gives us three equations which are valid for all values of $\tau$ and all $N_f=1,2,3$ with generic masses:
\begin{subequations}\label{3 equations generic ki}
    \begin{align}
        -iA^{-1}B\prod_{j,k}C_{jk}^{\frac14}&=\jt_4(\tau,\tfrac12\textstyle\sum_j v_j)~, \\
 \pi \Lambda_{N_f}\tfrac{da}{du}\sum_jH_{N_f,j}&=\jt_4^{(1)}(\tau,\tfrac12\textstyle\sum_j v_j)~, \\
 8 \left(\pi\Lambda_{N_f}\!\tfrac{da}{du}\right)^2G_{N_f}&=\jt_4^{(2)}(\tau,\tfrac12\textstyle\sum_j v_j)~.
    \end{align}
\end{subequations}
Here, we use the shorthand $\jt_4^{(n)}(\tau,z_0)\coloneqq \partial_z^n \log \jt_4(\tau,z)\big|_{z=z_0}$ for the logarithmic derivatives of $\jt_4$.\footnote{The logarithmic derivatives are $\jt_4^{(1)}(\tau,z)=\frac{\jt_4'(\tau,z)}{\jt_4(\tau,z)}$ and $\jt_4^{(2)}(\tau,z)=\frac{\jt_4''(\tau,z)}{\jt_4(\tau,z)}-\frac{\jt_4'(\tau,z)^2}{\jt_4(\tau,z)^2}$, where the derivatives are all w.r.t. the elliptic variable $z$.} 
It is customary to rewrite these derivatives using \ws{} elliptic functions, which in the above order correspond to the \ws{} $\sigma$, $\zeta$ and $\wp$ functions.\footnote{See Appendix~\ref{app:modularforms} for a summary, and~\cite{erdelyiII} for a comprehensive review} In Appendix~\ref{app:v as period integral}, we give an independent derivation of the third equation using period integrals, expressing $v_j$ in terms of the SW curve and differential.

We can specialise to equal masses, such that $C_{jk}$ are determined by two functions only.
There are $N_f$ diagonals $C_{ii}$ and $N_f(N_f-1)$ off-diagonals $C_{ij}$, $i\neq j$, which become the series $\tC_s$ and $\hat \tC_s$ at strong coupling.  Meanwhile, all contact terms $H_{N_f,j}$ become equal and are denoted by $H_{N_f}$. The same holds for the $v_j$, which we denote by $v$. Then the above equations become:
\begin{subequations} \label{3 equations}
    \begin{align}\label{3 equations a} 
-iA^{-1}BC_{11}^{\frac{N_f}{4}}C_{12}^{\frac{N_f(N_f-1)}{4}}&=\jt_4(\tau,\tfrac{N_f}{2}v)~, \\ \label{3 equations b}
 \pi  \Lambda_{N_f}\tfrac{da}{du}N_f H_{N_f}&=\jt_4^{(1)}(\tau,\tfrac{N_f}{2}v)~, \\
 \label{3 equations c}
 8 \left(\pi\Lambda_{N_f}\!\tfrac{da}{du}\right)^2G_{N_f}&=\jt_4^{(2)}(\tau,\tfrac{N_f}{2}v)~.
    \end{align}
\end{subequations}
Equations~\eqref{3 equations b}--\eqref{3 equations c} express the contact terms $G_{N_f}$ and $H_{N_f}$ for all $N_f=1,2,3$ through the coupling $v$:
\bea\label{GNf HNf solutions}
G_{N_f}&=\frac{\jt_4^{(2)}(\tau,\tfrac{N_f}{2}v)}{8 \left(\pi\Lambda_{N_f}\!\tfrac{da}{du}\right)^2}~, \\
H_{N_f}&=\frac{\jt_4^{(1)}(\tau,\tfrac{N_f}{2}v)}{\pi  \Lambda_{N_f}\tfrac{da}{du}N_f}~.
\eea

\sss{Couplings $C_{ij}$ from blowup equation.}
For $N_f=1$, $C_{12}$ does not exist, and~\eqref{3 equations a} can be solved for $C_{11}$.  For $N_f=2$ and $3$, we need another equation involving both $C_{11}$ and $C_{12}$. For this, we choose $\mu'=0$ with $k_i'=\frac12$ for all $i$ except $k_{N_f}'=-\frac12$. Then the blowup function has constant term 1 if 
\bea\label{1/2 -1/2}
N_f&=2\colon \qquad -iA^{-1}BC_{11}^{\frac12}C_{12}^{-\frac12}=\jt_4(\tau,0)~, \\
N_f&=3\colon \qquad-iA^{-1}BC_{11}^{\frac34}C_{12}^{-\frac12}=\jt_4(\tau,\tfrac v2)~.
\eea
Together with~\eqref{3 equations a}, these allow to solve for $C_{11}$ and $C_{12}$ individually. Although this explicitly expresses all the remaining couplings in terms of the Seiberg--Witten curve and $v$, we can simplify the result even further. In order to eliminate the dependence on the gravitational couplings $A^{-1}B$, we consider the flux $\mu'=\frac{1}{2}$. In line with~\eqref{constraints blowup equations}, we set $k_i'=1$ for only one $i$, while all others are $k_j'=0$. The constant term of the Taylor series~\eqref{CR Taylor} for $\CR_{\frac12}$ is $c$ if $-iA^{-1}BC_{ii}=c^{-1}\, \jt_1(\tau,v_i)$.
Verification with the prepotential demonstrates that this indeed holds for $N_f=1,2,3$ generic masses with $c=1/\sqrt{2}$.
Specialising to equal masses, all diagonal couplings $C_{ii}$ become $C_{11}$, and all couplings $v_i$ become equal to $v$. Then for equal masses we have
\bea\label{ABC=theta_1}
-iA^{-1}BC_{11}=\sqrt2 \jt_1(\tau,v)~,
\eea
for all $N_f=1,2,3$.
Combining this result~\eqref{ABC=theta_1} with~\eqref{3 equations a} and~\eqref{1/2 -1/2} gives for $N_f=1,2,3$ with equal masses:\footnote{Here, $4-N_f$ and $N_f-1$ are exponents}
\bea\label{C11 C12 Nf general}
C_{11}^{4-N_f}&=\frac{4\jt_1(\tau,v)^4}{\jt_4(\tau,\tfrac{N_f}{2}v)^{4-N_f}\jt_4(\tau,(\frac{N_f}{2}-1)v)^{N_f}}~,  \\
C_{12}^{N_f-1}&=\frac{\jt_4(\tau,\frac{N_f}{2}v)}{\jt_4(\tau,(\frac{N_f}{2}-1)v)}~.
\eea
We summarise all these relations:
\begin{subequations}\label{C11 C12 solutions}
    \begin{align} \label{C11 C12 Nf=1}
        N_f=1\colon \qquad C_{11}&=4^{\frac13}\frac{\jt_1(\tau,v)^{\frac43}}{\jt_4(\tau,\tfrac v2)^{\frac43}}~, \\
 \label{C11 C12 Nf=2}       
N_f=2\colon \qquad C_{11}&=2\frac{\jt_1(\tau,v)^2}{\jt_4(\tau,v)\jt_4(\tau,0)}~, &C_{12}&=\frac{\jt_4(\tau,v)}{\jt_4(\tau,0)}~, \\
\label{C11 C12 Nf=3}
N_f=3\colon \qquad C_{11}&=4\frac{\jt_1(\tau,v)^4}{\jt_4(\tau,\tfrac v2)^3\jt_4(\tau,\tfrac32v)}~, \!\!\!\!\!\!\!\! &C_{12}&=\frac{\jt_4(\tau,\tfrac32v)^{\frac12}}{\jt_4(\tau,\tfrac v2)^{\frac12}}~.
    \end{align}
\end{subequations}
Viewed as elliptic functions, both $C_{11}$ and $C_{12}$ for all $N_f=1,2,3$ are Jacobi forms of weight 0 and index $\frac12$.

Using~\eqref{dadu Delta} together with~\eqref{A B couplings}, we can use~\eqref{ABC=theta_1} to express $\frac{da}{du}$ in terms of modular forms:
\bea\label{dadu in terms of theta}
N_f=1\colon \qquad \frac{da}{du}&=\frac{2^{\frac16}i\eta^3 \jt_1(\tau,v)^{\frac13}}{\Lambda_1 \jt_4(\tau,\tfrac v2)^{\frac43}}~, \\
N_f=2\colon \qquad \frac{da}{du}&=\frac{2^{\frac12}i\eta^3 \jt_1(\tau,v)}{\Lambda_2 \jt_4(\tau,0)\jt_4(\tau,v)}~, \\
N_f=3\colon \qquad \frac{da}{du}&=\frac{2^{\frac32}i\eta^3 \jt_1(\tau,v)^3}{\Lambda_3 \jt_4(\tau,\frac12 v)^3\jt_4(\tau,\frac32v)}~,
\eea
with $\eta$ the Dedekind $\eta$-function~\eqref{dedekind eta}.
Inserting into~\eqref{GNf HNf solutions}, we find simple expressions for $G_{N_f}$ and $H_{N_f}$ in terms of theta functions of $v$ only. 

For $N_f=1$, the relations~\eqref{3 equations} were derived in~\cite[Section 6.4]{Gottsche:2010ig}, which determines the functions $H_1$, $G_1$ and $A^{-1}BC_{11}^{\frac14}$. 
Moreover for $N_f=2$, $C_{11}$ and $C_{12}$ were conjectured in~\cite[(E.16)]{Aspman:2023ate}. To the best of our knowledge, the other explicit expressions are new, \ie~\eqref{GNf HNf solutions} for $G_{N_f}$ and $H_{N_f}$, as well as~\eqref{C11 C12 Nf=1} and~\eqref{C11 C12 Nf=3} for $C_{11}$ and $C_{12}$, and Eq.~\eqref{dadu in terms of theta} for $\frac{da}{du}$.\footnote{Similar relations are known for the $\SU(2)$ $\nstar$ theory~\cite{Manschot:2021qqe} and the 4d $\CN=2$ Kaluza--Klein theory obtained from a circle compactification of 5d $\CN=1$ $\SU(2)$ SYM~\cite{Kim:2025fpz,Gottsche:2006}.}

It turns out that the number of inequivalent identities obtained from the blowup equation is precisely equal to the number of unknowns to be solved for. In general, we have obtained 5 equations for the 5 functions $A^{-1}B$, $G_{N_f}$, $H_{N_f}$, $C_{11}$ and $C_{12}$ to be determined for equal masses.\footnote{For the flux $\mu'=\frac12$ we only get one equation due to~\eqref{constraints blowup equations}, while for $\mu'=0$ we obtain the 3 equations~\eqref{3 equations}. The remaining equation comes from the fluxes $k_j'$ of distinct signs, $k_j'=\pm\frac12$. Due to the symmetry of $\jt_4$ in its elliptic argument, the distinct configurations with $k_j'=\pm\frac12$ are the number of $N_f$-tuples, up to permutation and reflection $k_j'\mapsto -k_j'$. As one easily checks, for $N_f=2$ and $N_f=3$ there are only two independent such configurations (with 0 and 1 minus signs, up to reflection). For $N_f=1$, $C_{12}$ does not exist, and accordingly there is only one choice of fluxes.} We have thus exhausted all possible equations from the blowup function.

\sss{The blowup formula to all orders.}
As a consequence of the above analysis, we are able to determine the blowup formula to all orders in $x'$.
In the equal mass case, we find by inserting~\eqref{GNf HNf solutions}, \eqref{ABC=theta_1} and \eqref{C11 C12 Nf general} into~\eqref{blowup R}:
\bea\label{equal mass generic ki}
\CR_{\mu'}(\tau,x',\{k_j'\})&=\exp\left(-\tfrac{2}{N_f}(k’)\jt_4^{(1)}(\tfrac{N_f}{2}v)\xi-\tfrac12\jt_4^{(2)}(\tfrac{N_f}{2}v)\xi^2\right)\left(\sqrt2\jt_1(v)\right)^{\frac{N_f-4(k’^2)}{4-N_f}} \\
&\times \jt_4(\tfrac{N_f}{2}v)^{(k’^2)-\frac{2\br{k’^2}}{N_f-1}-1} 
\jt_4\left((\tfrac{N_f}{2}-1)v\right)^{\frac{N_f((k’^2)-1)}{4-N_f}+\frac{2\br{k’^2}}{N_f-1}}\jt_\ell(\xi+v(k’))~,
\eea
where we rescale $\xi=x'/(2\pi \Lambda_{N_f}\tfrac{da}{du})$, and abbreviate the notation $(k’^p)\coloneqq k_1'^p+\dots+k_{N_f}'^p$ for $p=1,2$, and $\br{k’^2}\coloneqq \sum_{i<j}^{N_f}k_i'k_j'$, as before. For $N_f=1$, the terms $\br{k’^2}$ are absent. We also suppress the $\tau$ argument of $\jt_i$ for a cleaner presentation.

While the blowup function $\CR_{\mu'}$ is a well-defined function for any admissible fluxes $k_j'$ and $\mu'$, it simplifies substantially in the case where the virtual dimension of the $\CQ$-fixed equations is preserved under the blowup. By plugging in the equal mass solutions~\eqref{GNf HNf solutions} for $G_{N_f}$ and $H_{N_f}$ together with the combination of couplings $A$, $B$ and $C_{ij}$ being fixed by~\eqref{CR Taylor}, one can show that under the assumptions~\eqref{constraints blowup equations} for the $k_j'$, the blowup factor simplifies to\footnote{The contact term for pure $\CN=2$ $\SU(N)$ SYM was derived explicitly in~\cite{Edelstein:2000aj}. Since we can view $\SU(2)$ SQCD as a special rank $N_f+1$ theory, it is natural to ask whether our formulas can be derived from the $\SU(N)$ result. Due to the inclusion of non-trivial background fluxes $k_j'$ along the exceptional divisor, one can check however that our formula is not a special case of~\cite[(2.20)]{Edelstein:2000aj}.}
\bea\label{CR mu result}
\CR_{\mu'}(\tau, x', \{k_j'\})&= c_{\mu'}\exp\left(-\tfrac{2}{N_f}\textstyle\sum_jk_j'\jt_4^{(1)}(\tfrac{N_f}{2}v)\xi-\tfrac12\jt_4^{(2)}(\tfrac{N_f}{2}v)\xi^2\right) \\
&\times\frac{\jt_\ell(\xi+v\textstyle\sum_jk_j')}{\jt_\ell(v\textstyle\sum_jk_j')}~,
\eea
where $c_{\mu'}=1$ for $\mu'=0\mod 1$ and $c_{\mu'}=1/\sqrt2$ for $\mu'=\frac12\mod 1$, and $\ell=4$ for $\mu'=0$ and $\ell=1$ for $\mu'=\frac12$, as before.\footnote{Let us sketch the proof that~\eqref{CR mu result} is indeed a special case of~\eqref{equal mass generic ki}. For $\mu'=0$ we thus consider the case where $(k’^2)=\frac{N_f}{4}$, such that the factor $\sqrt2\jt_1(v)$ in~\eqref{equal mass generic ki} vanishes. Due to the symmetry~\eqref{CR symmetry} of $\CR_0$ in the fluxes $k_j'$, we can restrict to the case where the sum over fluxes is positive, $(k’)\geq 0$. These are at most two cases, and the other cases follow by symmetry. When $(k’)=\frac{N_f}{2}$ we have $\br{k’^2}=\frac{1}{8}N_f(N_f-1)$, such that $\jt_4\left((\tfrac{N_f}{2}-1)v\right)$ disappears, and we obtain~\eqref{CR mu result}. When $(k’)=\tfrac{N_f}{2}-1$, then $\br{k’^2}=\frac 18(N_f-1)(N_f-4)$, such that the $\jt_4(\tfrac{N_f}{2}v)$ factor disappears. An analogous argument holds for $\mu'=\frac12$.}

This elucidates the statements~\eqref{CR Taylor} and~\eqref{Vanishing theorem} about the leading coefficients of $\CR_{\mu'}$: Clearly, $\CR_{0}$ has vanishing linear term only if $\sum_jk_j'=\frac{N_f}{2}$, that is, $k_j'=\frac12$ for all $j=1,\dots, N_f$. 
In that most symmetric case, we are able to test if this result can be improved. For $N_f=1$, using \eg~\eqref{dadu in terms of theta} we compute
$\CR_0(\tau,x',\frac12)=1+\frac{i}{6\sqrt2}x'^3+\CO(x'^4)$. This demonstrates that the statement~\eqref{Vanishing theorem} is indeed sharp.
In the case $\mu'=\frac12$, the linear term is $\jt_1^{(1)}(\tfrac{N_f}{2}v)-\jt_4^{(1)}(\tfrac{N_f}{2}v)$, which does not vanish. We conclude that these statements are indeed sharp:  $\CR_{\mu'}(\tau,x',k_j')=1+\CO(x'^3)$ if $\mu'=0$ and $k_j'=\frac12$ for all $j=1,\dots, N_f$, and it is $c_{\frac12}+\CO(x')$ otherwise.\footnote{An accidental improvement is the equal mass case for $N_f=2$, where $\CR_{0}(\tau,x',\{\tfrac12,-\tfrac12\})=1+\CO(x'^2)$ due to a trivial linear coefficient $\jt_4'(0)=0$. Similarly, for $N_f=0$ we have $\CR_{0}(\tau,x')=1+\CO(x'^4)$ since $\jt_4^{'''}(0)=0$, in agreement with~\cite[Theorem 2.1]{Nakajima2011}.
}

The result~\eqref{equal mass generic ki} is valid for arbitrary fluxes $\mu'$ and $k_j'$, but only for equal masses. For $N_f=1,2,3$ generic masses, we can write down a similar result, for instance with $\mu'=0$ and background fluxes $k_i=\frac12$. By inserting~\eqref{3 equations generic ki} into~\eqref{blowup R}, we find
\bea\label{CR0 result}
\CR_0(\tau,x', \{\tfrac12\})=\exp\left(-\jt_4^{(1)}(\tfrac12\textstyle\sum_jv_j)\xi-\tfrac12\jt_4^{(2)}(\tfrac12\textstyle\sum_jv_j)\xi^2\right)\frac{\jt_4(x'+\tfrac12\textstyle\sum_jv_j)}{\jt_4(\tfrac12\textstyle\sum_jv_j)}~.
\eea
 The gap condition~\eqref{Vanishing theorem} then follows by direct inspection.

\section{Universal functions from topological partition functions}
\label{sec:universal functions from PF}
In this section, we relate the topological partition functions of four-manifolds to the generating function of virtual Segre numbers for algebraic varieties. To this end, we specialise the four-manifold $X$ to a smooth projective surface $S$ over $\mathbb{C}$. Section \ref{sec: segre numbers} reviews the results of~\cite{Gottsche:2020ass,Gottsche:2019meh,Gttsche2024} for Segre numbers of $S$, in particular those corresponding to rank 2 sheaves, or gauge group $\SU(2)$. Section \ref{sec:comparison} identifies the algebraic data, such as Chern classes, in terms of the physical data, such as background fluxes. The rest of the section further brings the generating functions of Ref.~\cite{Gottsche:2020ass} to a form which makes comparison with the topological partition functions in Section \ref{sec:deriving_alg_functions} straightforward.

\subsection{Segre numbers and universal functions}
\label{sec: segre numbers}

\sss{Virtual Segre numbers.}
Let $S$ be a smooth projective surface over $\mathbb C$, with $b_1(S)=0$ and geometric genus $p_g(S)>0$.\footnote{We follow the definitions and notation of Göttsche and Kool~\cite{Gottsche:2020ass,Gottsche:2019meh,Gttsche2024}.} The geometric genus counts the number of independent holomorphic 2-forms on $S$, which is non-zero if  $b_2^+(S)>1$. The Grothendieck group $K(S)$ of coherent sheaves on $S$ is the free abelian group generated by isomorphism classes of coherent sheaves on $S$, with relations coming from short exact sequences.
The stability condition for sheaves on $S$ is determined through a polarisation $H$ on $S$.

Consider an integer $\rho\in\mathbb Z_{>0}$ and algebraic classes $c_1\in H^2(S,\mathbb Z)$ and $c_2\in H^4(S,\mathbb Z)$. We denote by $M\coloneqq M_S^H(\rho,c_1,c_2)$ the moduli space of rank $\rho$ Gieseker $H$-semistable sheaves on $S$ with Chern classes $c_1$, $c_2$. Ref.~\cite{Gottsche:2020ass} moreover requires that $M_S^H$ does not  contain strictly semistable sheaves with respect to $H$, such that $(\rho,c_1,c_2)$ is required to be a primitive vector. As $M$ may be singular, from obstruction theory one obtains a virtual (or expected) dimension
\begin{equation} \label{vdim M}
    \vd M=2\rho c_2-(\rho-1)c_1^2-(\rho^2-1)\chih~.
\end{equation}
On $M$, there exists a well-defined virtual fundamental class $M^{[\rm vir]}\in H_{2\vd M}(M)$. 

In analogy with Donaldson theory, one defines Chern classes on $M$ using a universal sheaf on $S\times M$. Given a class $\alpha \in K(S)$, one defines a tautological class $\alpha_M$ on the moduli space $M$, with Chern character ${\rm ch}(\alpha_M)$.\footnote{Explicitly, ${\rm ch}(\alpha_M)={\rm ch}(-\pi_{M!}(\pi^*_S\alpha\cdot E \cdot {\rm det}(E)^{-1/\rho}))$, with $E$ the sheaf with Chern character $(\rho,c_1,c_2)$ associated to the gauge principal bundle~\cite{Gottsche:2020ass}.} From this Chern character, we require the total Chern class $c(\alpha_M)=\sum_i c_i(\alpha_M)$ corresponding to ${\rm ch}(\alpha_M)$. Virtual Segre numbers are then defined as the integrals of $c(\alpha_M)$ over $M^{[\rm vir]}$.

For the point and surface classes familiar from Donaldson theory, write $\point\in H^4(S,\mathbb Z)$ for the Poincaré dual of the point class, and $p'$ a formal variable. For the surface observable, we pick $L\in\mathrm{Pic}(S)$, and keep track of insertions using the formal variable $x'$.
The main objective is then the computation of virtual Segre numbers of $M$ with fixed $(\rho,c_1,c_2)$,
\begin{equation}\label{virtual Segre def}
    \segre_\alpha[p',x']\coloneqq\int_{M^{[\rm vir]}}c(\alpha_M)  \exp(\mud(\point)p'+\mud(L)x')~,
\end{equation}
with $\mud$ the usual Donaldson map.
Fixing the rank $\rho$ and the first Chern class $c_1$ but summing over $c_2$ furnishes the generating function of virtual Segre numbers
\bea\label{segre gen}
\segregen_\segre[p',x']=\sum_{c_2}z^{\frac{1}{2} \vd M}\segre_\alpha[p',x']~.
\eea

For $\alpha=0$, the virtual Segre numbers reduce to ${\rm SU}(\rho)$ Donaldson invariants, and the conjecture below reduces to an algebraic version of the Mariño--Moore conjecture~\cite{Marino:1998bm}. 
On the other hand for $\rho=1$ and $c_1=0$, $M=M_S^H(1,0,n)\cong S^{[n]}$ becomes the Hilbert scheme of $n$ points on $S$ with total Chern class  $c(\alpha_M)=c(\alpha^{[n]})$, and the virtual Segre numbers of $S$ as a consequence are ordinary Segre numbers.\footnote{\label{footnote Segre}The relation to Segre classes is as follows. On the Hilbert scheme $S^{[n]}$ of $n$ points on $S$, we have a tautological class $\alpha^{[n]}\in K(S^{[n]})$. The coefficients of the generating series $\sum_{n=0}^\infty z^n \int_{S^{[n]}} c(\alpha^{[n]})\in \mathbb Q[[z]]$ are called \emph{Segre numbers} of $S$. 
Here, $c$ denotes the total Chern class, and $ \mathbb Q[[z]]$ refers to the ring of formal power series in the variable $z$ with coefficients in $\BQ$. For $\alpha=-V$ , where $V$ is the class of a vector bundle, $c(\alpha^{[n]})=s(V^{[n]})$, where $s$ denotes the total Segre class. }

Before we state the main conjecture, recall that there are two conventions for the Seiberg--Witten invariants. Mochizuki's convention~\cite{Mochizuki2009} is used in algebraic geometry, which we denote by $\sw(a)$ for some class $a\in H^2(S,\mathbb Z)$. The finitely many elements $a\in H^2(S,\mathbb Z)$ such that $\sw(a)\neq 0$ are called SW basic classes, and they satisfy $a^2=aK$, where $K$ is the canonical class of $S$. The convention used in differential geometry and physics we denote by $\SW(c)$ with $c\in H^2(S,\mathbb Z)$. The SW basic classes satisfy $c^2=K^2$, and they are related by (see \eg~\cite[Proposition 1.4.5]{Mochizuki2009})
\begin{equation}\label{SW shift}
    \sw(a)= \SW(2a-K)=\SW(c)~.
\end{equation}

Let us state the Göttsche--Kool conjecture, adjusted to rank $\rho=2$~\cite[Conjecture 2.8]{Gottsche:2020ass}:
\begin{conjecture}[Göttsche--Kool]\label{conj_GK}
Let $M$ be as above, and $s\in\mathbb Z$. There exist universal functions
\bea\label{Cz Cz12}
Q_s, R_s, T_s, V_s, W_s, X_s  &\in \mathbb C[[z]]~, \\
S_s, S_{1,s}, Y_s, Y_{1,s}, Z_s,  Z_{11,s} &\in \mathbb C[[z^{\frac{1}{2}}]]~,
\eea
such that for any $\alpha\in K(S)$ with $\rk(\alpha)=s$ we have 
{\normalfont
\bea\label{GK conjecture}
\segre_\alpha[p',x']=\universal_\alpha[p',x'] \Big|_{z^{\frac12\vd M}}~,
\eea
}
where 
{\normalfont
\bea\label{universal series}
\universal_\alpha[p',x']& \coloneqq 2^{2-\chih+K^2}V_s^{c_2(\alpha)}W_s^{c_1(\alpha)^2}X_s^{\chih} Y_s^{c_1(\alpha)K}Z_s^{K^2}e^{x'^2 Q_s + c_1(\alpha)x' R_s + x'K S_{s}  +  p'\, T_s} \\
&\quad \times \sum_a (-1)^{a c_1}\, \sw(a) Y_{1,s}^{c_1(\alpha)a}e^{a x' S_{1,s}} Z_{11,s}^{a^2}~,
\eea
}
and 
\bea\label{changeofvar}
z = t (1+(1-\tfrac{s}{2}) t)^{1-\frac{s}{2}}~.
\eea
Here, 
\bea\label{Vs Ws Xs Qs Rs Ts} 
V_s(t) &= (1+(1-\tfrac{s}{2})t)^{2-s} (1+(2-\tfrac{s}{2})t)^s~, \\
W_{s}(t)&= 2\left(2+(2-s)t\right)^{-\frac32+\frac12s}\left(2+(4-s)t\right)^{\frac12-\frac12s}~, \\
X_{s}(t)&= (1+(1-\tfrac{s}{2})t)^{\frac12s^2-\frac54s}(1+(2-\tfrac{s}{2})t)^{-\frac12 s^2+\frac12}(1+(1-\tfrac{s}{2})(2-\tfrac{s}{2})t)^{-\frac12}~, \\
Q_s(t) &= \tfrac{1}{2}t(1+(1-\tfrac{s}{2}) t)~, \quad  R_s(t) = t~, \quad 
T_{s}(t) = 2 t (1+ \tfrac{1}{2}(1-\tfrac{s}{2})(2-\tfrac{s}{2})t)~, \\
\eea
and moreover $Y_s$, $Y_{1,s}$,  $Z_s$, $Z_{11,s}$, $S_s$, $S_{1,s}$ are all algebraic functions.
\end{conjecture}

For $s=1,2,3$, they conjecture the functions~\cite[Section 4.2]{Gottsche:2020ass}\footnote{In contrast to~\cite{Gottsche:2020ass}, we define the algebraic functions to have arguments $t$ rather than $z$.}:
\bea\label{Ys Zs Ss}
Y_1(t)&=(1+t)+t^{\frac{1}{2}}(1+\tfrac{3}{4}t)^{\frac{1}{2}}~,\,
Z_1(t)=\frac{1+\tfrac{3}{4}t -\tfrac{1}{2} t^{\frac{1}{2}}(1+\tfrac{3}{4}t)^{\frac{1}{2}}}{1+\tfrac{1}{2}t}~, \,
S_1(t)=-\tfrac{1}{2}t+t^{\frac{1}{2}}(1+\tfrac{3}{4}t)^{\frac{1}{2}}~, \\
Y_2(t)&=t^{\frac{1}{2}} +(1+t)^{\frac{1}{2}}~,\quad
Z_2(t)=1+t-t^{\frac{1}{2}}(1+t)^{\frac{1}{2}}~,\quad
S_2(t)=-t+ t^{\frac{1}{2}}(1+t)^{\frac{1}{2}}. \\
Y_{3}(t)&=1+t^{\frac{1}{2}}(1-\tfrac{1}{4}t)^{\frac{1}{2}}~, \quad 
Z_3(t)=\frac{(1+\tfrac{1}{2}t)((1-\tfrac{1}{4}t)(1+\tfrac{1}{2}t)-\tfrac{3}{2}t^{\frac{1}{2}}(1-\tfrac{1}{4}t)^{\frac{1}{2}}(1-\tfrac{1}{6}t))}{(1-\tfrac{1}{2}t)^3}~,\\
S_3(t)&=\frac{-\tfrac{3}{2}t(1-\tfrac{1}{6}t)+t^{\frac{1}{2}} (1-\tfrac{1}{4}t)^{\frac{1}{2}}(1+\tfrac{1}{2}t)}{1-\frac{1}{2}t}~.
\eea
The remaining functions  $Y_{1,s}$, $Z_{11,s}$ and $S_{1,s}$, which couple to the basic classes $a$, are determined from $Y_s$, $Z_s$ and $S_s$ through a reflection on the argument. With the change of variables~\eqref{changeofvar}, the three functions  $Y_s$, $Z_s$ and $S_s$ admit power series in $z^{\frac12}$. We denote by $\t Y_s$, $\t Z_s$ and $\t S_s$ the power series obtained from those of $Y_s$, $Z_s$ and $S_s$, where we flip the sign of the coefficients for the half-integer exponents.\footnote{For this we can first rescale to obtain integer exponents,  $\Upsilon_s(z)\coloneqq Y_s(t(z^2))\in \mathbb C[[z]]$, and then define $\t Y_s(z)\coloneqq \Upsilon_s(- z^{\frac12})$. See~\cite[Remark 2.10]{Gottsche:2020ass}. In their notation, we have \eg~$Y_s(z^{\frac12})Y_{1,s}(z^{\frac12})=Y_{\{1\},s}(z^{\frac12})=Y_{\varnothing,s}(-z^{\frac12})=Y_s(-z^{\frac12})$, while similarly  $Z_s(z^{\frac12})Z_{11,s}(z^{\frac12})=Z_{\{1\},s}(z^{\frac12})=Z_{\varnothing,s}(-z^{\frac12})=Z_s(-z^{\frac12})$, and $S_s(z^{\frac12})+S_{1,s}(z^{\frac12})=S_{\{1\},s}(z^{\frac12})=S_{\varnothing,s}(-z^{\frac12})=S_s(-z^{\frac12})$. We omit this  notation in the main text. } 
Then the three remaining functions are conjectured to be
\bea\label{Z11s Y1s S1s}
Y_{1,s}=\frac{\t Y_s}{ Y_s}~, \qquad Z_{11,s}=\frac{\t Z_s}{ Z_s}~, \qquad S_{1,s}=\t S_s-S_s~.
\eea

Previous results for projective surfaces $S$ were known for rank $\rho=1$, where the moduli space $M$ reduces to the Hilbert scheme on $S$: The generating series of Segre numbers for $S$ was determined using the universal power series $V_s$, $W_s$, $X_s$, $Y_s$, $Z_s$~\cite{Marian2021}. 
The formulas for $V_s$, $W_s$, $X_s$ were proven by Marian--Oprea--Pandharipande, based on earlier work~\cite{Ellingsrud:1999iv}.

\sss{Generating function.}
The virtual Segre numbers, which are conjecturally determined above, are given in terms of coefficients of the universal series $\universal_\alpha=\universal_\alpha(z)$~\eqref{universal series} for a given $K$-theory class $\alpha$. As is clear from this expression, the universal series depends on the Chern character data $(\rk(\alpha),c_1(\alpha),c_2(\alpha))$, while it does \emph{not} depend on the Chern class $c_2$ characterising sheaves on the moduli space $M$: it can be understood as a generating function that includes the contribution from all sheaves.\footnote{We always fix the rank $\rho=2$ of the sheaves on $S$, and the dependence on $c_1$ is only through a sign} However, not all the coefficients in the $z$-series of $\universal_\alpha$ are realised as virtual Segre numbers. That is, the generating function $\segregen_\segre$~\eqref{segre gen} is not equal to the whole series  $\universal_\alpha$.

In $\segregen_\segre$, we sum over $c_2$ while keeping $\rho=2$ and $c_1$ fixed. With the fugacity $z$, this selects all coefficients in the series $\universal_\alpha$ whose exponents are realised as a virtual dimension~\eqref{vdim M} for some value of $c_2$. Hence, we keep only the terms in $\universal_\alpha$ whose exponent is of the form $\frac12 \vd M$: 
\bea\label{segre generating}
\segregen_{\universal}[p',x']\coloneqq \universal_\alpha[p',x'] \Big|_{z^{\{\frac12\vd M\mid c_2\in\mathbb Z\}}}~.
\eea
Conjecture~\ref{conj_GK} (and in particular~\eqref{GK conjecture}) is then equivalent to the statement that $\segregen_{\universal}=\segregen_{\segre}$, for the case where $X$ is smooth projective. We are able to determine partition functions on more general four-manifolds $X$, and express the result in terms of the generating function $\segregen_{\universal}$, which is defined using the power series~$\universal_\alpha$.

\subsection{Comparison for all \texorpdfstring{$N_f$}{Nf}}
\label{sec:comparison}

In this section, we compare the generating function of virtual Segre numbers with SQCD partition functions, for all $N_f=1,2,3$. 

\sss{Motivation.}
Before attempting to do so, let us provide some motivation for the proposal. For this purpose, assume that $X$ is smooth projective. By the Hitchin--Kobayashi correspondence, anti-self-dual connections on $X$ are equivalent to holomorphic vector bundles with Hermitian--Einstein metrics.\footnote{see \eg~\cite{Kobayashi1982,Donaldson1985,Uhlenbeck1986,Huybrechts2010,Mochizuki2009} for a selection of the vast literature} As a consequence, Gieseker semistable sheaves on $X$ correspond to instanton solutions, and the moduli space $M$ introduced in Section~\ref{sec: segre numbers} is isomorphic to the $\SU(2)$ (or $\SO(3)$) instanton moduli space introduced in Section~\ref{top twist}. More precisely, by identifying the (integral) Chern character $(\rho,c_1,c_2)=(2,2\bfmu,k+\bfmu^2)$, the virtual dimension $\vd\CM_{k}^{\CQ,0}$ of the $\SU(2)$ instanton moduli space~\eqref{vdimML} matches with that~\eqref{vdim M} of the moduli space $M$ of rank 2 semistable sheaves, $\vd M=\vd \CM_{k}^{\CQ,0}=\vd \Mi$.

While the topological theory automatically includes all instanton sectors, the Segre numbers are defined for fixed values of $c_2$. As in~\eqref{segre gen}, we consider the sum over $c_2$, momentarily in the absence of point or surface classes,
\begin{equation}
  \segregen_{\segre}=\sum_{c_2}z^{\frac{1}{2} \vd M} \int_{M^{[\rm vir]}}c(\alpha_M)~.
\end{equation}
By identifying $z$ with $\frac{\Lambda_0^2}{m^2}$, the instanton component~\eqref{DZSWi} to the topological partition function---including decoupling factor---admits such a series as well:
\bea
\decoup Z_\SW^{\rm i}= \sum_k z^{\frac12\vd\Mi}\int_{\Mi} c(W_k)~.
\eea
Thus, we deduce that the counterpart of the tautological class $\alpha_M$ in topological QFT is the sum $W_k$ \eqref{W_k} of index bundles for the hypermultiplets~\eqref{W_k}, such that
\bea\label{prediction Z=Segre}
\decoup Z_\SW^{\rm i}= \segregen_{\segre}~.
\eea
This is the main claim of this paper.

\sss{Comparison to $K$-theory classes.}
In order to formulate the statement more precisely, we need to relate the $K$-theory classes $\alpha\in K(S)$ of Conjecture~\ref{conj_GK} to the line bundles $\CL_j$ parametrising the topological twist. This is best understood in the case where the 't Hooft flux $\bfmu$ vanishes, which we treat first. In this case, the virtual bundle $\alpha\cong\bigoplus_{j=1}^s [L_j]$ is the direct sum of $s$ classes of line bundles, where $[L_j]$ is the class of $L_j$ in $K$-theory. 
The splitting principle dictates that
\bea\label{c1c2 alpha}
c_1(\alpha)=\sum_{j=1}^s c_1(L_i)~,\qquad 
c_2(\alpha)=\sum_{i<j}^s c_1(L_i)c_1(L_j)~.
\eea
Recall from section~\ref{top twist} that in the topological twist of $\CN=2$ SQCD, we couple the $j$'th hypermultiplet to a line bundle $\CL_j$ with flux $c_1(\CL_j)=2\bfk_j$. In the case $\bfmu=0$, the bundles $L_j$ are related to $\CL_j$ as  
\begin{equation}
    L_j\cong K_X^{\frac12} \otimes \CL_j^{\frac12}~,
\end{equation}
with $K_X$ the canonical bundle of $X$. As a consequence, the Chern classes are related as
\begin{equation}\label{c1 L_j}
    c_1(L_j)=\frac 12K+\bfk_j~.
\end{equation}
Note that due to~\eqref{constraint_c1CL}, the rhs is indeed integral for $\mu=0$. 
Inserting into~\eqref{c1c2 alpha} gives 
\bea\label{c1 c2 definition}
c_1(\alpha) &= \tfrac s2K+(\bfk)~, \\
c_2(\alpha)&=\tfrac{s(s-1)}{8}K^2+\tfrac{s-1}{2}K(\bfk)+\br{\bfk^2}~,
\eea
where again $(\bfk^p)\coloneqq\bfk_1^p+\dots+\bfk_s^p$ for $p=1,2$, and $\br{\bfk^2}\coloneqq \sum_{i<j}^s\bfk_i\bfk_j$.\footnote{Note that for rank $s=1$ we have trivial $c_2(\alpha)=0$.}
By construction, the rhs is integral as required by the lhs being a Chern class. 

In the general case, $\bfmu$ is an element of $\frac12 L$ rather than congruent to 0, and the combination $\frac12K+\bfk_j$~\eqref{c1 L_j} is half-integral rather than integral, and thus is a formal Chern root rather than the Chern class of an honest line bundle. While~\eqref{c1 c2 definition} remain the correct identifications, whenever $\bfmu\neq 0$, $\alpha$ does not split into a direct sum of line bundles, and $c_1(\alpha)$ is formally an element of $\frac12 L$, while $c_2(\alpha)$ is formally an element of $\frac14\mathbb Z$. This is understood from the physical perspective, since the monopole fields $M^j$ are sections of a tensor product of the spinor bundle $S^+$, a bundle associated to the fundamental representation of the gauge group, and $\alpha$. For the combined bundle to be well-defined, $\alpha$ is not necessarily an integral $K$-theory class. See also Eq. (\ref{constraint_c1CL}). 

Inserting~\eqref{c1 c2 definition} into the conjecture~\eqref{GK conjecture}, using the simple type condition $c^2=K^2$, changing the convention~\eqref{SW shift} of $\spinc$ structures, using  $(\bfk)^2=(\bfk^2)+2\br{\bfk^2}$, and sorting by exponent gives
\begin{equation}\label{All Nf Segre}
\begin{aligned}
\universal_\alpha[p',x']&=4W_s^{(\bfk^2)}(V_sW_s^2)^{\br{\bfk^2}}(\tfrac12 X_s)^{\chih} 
(2 V_s^{\frac{s(s-1)}{8}}W_s^{\frac{s^2}{4}} Y_s^{\frac s2} Y_{1,s}^{\frac s4} Z_sZ_{11,s}^{\frac12})^{ K^2}   \\ &\quad
\times (V_s^{s-1}W_s^{2s}Y_s^2 Y_{1,s})^{\frac12  K(\bfk)}e^{x'^2Q_s+(\bfk)x'R_s+\frac12x'K(sR_s+2S_s+S_{1,s})+p' T_s} \\ &\quad
\times 
\sum_c (-1)^{\frac{c_1}{2}(c+K)}\SW(c) Y_{1,s}^{\frac12 c(\bfk) }(Y_{1,s}^sZ_{11,s}^2)^{\frac14 c K}e^{\frac12 c x'S_{1,s}}~.
\end{aligned}
\end{equation}

\subsection{Contribution from monopole singularity}
We want to compare this expression to the contribution~\eqref{D ZSW p x} from the monopole singularity $\tu_{N_f}^-$. For this, we need to include the decoupling factor for the following reason. For $z\to 0$, the virtual Segre numbers reduce to $\SU(2)$ Donaldson invariants. Since this is guaranteed for the SQCD partition function only when the decoupling factor is included, it is a necessary condition for the functions to match.

The decoupling factor $\decoup$ also involves a rescaling of the point and surface observables $p$ and $x$, respectively~\cite{Manschot:2021qqe,Aspman:2022sfj}. We therefore aim to compare to~\eqref{All Nf Segre} to the rescaled correlation function~\eqref{D ZSW p x}:
\begin{equation}\label{D Z_SW}
\begin{aligned}
    \decoup Z_{\SW,-} \left[ \left( \tfrac{\Lambda_{s}}{\Lambda_0} \right)^2 p,\tfrac{\Lambda_{s}}{\Lambda_0}x\right]&=2\, (\tfrac{m}{\Lambda_{s}}\tC_s)^{(\bfk^2)}\hat\tC_s^{2\br{\bfk^2}}\left(-2i\Lambda_{0}^5\tm_s\tx_s^7\right)^{\chih}\left(\Lambda_{0}^{-1}\tx_s^{-1}\right)^{K^2} \\ 
&\quad \times e^{x^2 \left(\frac{\Lambda_{s}}{\Lambda_0}\right)^2 \tG_{s}+(\bfk)x \frac{\Lambda_{s}}{\Lambda_0}\tH_{s}+2p\tfrac{\tu_s}{\Lambda_{0}^2}} \\
&\quad \times \sum_c(-1)^{\bfmu(c+K)}\SW(c) \tD_s^{c (\bfk)}e^{(2i\Lambda_{0}\tx_s)^{-1}cx}~,
\end{aligned}
\end{equation}
where now $s=N_f$.
We can then formulate the
\begin{proposition} \label{prop:Segre=ZSW}
Let $X$ be a smooth, oriented, compact and simply-connected four-manifold of Seiberg--Witten simple type, with $b_2^+(X)>1$. Let $ Z_{\SW,-}$ \eqref{SW all Nf} be the contribution of the monopole singularity $\tu_{N_f}^-$ to the topological partition function of $\CN=2$ $\rm{SU}(2)$ SQCD with $N_f=1,2,3$ hypermultiplets with large equal mass $m$ on $X$, and $\decoup$ the decoupling factor~\eqref{decoup equal mass}. Then, $\decoup Z_{\SW,-}$  agrees  with half of the universal function $\universal_\alpha$~\eqref{universal series} with $c_1=2\bfmu$, $\rk(\alpha)=s=N_f$, and Chern classes~\eqref{c1 c2 definition}:
{\normalfont
\begin{equation}\label{DZ=1/2U}
    \decoup Z_{\SW,-} \left[p,x\right]=\tfrac12\universal_\alpha[p',x']~.
\end{equation}
}\noindent
The functions agree in the decoupling limit $m^{s}\Lambda_{s}^{4-s}=\Lambda_0^4$ as series in 
{\normalfont
\bea\label{zpm_def}
z=\tfrac12\left(\tfrac{\Lambda_{s}}{m}\right)^{\frac{4-s}{2}}= \tfrac{\Lambda_0^2}{2m^2}~,
\eea}\noindent 
with the change of variables $z=t(1+(1-\tfrac s2)t)^{1-\frac s2}$. 
The observables are related as $x'=\sqrt2i\tfrac{m}{\Lambda_{s}}x$, $p'=-2\big(\tfrac{m}{\Lambda_{s}}\big)^2$p, and the universal functions are given in terms of Coulomb branch functions as
{\normalfont
\begin{equation}\label{universal functions from SW}
        \begin{aligned}
        Q_s&=-\tfrac12 (\tfrac{\Lambda_{s}}{m})^{2}\tG_{s}~, \qquad 
        &R_s&=-\tfrac{i}{\sqrt2}\tfrac{\Lambda_{s}}{m}\tH_s~, \\
        S_s&=\tfrac{is}{2\sqrt2}\tfrac{\Lambda_{s}}{m}\tH_s+\tfrac{1}{2\sqrt2m\tx_s}~, \qquad 
        &S_{1,s}&=-\tfrac{1}{\sqrt2 m\tx_s}~, \\
            T_s&=-\tfrac{1}{m^2}\tu_s~, \qquad 
&V_{s\geq 2}&= (\tfrac{m}{\Lambda_{s}}\tC_s)^{-2} \hat \tC_s^2~, \\ 
W_s&=\tfrac{m}{\Lambda_{s}}\tC_s~, \qquad
&X_s&= -4\Lambda_{0}^5\tm_s\tx_s^7~,  \\ 
Y_s&=(\tfrac{m}{\Lambda_{s}}\tC_s)^{-1}\hat \tC_s^{1-s}\tD_s^{-1}~,\qquad
&Y_{1,s}&=\tD_s^2~, \\
Z_s&=\tfrac 12 (\tfrac{m}{\Lambda_{s}}\tC_s)^{\frac s4}\hat \tC_s^{\frac s4(s-1)}\tD_s^{\frac s2} (\Lambda_0 \tx_s)^{-1}~, \, \,  &Z_{11,s}&=\tD_s^{-s}~. 
        \end{aligned}
\end{equation}
}
\end{proposition}

Partial results on Proposition~\ref{prop:Segre=ZSW} have been obtained in~\cite{Aspman:2023ate}, where up to overall factors the series $W_1$, $X_1$, $T_1$, $X_2$ have been shown to match to low order with partition functions on the example of $K3$, and in particular with the identification~\eqref{zpm_def} for $N_f=1,2$.

The dictionary follows by a straightforward comparison of~\eqref{All Nf Segre} with~\eqref{D Z_SW}: We can consider the fluxes $\bfk_j$ and the various products and combinations with $c$, $K$ and $\chih$ to be $\mathbb Q$-linearly independent, which allows us to compare the bases of the two expressions. The factor of $\frac12$ in~\eqref{DZ=1/2U} is important: it indicates that $\decoup Z_{\SW_,-}$ is only half of the contribution to the Segre numbers, which we will discuss in Section~\ref{sec:contribution instanton} below.

The proof of Proposition~\ref{prop:Segre=ZSW} is somewhat different for each theory with $N_f=1,2,3$ flavours. Since the equality~\eqref{DZ=1/2U} is a straightforward comparison of~\eqref{All Nf Segre} with~\eqref{D Z_SW}, the proof amounts to demonstrating the dictionary~\eqref{universal functions from SW}. We postpone this to Section~\ref{sec:deriving_alg_functions}, and explain in the rest of the Section the Proposition in more detail, and elaborate on its consequences.

\sss{Function counting.}
Proposition~\ref{prop:Segre=ZSW} states that all universal functions are determined uniquely in terms of quantities obtained from the SW curve. In order to see how this matching is possible with the different number equations and unknowns, let us do some bookkeeping. The Segre numbers $\segre$~\eqref{GK conjecture} are conjecturally expressed in terms of 9 universal functions $Q_s$, $R_s$, $S_s$, $T_s$, $V_s$, $W_s$, $X_s$, $Y_s$, $Z_s$, together with 3 a priori unrelated functions  $S_{1,s}$, $Y_{1,s}$ and $Z_{11,s}$. The partition function~\eqref{D Z_SW} requires only 8 series $\tC_s$, $\hat \tC_s$, $\tD_s$, $\tG_s$, $\tH_s$, $\tm_s$, $\tx_s$, $\tu_s$.\footnote{The reason why there is one function less required is that the coupling of the SW basic class $c$ to the surface $x$ is determined by the same function $\tx_s$, which is due to a version of Matone's formula~\cite{Matone:1995rx,Aspman:2021vhs} (see also Appendix~\ref{app:coupling bf}} Thus, in order to solve for the 12 universal functions, we need 4 more equations. These indeed exist:\footnote{One can also check that $Y_s^sZ_s^4Y_{1,s}^{-\frac s2}=W_s^{4s}X_s^{-4}$. We do not use these identities.}
\bea\label{relations algebraic}
V_s^{s-1}W_s^{2s}Y_s^2 Y_{1,s}&=1~, \\
Y_{1,s}^s Z_{11,s}^2&=1~, \\
sR_s+2S_s+ S_{1,s}&=0~, \\
S_{1,s}Y_{1,s}^{\frac s8}Y_s^{-\frac s4}+2z^{\frac12}Z_s&=0~.
\eea
The first three originate from the absence of terms in~\eqref{D Z_SW} with exponents $K(\bfk)$, $cK$ and $x'K$ which are present in~\eqref{All Nf Segre}. The fourth one is a consequence of the bases for $cx$ and $K^2$ being determined by the single function $\tx_s$ in~\eqref{D Z_SW}, while the terms for $cx'$ and $K^2$ in~\eqref{All Nf Segre} are a priori distinct. The fact that there are 12 equations is also apparent from~\eqref{All Nf Segre} in that there are 12 bases for different exponents to be matched: $(\bfk^2)$, $\br{\bfk^2}$, $\chih$, $K(\bfk)$, $K^2$, $x'^2$, $(\bfk)x'$, $cx'$, $x'K$, $p'$, $c(\bfk)$, and $cK$, giving 12 equations for the 12 unknown series.

Since the solution to the system of 12 equations exists, we can alternatively solve~\eqref{universal functions from SW} for the CB functions in terms of algebraic functions:
\begin{equation}\begin{aligned}\label{CB functions from universal}
          \tC_s&= \tfrac{\Lambda_s}{m}W_s~, \qquad 
&\hat \tC_s&= V_s^{\frac12}W_s~, 
\\
\tD_s&= Y_{1,s}^{\frac12}~, \qquad
&\tG_s&= -2\left(\tfrac{m}{\Lambda_{s}}\right)^2 Q_s~, \\
\tH_s&=\sqrt2i\tfrac{m}{\Lambda_s}R_s~, \qquad
&\tm_s&=2\sqrt2\tfrac{m^7}{\Lambda_0^5}S_{1,s}^7X_s~, \\
\tx_s&=-\tfrac{1}{\sqrt2m}S_{1,s}^{-1}~, \qquad
&\tu_s&= -m^2T_s~.
    \end{aligned}
\end{equation}
In Section~\ref{sec:deriving_alg_functions}, we calculate all 8 functions for $s=N_f=1,2,3$, and they agree to all orders in $z$, including prefactors and signs, with the universal functions~\eqref{Vs Ws Xs Qs Rs Ts} as proposed by Göttsche--Kool. 
It is straightforward to check that the
solutions~\eqref{universal functions from SW} satisfy~\eqref{Z11s Y1s S1s}. That is, the series $S_{1,s}$, $Y_{1,s}$ and $Z_{11,s}$ are indeed obtained from $S_s$, $Y_s$ and $Z_s$ by a quotients or sums with their related functions $\t S_s$, $\t Y_s$ and $\t Z_s$ with flipped signs, and satisfy~\eqref{Z11s Y1s S1s} as conjectured in~\cite{Gottsche:2020ass}. This gives us the functions with flipped signs~,
\bea
\t S_s&=\tfrac{is}{2\sqrt2}\tfrac{\Lambda_{N_f}}{m}\tH_s-\tfrac{1}{2\sqrt2m\tx_s}~, \\
\t Y_s&=(\tfrac{m}{\Lambda_{N_f}}\tC_s)^{-1}\t \tC_s^{1-s}\tD_s\\
\t Z_s&=\tfrac 12 (\tfrac{m}{\Lambda_{N_f}}\tC_s)^{\frac s4}\hat \tC_s^{\frac s4(s-1)}\tD_s^{-\frac s2} (\Lambda_0 \tx_s)^{-1}~.
\eea
We list in Table~\ref{tab:alg_functions} the algebraic functions that this dictionary derives.\footnote{\label{ft:R3} All the functions in this table agree with the ones proposed by~\cite{Gottsche:2020ass}, except for $R_3$. With their functions~\eqref{Ys Zs Ss}, we find $sR_s+2S_s+S_{1,s}=-z^2$ rather than $=0$, which is required in~\eqref{relations algebraic} to eliminate a coupling of the surface $x'$ to the canonical class $K$. Our proposed correction fixes this relation, while leaving the other important relations unchanged. We note further that all other identities~\eqref{relations algebraic} are satisfied by the functions proposed in~\cite{Gottsche:2020ass}.}

\begin{table}
\centering 
\renewcommand{\arraystretch}{1.2}
\begin{tabular}{|Sc||Sc|Sc|Sc|}
\hline
$s$ & 1& 2& 3
\\ \hline\hline
$z_s$ & $t(1+\frac12t)^{\frac12}$ & $t$ & $t(1-\frac12t)^{-\frac12}$ \\ \hline
$Q_s$ &$\frac12t(1+\frac12t)$& $\frac12 t$ & $\frac12t(1-\frac12t)$  \\ \hline
$R_s$ &$t$ & $t$ & $t(1-\frac16t)(1-\frac12t)^{-1}$ \\ \hline
$S_s$& $-\frac{1}{2}t+t^{\frac{1}{2}}(1+\frac{3}{4}t)^{\frac{1}{2}}$ & $-t+ t^{\frac{1}{2}}(1+t)^{\frac{1}{2}}$ & $\frac{-\frac{3}{2}t(1-\frac{1}{6}t)+t^{\frac{1}{2}} (1-\frac{1}{4}t)^{\frac{1}{2}}(1+\frac{1}{2}t)}{1-\frac{1}{2}t}$ \\ \hline
$T_s$ & $2t(1+\frac38t)$ & $2t$ & $2t(1-\frac18t)$ \\ \hline
$V_s$ &  & $(1+t)^2$ & $(1+\frac12t)^3(1-\frac12t)^{-1}$ \\ \hline
$W_s$& $(1+\frac12t)^{-1}$  & $(1+t)^{-\frac12}$ & $(1+\frac12t)^{-1}$
\\ \hline
$X_s$ &$(1+\frac12t)^{-\frac34}(1+\frac34t)^{-\frac12}$ & $(1+t)^{-\frac32}$ & 
$(1-\frac12t)^{\frac34}(1-\frac14t)^{-\frac12}(1+\frac12 t)^{-4}$ \\ \hline
$Y_s$ & $ 1+t+t^{\frac12}(1+\frac34t)^{\frac12}$ & $t^{\frac12}+(1+t)^{\frac12}$ & $1+t^{\frac12}(1-\frac14t)^{\frac12}$ \\ \hline
$Z_s$ & $\frac{1+\frac34t-\frac12t^{\frac12}(1+\frac34t)^{\frac12}}{1+\frac12 t}$ & $1+t-t^{\frac12}(1+t)^{\frac12}$ & $\frac{(1+\frac{1}{2}t)((1-\frac{1}{4}t)(1+\frac{1}{2}t)-\frac{3}{2}t^{\frac12}(1-\frac{1}{4}t)^{\frac{1}{2}}(1-\frac{1}{6}t))}{(1-\frac{1}{2}t)^3}$ \\ \hline
\end{tabular}
\caption{Algebraic functions~\eqref{Vs Ws Xs Qs Rs Ts} and~\eqref{Ys Zs Ss} for $s=1,2,3$. For $R_3$, we propose a correction compared to~\cite[Section 4.2]{Gottsche:2020ass}. The function $V_s$ encodes the coupling $\partial^2 \CF/\partial m_i\partial m_j$ for $i\neq j$, which does not exist for $N_f=1$, and the topological correlators consequently do not depend on it.} \label{tab:alg_functions}
\end{table}

\sss{Change of variables.}
The correspondence of Proposition~\ref{prop:Segre=ZSW} involves a rather peculiar change of variables, from the argument $t$ of the algebraic functions to the fugacity $z$  for the virtual dimension of the moduli space $M$. An important statement of Conjecture~\ref{conj_GK} is $R_s(t)=t$, that is, $z=z(t)$ is the inverse series of $R_s(t)$.

Proposition~\ref{prop:Segre=ZSW} identifies $R_s$ with $-\frac{i}{\sqrt2}\frac{\Lambda_s}{m}\tH_s$, where $\tH_s$ is equal to $\frac{4\sqrt\pi}{4-N_f}\frac{\partial^2 \CF}{\partial\Lambda_{N_f}\partial m_j}$ evaluated for equal masses at the monopole singularity. This gives
\bea\label{R_s in terms of F}
R_s=-\frac{4\pi i}{4-N_f}\frac{\Lambda_s}{m}\frac{\partial^2 \CF}{\partial\Lambda_{s}\partial m_i}\Big|_{\substack{ m_i=m_j=m \\\tau=0}}~.
\eea
such that $R_s$ can be identified with the mass derivative of $u$. It is remarkable that with the identification $R_s=t$, $T_s$ (or equivalently $u$) becomes a simple polynomial in $t$.

Since we identify~$z=\frac{\Lambda_0^2}{2m^2}$, the statement~$z=t(1+(1-\tfrac s2)t)^{1-\frac s2}$ is equivalent to $R_s$ of~\eqref{R_s in terms of F} satisfying\footnote{For $s=N_f=3$, we find $R_3=t+\frac 13z^2$ instead of $R_3=t$, such that we need to shift it by $-\frac 13z^2$ to obtain the correct formula.}
\bea
R_s(1+(1-\tfrac s2)R_s)^{1-\frac s2}=\tfrac{\Lambda_0^2}{2m^2}~.
\eea
As a consequence, $R_s$ satisfies a polynomial equation of degree $3,1,2$ for $s=1,2,3$.
Moreover, all universal functions of Conjecture~\ref{conj_GK} are algebraic in $t=R_s(t)$. As a consequence of Proposition~\ref{prop:Segre=ZSW}, it follows that all eight functions $\tC_s$, $\hat \tC_s$, $\tD_s$, $\tG_s$, $\tH_s$, $\tm_s$, $\tx_s$, $\tu_s$ satisfy algebraic relations among each other.

\sss{Discussion.}
There are two small differences between our solution and the mathematical setup, as highlighted in Table~\ref{tab:alg_functions}.

First, our formalism does not produce the algebraic function $V_1$ for $s=N_f=1$, as it appears only as the universal function associated to $c_2(\alpha)$. Since we only consider a proper line bundle for the $N_f=1$ flavour bundle, the corresponding rank-1 bundle $\alpha$ automatically satisfies $c_2(\alpha)=0$. One possible approach to determine $V_1$ would be to consider a flavour line bundle (or sheaf) whose curvature is concentrated in a single point. This is analogous to a pointlike instanton for a gauge bundle as considered in~\cite{LoNeSha} for example. It would be interesting to explore if such effects can be incorporated in the effective field theory of the TQFT.

Finally, we propose a slightly different formula for the function $R_3$, which we derive in~\eqref{R3 correction} in the following Section. While the formalism is agnostic about the precise form of the SW prepotential, using~\eqref{universal functions from SW} and~\eqref{shift functions}, we can trace back the expression responsible for the shift in $R_3$ to the term $-i/(32\pi) \Lambda_3\sum_j m_j$ in the prepotential~\eqref{F shift}, which is independent of the local coordinate $a$.
A strong constraint on any contact term comes from the fact that the $u$-plane integral must be single-valued under monodromies. Since the term in the prepotential does not affect this, we could in principle modify the coupling $H_{N_f}$ to agree with $R_3=t$. However, this would no longer agree with the vanishing of the quadratic term in the blowup equation~\eqref{Vanishing theorem UV} for $k'_j=\frac{1}{2}$.

\subsection{Contribution from instanton branch}
\label{sec:contribution instanton}

Proposition~\ref{prop:Segre=ZSW} establishes the matching~\eqref{zpm_def} of the fugacity $z$ with the mass $m$ and scale $\Lambda_{N_f}$ of the theory, and identifies the contribution~$Z_{\SW,-}$ of the monopole singularity $\tu_{N_f}^-$ to the universal function $\universal_\alpha$. This does not capture the full instanton contribution $Z_\SW^{\rm i}$~\eqref{Z_inst} to the partition function, which furthermore requires the contribution from the dyon singularity $\tu_{N_f}^+$. As described in Section~\ref{sec:CB geometry}, the singularities $u_{N_f}^\pm$ are exchanged by replacing $z\mapsto -z$. Schematically, the instanton contribution is then given by $\universal_\alpha + (z\mapsto -z)$. However, due to the half-integer powers of $z$ that can occur in $\universal_\alpha$, this expression requires some care.

First, using the solutions~\eqref{universal functions from SW} and constraints~\eqref{relations algebraic}, we simplify~\eqref{All Nf Segre}: 
\begin{equation}\label{universal simplified}
\begin{aligned}
\universal_\alpha[p',x']&=4W_s^{(\bfk^2)}(V_sW_s^2)^{\br{\bfk^2}}(\tfrac12 X_s)^{\chih} 
(-z^{-\frac12}S_{1,s})^{ K^2} e^{x'^2Q_s+(\bfk)x'R_s+p' T_s} \\ &\quad
\times 
\sum_c (-1)^{\bfmu(c+K)}\SW(c) Y_{1,s}^{\frac12 c(\bfk) }e^{\frac12 c x'S_{1,s}}~.
\end{aligned}
\end{equation}
While $S_s\in \mathbb C[[z^{\frac12}]]$, it is clear from~\eqref{Z11s Y1s S1s} that $S_{1,s}\in z^{\frac12}\mathbb C[[z]]$, and as a consequence $z^{–\frac12}S_{1,s}\in \mathbb C[[z]]$. Therefore, the first line in~\eqref{universal simplified} is a series in $\mathbb C[[z]]$, since by~\eqref{Cz Cz12}, all the functions $Q_s,R_s,T_s,V_s,W_s,X_s$ are. 
In order to determine the nature of the generating series $\universal_\alpha$, we probe its behaviour under the formal change $z\mapsto e^{2\pi i}z$. Viewing the universal series a functions of $z$, we have again due to~\eqref{Z11s Y1s S1s} that $Y_{1,s}(e^{2\pi i}z)=Y_{1,s}(z)^{-1}$, while $S_{1,s}(e^{2\pi i}z)=-S_{1,s}(z)$. Using the (anti)-symmetry of the $\SW$ invariant
$\SW(-c)=(-1)^{\chih}\SW(c)$~\cite{Witten:1994cg}, we then find
\bea
   \universal_\alpha(e^{2\pi i}z)=(-1)^{\kappa}\universal_\alpha(z)~,
\eea
where we define $\kappa=3\chih+4\bfmu^2$.\footnote{We also use that $K$ and $c$ are characteristic vectors of $L$, and as a consequence $2\mu K+4\mu^2\in2\mathbb Z$, and $(-1)^{2\mu(c+K)}=1$} It follows that $\universal_\alpha(z)$ is a power series in $z$ with (half)-integer exponents,
\bea\label{universal (half)integer}
 \universal_\alpha(z)\in \begin{cases}
    \mathbb C[[z]]~, \qquad &\kappa\in 2\mathbb Z~, \\
   z^{\frac12}\mathbb C[[z]]~, \qquad &\kappa\in 2\mathbb Z+1~. \\
\end{cases}
\eea

\sss{The dyon singularity.}
This analysis is crucial for combining the contribution from the monopole singularity $\tu_{N_f}^-$ with the dyon singularity $\tu_{N_f}^+$. The latter is obtained similarly to~\eqref{SWuniversal}, and is related to $Z_{\SW,-}$ simply by flipping the sign $z\mapsto -z$. Additionally, it receives an overall phase due to the $T$-transformation of the effective coupling acting on the measure factor and the photon partition function before obtaining the contribution from wall-crossing. It can be understood as a global anomaly that exchanges the two vacua $\tu_{N_f}^\pm$, as in Witten's formula~\cite{Witten:1994cg}. 
Dropping the dependence on the observables in favour of the dependence on $z$, this gives
\bea\label{Z+ in terms of Z-}
Z_{\SW,+}(z)=i^\kappa Z_{\SW,-}(-z)~,
\eea
where the phase spells out $i^\kappa=e^{\frac 32\pi i\chih}e^{2\pi i\bfmu^2}$.\footnote{The prefactor is the inverse of $i^{\Delta-z^2}$ of~\cite[(2.17)]{Witten:1994cg}, which is due to an exchange of the labels for $\tu_{N_f}^\pm$. See also~\cite[(12.23)]{Aspman:2023ate}} 
Combining both contributions, we obtain $
Z_\SW^{\rm i}(z)=Z_{\SW,-}(z)+i^\kappa Z_{\SW,-}(-z)$.
Multiplication with $\decoup$ and using~\eqref{DZ=1/2U} of Proposition~\ref{prop:Segre=ZSW} gives
\bea\label{u+u(-z)}
\decoup Z_\SW^{\rm i}(z)=\tfrac12\left(\universal_\alpha(z)+i^\kappa \universal_\alpha(-z)\right)~.
\eea
Let us now prove that this projection precisely keeps the relevant coefficients in the series $\universal_\alpha$, as required in the generating function~\eqref{segre generating}.

The integration over the virtual fundamental class results in a coefficient of $z^{\frac12\vd M}$, with the virtual dimension~\eqref{vdim M}  being $\vd M=4c_2-\kappa$. Since $c_2$ runs over the integers, the generating function~\eqref{segre generating}
in our case is a series $\segregen_{\universal}\in z^{-\frac \kappa2}\mathbb C[[z^2]]$.\footnote{In order to see that the restriction is indeed necessary, we can resort to the arguably simplest example $X=K3$, which has $\chih=2$ and $K^2=0$ with the only basic class $c=0$ having $\SW(0)=1$. The universal series for $N_f=1$ and $\bfmu=\bfk_1=0$ becomes $\universal_\alpha=X_1^2$, which has the power series $X_1(t)^2=1-\frac32 z+\frac{63}{32}z^2-\frac52z^3+\CO(z^4)$. Since $\kappa\equiv 2\mod4$, we have $\segregen_{\universal}\in z\mathbb C[[z^2]]$, while $\universal_\alpha\in \mathbb C[[z]]$. In this case, $\segregen_{\universal}$ is simply the (non-trivial) odd part of $\universal_\alpha$.} The remainder of $\kappa$ modulo 4 determines the powers of $z$ that can occur:
\bea\label{vdM cases}
\tfrac12\vd M\in \begin{cases}
    0+2\mathbb Z~, \qquad &\kappa\equiv 0\mod 4~, \\
    \frac32+2\mathbb Z~, \qquad &\kappa\equiv 1\mod 4~, \\
    1+2\mathbb Z~, \qquad &\kappa\equiv 2\mod 4~, \\
    \tfrac12 +2\mathbb Z~, \qquad &\kappa\equiv 3\mod 4~.
\end{cases}
\eea
Using~\eqref{universal (half)integer}, it is then straightforward to check that~\eqref{u+u(-z)} is the correct projection map: If $\kappa$ is even, then $\universal_\alpha(z)\in\mathbb C[[z]]$ and $\universal_\alpha(z)\pm \universal_\alpha (-z)$ gives the even/odd part of the series, depending on whether $\kappa\equiv 0$ or $2\mod 4$. 
If $\kappa$ is odd on the other hand, the series $\universal_\alpha(z)\in z^{\frac12}\mathbb C[[z]]$ is shifted by $\frac12$, and $\universal_\alpha(z)\pm i \universal_\alpha (-z)$ only keeps the exponents $\frac32$ or $\frac12$ modulo $2$. In all cases, the non-zero coefficients of $\universal_\alpha(z)\pm i^\kappa \universal_\alpha (-z)$ agree with those of $\universal_\alpha(z)$ up to a factor of 2, which cancels with the $\frac12$ of~\eqref{u+u(-z)}.
This then establishes the generating function~\eqref{segre generating}, 
\bea\label{segre generating kappa}
\segregen_{\universal}(z)= \tfrac12\left(\universal_\alpha(z)+i^\kappa \universal_\alpha(-z)\right)~.
\eea
Retaining the notation and identifications from Proposition~\ref{prop:Segre=ZSW}, we have proven the 
\begin{theorem}\label{thm: DZ=G}
The contribution $Z_{\SW}^{\rm i}=Z_{{\rm SW},+}+Z_{{\rm SW},-}$ of the instanton component to the topological SQCD partition function on $X$, with $Z_{{\rm SW},-}$ as in \eqref{SW all Nf} and $Z_{{\rm SW},+}$ as in  \eqref{Z+ in terms of Z-}, agrees with the generating function~\eqref{segre generating kappa}:
{\normalfont
    \bea
    \decoup Z_{\SW}^{\rm i}[p,x]&=\segregen_{\universal}[p',x']~.
    \eea
    }
That is, 
{\normalfont
    \bea
    \decoup Z_{\SW}^{\rm i}[p,x]&=4W_s^{(\bfk^2)}(V_sW_s^2)^{\br{\bfk^2}}(\tfrac12 X_s)^{\chih} 
(-z^{-\frac12}S_{1,s})^{ K^2} e^{x'^2Q_s+(\bfk)x'R_s+p' T_s} \\ &\quad
\times 
\sum_c (-1)^{\bfmu(c+K)}\SW(c) Y_{1,s}^{\frac12 c(\bfk) }e^{\frac12 c x'S_{1,s}}\Big|_{z^{-\frac{3\chih+4\bfmu^2}{2}+2\mathbb Z}}~.
    \eea
}
\end{theorem}
Specialising to the algebraic setup, with only minor modifications the  prediction~\eqref{prediction Z=Segre} follows. Indeed, when $S$ is algebraic,  Conjecture~\ref{conj_GK} states that $\segregen_\universal=\segregen_\segre$, and the instanton component to the topological SQCD partition function on $S$ conjecturally agrees with the generating function~\eqref{segre gen} of virtual Segre numbers for $S$,
 \bea
    \decoup Z_{\SW}^{\rm i}[p,x]&=\segregen_{\segre}[p',x']~.
\eea
In particular, the universal functions determined in~\cite{Gottsche:2020ass} `predict' the partition functions even when the general four-manifold $X$ is not smooth projective, and the algebro-geometric techniques for the construction of virtual Segre numbers---such as a perfect obstruction theory and the universal sheaf---are not available. 

We summarise the above results in an approximate math-physics dictionary:
\bea
\centering
\begin{tabular}{|c| c|c|c|c|} 
\hline
Geometry & Physics \\ \hline\hline
$\alpha$ & Sum of line bundles $\CL_j$ of topological twist \\ 
$\alpha_M$ & Index bundle $W_k$ \\ 
$M$ & Moduli space of instantons $\Mi$\\
$z$ & Inverse mass $1/m$ \\
$c_1$ & 't Hooft flux $\mu$ \\
$c_2$ & Instanton charge $k$ \\
$\rho$ & Rank ($+1$) of gauge group $\SU(\rho)$ \\
$\segre_\alpha$ & Contribution from fixed instanton charge $k$ to $\decoup Z_{\SW,-}$ \\
$\universal_\alpha$ & Contribution  from monopole singularity $\decoup Z_{\SW,-}$ \\
$\segregen_{\universal}$& Contribution  from instanton branch $\decoup Z_{\SW}^{\rm i}$\\ \hline
\end{tabular}
\eea

\section{Universal functions from SW geometry}
\label{sec:deriving_alg_functions}

In this final section, give a full proof of Proposition~\ref{prop:Segre=ZSW} by demonstrating the complete correspondence~\eqref{universal functions from SW} between Coulomb branch couplings and universal functions.
For each case $s=N_f=1,2,3$,  Proposition~\ref{prop:Segre=ZSW} expresses the 8 series $\tC_s$, $\hat \tC_s$, $\tD_s$, $\tG_s$, $\tH_s$, $\tm_s$, $\tx_s$, $\tu_s$ obtained from Coulomb branch functions in terms of the 12 universal functions $Q_s$, $R_s$, $S_s$, $T_s$, $V_s$, $W_s$, $X_s$, $Y_s$, $Z_s$, $S_{1,s}$, $Y_{1,s}$, $Z_{11,s}$. For this, we use the exact, closed-form expressions for $\tG_s$, $\tm_s$, $\tx_s$ and $\tu_s$, obtained from the Seiberg--Witten curves, which allow us to derive the universal series $Q_s$, $T_s$ and $X_s$. The remaining series are determined using combinations of identities originating from the blowup formula, as explained in Section~\ref{sec:blowup}. An alternative strategy to match the couplings with universal functions is to determine the various Coulomb branch functions by explicit integration of the SW curve, to obtain large mass expansions of the prepotential. These calculations serve as important independent checks of the blowup formulas, which we list in Appendix~\ref{app: large mass expansions}.

\sss{Strategy.}
While much of Section~\ref{sec:universal functions from PF} can be discussed uniformly for all $N_f=1,2,3$, the explicit calculation and proofs take a rather distinct flavour for each case. This is due to the specific symmetries organising the singularities on the Coulomb branch in the equal mass case: it can be shown that the singularities $\tu_{N_f}^\pm$ for $N_f=1,2,3$ can be expressed as roots of a polynomial of degree $3$, $1$ and $2$ (see~\eqref{upm_exact}). The simplicity of the case $N_f=2$ is reflected in the fact that the change of variables~\eqref{changeofvar} proposed in Conjecture~\ref{conj_GK} in this case becomes trivial, $z=t$.

As arguably the simplest theory after pure $\CN=2$ $\SU(2)$ SYM, the (massive) $N_f=1$ theory has been of significant interest in the past~\cite{Nahm:1996di,Aspman:2021vhs,Aspman:2020lmf,Huang:2009md,Dolan:2005dw,AlvarezGaume:1997ek,AlvarezGaume:1997fg,Gottsche:2010ig,Malmendier:2008db}. However, partially due to the more involved mathematical structure, obtaining exact results have been more challenging. As with the other cases $N_f=2,3$, it turns out that a combination with the large and equal mass limit and the change of variables simplifies the equations substantially.
This is among other things enabled by the fact that the singularities $\tu_{N_f}^\pm$ in the theory with $N_f=0,\dots, 3$ equal masses $m$ can be expressed in terms of $m$ through radicals, which is not true in particular for generic masses (\eg~for $N_f=3$ and $4$)~\cite{Aspman:2021vhs}. 
These exact expressions are given in~\eqref{upm_exact} in Appendix~\ref{app:largemassSW}. 

The change of variables from $z$ to $t$ can be motivated by the following fact: While $\tu_{N_f}^\pm$ are algebraic in $m$,  they are \emph{polynomial} in $t$.  In some sense, $z=z(t)$ is the change of variables that facilitates the factorisation of the equation $\Delta_{N_f}(\tu_{N_f}^\pm)=0$ into a product including a linear factor $(u-\tu_{N_f}^\pm)$. This is the crucial trick to derive the algebraic function $T_s$. From this, all other universal functions can be derived, as will be explicitly demonstrated.

\subsection{\texorpdfstring{$N_f=1$}{Nf=1}}
\label{sec:Nf1}
Let us first derive the functions $T_1$, $Q_1$ and $X_1$ for $N_f=1$ from the SW curve. 

\sss{The functions $T_1$, $Q_1$ and $X_1$.}
We start with $T_1$. Consider the \ws{} invariants~\eqref{g2g3 Nf=1} of the $N_f=1$ theory, and form the physical discriminant $\Delta_1=g_2^3-27g_3^2$. 
The change of variables is $z=t(1+\tfrac12t)^{\frac12}$, where we identify $z=\frac{\Lambda_0^2}{2m^2}$ in the decoupling limit $m\Lambda_1^3=\Lambda_0^3$. From~\eqref{u/m2} it is clear that $\tu_1/m^2$ admits a Taylor series in $z$, where $\tu_1=\tu_{1}^-$. Let us denote this series by $\tu_1=-m^2 T_1$, and we aim to derive $T_1$. 
Eliminating the mass using $m=\frac{\Lambda_0}{\sqrt{2z}}$, we plug the ansatz for $\tu_1$ into the discriminant to obtain
\bea\label{delta u1}
\Delta_1(\tu_1)=\frac{ (t (3
   t+8)-4 T_1)\left(16 T_1^2+4 T_1 (t+2) (3 t+2)+t^2 (3 t+2)^2+32 t\right)}{128 \sqrt{2} t^3 (t+2)^{3/2}}~.
\eea
Among the three solutions, there is one particularly simple solution to $\Delta_1(\tu_1)=0$ from the first factor in the numerator, which gives 
\bea
T_1=2t(1+\tfrac38t)~.
\eea
Taking $m\to \infty$ shows that $-m^2T_1\to -\Lambda_0^2$, which is not the case for the other two solutions. We therefore obtain $\tu_1=-m^2T_1$, with $T_1$ as above. This derives the universal function $T_1$ from the SW geometry. 

Next, the two functions $Q_1$ and $X_1$ can be derived by substituting $\tu_1=-m^2T_1$ into the definitions~\eqref{mu def} and~\eqref{dadu def} of $\tm_1$ and $\tx_1$. The contact term $\tG_1$ can be determined from those using~\eqref{contact term series}. Then, from the dictionary~\eqref{universal functions from SW}, we have $X_1=-4\Lambda_{0}^5\tm_1\tx_1^7$, while $Q_1=-\tfrac12 (\tfrac{\Lambda_{1}}{m})^{2}\tG_{1}$. By direct computation, we find
\bea
Q_1=\tfrac12t(1+\tfrac12t), \qquad X_1=(1+\tfrac12t)^{-\frac34}(1+\tfrac34t)^{-\frac12}~.
\eea
We can furthermore determine $S_{1,1}$ using $S_{1,1}=-1/\sqrt2 m \tx_1$~\eqref{universal functions from SW}, which gives
\bea\label{S11}
S_{1,1}=-2t^{\frac12}(1+\tfrac34t)^{\frac12}~.
\eea
This analysis derives the functions $Q_1$, $T_1$, $X_1$ and $S_{1,1}$ of Conjecture~\ref{conj_GK}.

\sss{The remaining couplings.}
In order to derive the remaining universal functions, we use the blowup equations~\eqref{3 equations}. First, we aim to evaluate~\eqref{3 equations c} at the strong coupling point $\tau=0$. For this, we take the modular $S$-transformation~\eqref{jt_S-transformation} of the elliptic theta functions, compute the new $q$-series and only keep the constant term. On the left hand side, the period $\frac{da}{du}$ then becomes the series $\tx_1$, which determines the function $S_{1,1}=-(\sqrt2 m \tx_1)^{-1}$. The contact term $G_{1}$ becomes the series $\tG_1$, which determines the function $Q_1=-\frac12\left(\frac{\Lambda_1}{m}\right)^2 \tG_1$. As a consequence, at strong coupling the identity~\eqref{3 equations c} becomes
\bea
8Q_1S_{1,1}^{-2}=\sec(\tfrac{\pi \tv_1}{2})^2~.
\eea
With $Y_{1,1}=\tD_1^2=e^{-2\pi i\tv_1}$, and $S_{1,1}$ and $Q_1$ from above, this gives a quadratic equation in $Y_{1,1}^{\frac12}$. 
We pick the solution with  $Y_{1,1}^{\frac12}=1+\CO(t^{\frac12})$, which is in line with $\tv_1\to 0$ in the decoupling limit $m\to\infty$. This gives
\bea\label{Y11 sqrt}
Y_{1,1}^{\frac12}=2\frac{1+t-t^{\frac12}(1+\tfrac34t)^{\frac12}}{2+t}~,
\eea
or finally,
\bea\label{Y11}
Y_{1,1}=\frac{(1+t)-t^{\frac12}(1+\tfrac34t)^{\frac12}}{(1+t)+t^{\frac12}(1+\tfrac34t)^{\frac12}}~.
\eea

Next, we use~\eqref{3 equations b} to determine the universal function $R_1$. At $\tau=0$, the coupling $H_1$ becomes a series $\tH_1$, which by~\eqref{universal functions from SW} determines $R_1=-\frac{i}{\sqrt2}\frac{\Lambda_1}{m}\tH_1$. Using this, the equation~\eqref{3 equations b} evaluates at $\tau=0$ to 
\bea
i R_1S_{1,1}^{-1}=\tan(\tfrac{\pi \tv_1}{2})~.
\eea
Plugging in~\eqref{Y11 sqrt} and $S_{1,1}$ gives 
\bea
R_1=t~.
\eea
This simple relation can be seen as the motivation for the change of variables from $z$ to $t$~in\cite{Gottsche:2020ass}.

Finally, we determine the function $W_1$. The coupling $C_{11}$ at the monopole singularity becomes a series $\tC_1$, which gives $W_1=\frac{m}{\Lambda_1}\tC_1$. From~\eqref{C11 C12 Nf=1}, we obtain at strong coupling,
\bea
\tC_1=4\sin(\tfrac{\pi \tv_1}{2})^{\frac43}~.
\eea
We eliminate the mass prefactor in~$W_1=\frac{m}{\Lambda_1}\tC_1$ by remembering that $z=\frac12(\tfrac{\Lambda_1}{m})^{\frac32}$, which relates to $t$ by $z=t(1+\frac12t)^{\frac12}$. A small calculation then gives
\bea
W_{1}=2(2+t)^{-1}~.
\eea

We have therefore derived the 7 universal algebraic functions $R_1$, $Q_1$, $S_{1,1}$,  $T_1$, $W_1$,  $X_1$, $Y_{1,1}$. However, for $N_f=1$ there are in total 11 functions to be determined. The remaining 4 functions $S_1$, $Y_1$, $Z_1$ and $Z_{11,1}$ are not independent, as can be seen by comparing~\eqref{universal functions from SW} with~\eqref{CB functions from universal}. It is then straightforward to plug the known answers into the remaining relations in~\eqref{universal functions from SW}. This derives the complete collection of functions, and in particular the 8 `single-subscript' algebraic functions $Q_1$, $R_1$, $S_1$, $T_1$, $W_1$, $X_1$, $Y_1$ and $Z_1$, which we collected in Table~\ref{tab:alg_functions}.

\subsection{\texorpdfstring{$N_f=2$}{Nf=2}}
\label{sec:Nf2}
Moving on to the equal mass $N_f=2$ theory, we have $z=\frac{\Lambda_0^2}{2m^2}=t$ and $m\Lambda_2=\Lambda_0^2$.

\sss{Confirming the dictionary.}
As apparent from~\eqref{upm_exact}, the two singularities $\tu_{2}^\pm$ trivially depend on $m$ in this case. We consider $\tu_2=\tu_2^-$ and incorporate the shift~\eqref{shift functions}, such that $\tu_2=-m\Lambda_2=-\Lambda_0^2$.
As before, we write $\tu_2=-m^2 T_2$, such that $T_2=\Lambda_0^2/m^2=2t$, so
\bea
T_2=2t~.
\eea
Using this first universal function, we substitute it into the expressions for $\tm_2$~\eqref{mu def}, $\tx_2$~\eqref{dadu def} and $\tG_{2}$~\eqref{contact term series}. With $Q_2=-\tfrac12 (\tfrac{\Lambda_{2}}{m})^{2}\tG_{2}$, $X_2= -4\Lambda_{0}^5\tm_2\tx_2^7$ and $S_{1,2}=-1/\sqrt2 m \tx_2$, keeping in mind the shift~\eqref{shift functions} of the contact term $\tG_2$, we then find
\bea
Q_2=\tfrac12t~, \qquad X_2=(1+t)^{-\frac32}~, \qquad S_{1,2}=-2t^{\frac12}(1+t)^{\frac12}~.
\eea

In order to derive the remaining algebraic series, we start as for $N_f=1$ with the identity relating the surface observable contact term $G_2$ to the coupling $v$, which is~\eqref{3 equations c}. By using the dictionary~\eqref{universal functions from SW}, and evaluating at $\tau=0$ we find
\bea
8Q_2S_{1,2}^{-2}=\sec(\pi \tv_2)^2~.
\eea
Inserting $Q_2$ and $S_{1,2}$ from above, together with $Y_{1,2}=\tD_2^2=e^{-2\pi i\tv_2}$ gives
\bea
Y_{1,2}=1+2t-2t^{\frac12}(1+t)^{\frac12}~.
\eea
This allows us to solve for $R_2$. Equation~\eqref{3 equations b} becomes at strong coupling
\bea
2i R_2S_{1,2}^{-1}=\tan(\pi \tv_2)~,
\eea
which we easily solve for 
\bea
R_2=t~.
\eea

Finally, we determine the functions $V_2$ and $W_2$, where $\tC_2=\frac{\Lambda_2}{m}W_2$ and $\hat \tC_2=V_2^{\frac12}W_2$. First, from~\eqref{C11 C12 Nf=2} we compute\footnote{As a consequence, $\tC_2$ and $\hat\tC_2$ and are related as $\hat \tC_2(\hat \tC_2-\tfrac12\tC_2)=1$, or $\hat \tC_2^{-1}=\frac{m}{\Lambda_2}\tC_2$.}
\bea
\tC_2&=-2\sin(\pi \tv_2)\tan(\pi \tv_2)~, \\
\hat \tC_2&=\cos(\pi \tv_2)~.
\eea
Then using $z=\frac{\Lambda_0^2}{2m^2}$, $m\Lambda_2=\Lambda_0^2$ and $z=t$ we find
\bea
V_2=(1+t)^2~, \qquad W_2=(1+t)^{-\frac12}~. 
\eea
As before, using the dictionary~\eqref{CB functions from universal}, this derives all 12 algebraic functions.

\subsection{\texorpdfstring{$N_f=3$}{Nf=3}}
\label{sec:Nf3}
The calculation for $N_f=3$ follows similarly from that of the case $N_f=1$, which is however simpler since the discriminant factors into a quadratic factor in the equal mass case. 
Writing again $\tu_{3}^-=\tu_3$ with $\tu_3=-m^2 T_3$, with $z=(1-\frac12t)^{-\frac12}$ and $m^3\Lambda_3=\Lambda_0^4$, we compute $\Delta_3(\tu_3)=0$ and solve for $T_3$, which gives
\bea
T_3=2t(1-\tfrac18t)~.
\eea
For $S_{1,3}$, we compare with $\tx_3$, and  $Q_3$ and $X_3$ can be found as in the cases above. We obtain:
\bea\label{Q3 S13 X3}
Q_3=\frac t4(2-t)~, \quad 
S_{1,3}=\frac{(2+t)t^{\frac12}(4-t)^{\frac12}}{t-2}~, \quad X_3=(1-\tfrac12t)^{\frac34}(1-\tfrac14t)^{-\frac12}(1+\tfrac12 t)^{-4}~,
\eea
which agree with those given in Table~\ref{tab:alg_functions}.

In order to derive $Y_{1,3}=e^{-2\pi i \tv_3}$, we compare with the surface observable contact term from~\eqref{3 equations c}, which is 
\bea
8Q_3S_{1,3}^{-2}=\sec(\tfrac{3\pi \tv_3}{2})^2~,
\eea
such that
\bea
Y_{1,3}=\frac{4+4t-t^2-4t^{\frac12}(4-t)^\frac12}{(2-t)^2}~.
\eea
In order to find $R_3$ from this, we solve~\eqref{3 equations b}, which gives at $\tau=0$: 
\bea
3i R_3S_{1,3}^{-1}=\tan(\tfrac{3\pi \tv_3}{2})~,
\eea
such that 
\bea\label{R3}
R_3=\frac{t(6-t)}{3(2-t)}~.
\eea
This differs from the function proposed in~\eqref{Vs Ws Xs Qs Rs Ts} in Conjecture~\ref{conj_GK}.
Using the change of variables, we may express it as
\bea\label{R3 correction}
R_3=t+\tfrac13z^2~,
\eea
so it differs from the proposed function $R_3=t$ in the $z$-expansion only by a single coefficient. 
To determine $S_3$, we combine the relations for $R_3$ and $S_{1,3}$ of~\eqref{universal functions from SW} with that of $S_{1,3}$, which gives $S_{3}=-\frac12(3R_3+S_{1,3})$. Inserting $R_3$~\eqref{R3} and $S_{1,3}$~\eqref{Q3 S13 X3} gives
\bea
S_3=\frac{-\frac{3}{2}t(1-\frac{1}{6}t)+t^{\frac{1}{2}} (1-\frac{1}{4}t)^{\frac{1}{2}}(1+\frac{1}{2}t)}{1-\frac{1}{2}t}~.
\eea
It agrees perfectly with the function~\eqref{Ys Zs Ss} proposed in Conjecture~\ref{conj_GK}.
Moreover, it satisfies the relation $3R_3+2S_3+S_{1,3}=0$, as required by~\eqref{relations algebraic}. The function $R_3=t$ of Conjecture~\ref{conj_GK} does \emph{not} satisfy it (see also footnote~\ref{ft:R3}).

Finally, from~\eqref{C11 C12 Nf=3} we obtain
\bea
\tC_3&=\frac{64\sin(\tfrac{\pi\tv_3}{2})^4}{2\cos(\pi\tv_3)-1}~, \\
\hat\tC_3&=\frac{\cos(\tfrac{3\pi \tv_3}{2})^{\frac12}}{\cos(\tfrac{\pi \tv_3}{2})^{\frac12}}~,
\eea
from which we get $V_3$ and $W_3$, using $z=\frac{\Lambda_0^2}{2m^2}$ together with $m^3\Lambda_3=\Lambda_0^4$ and $z=t(1-\frac12)^{-\frac12}$:
\bea
V_3=\frac{(2+t)^3}{4(2-t)}~, \quad W_3=\frac{2}{2+t}~.
\eea
Again using the dictionary, we can compute the remaining functions $S_3$, $Y_3$, and $Z_3$, which we print in Table~\ref{tab:alg_functions}. For $N_f=3$, all universal functions we have derived except $R_3$ agree with the ones proposed in Conjecture~\ref{conj_GK}. As a consistency check, they satisfy the four conditions~\eqref{relations algebraic}. We list the remaining function $Z_{11,3}$ for completeness:
\bea
Z_{11,3}=\frac{1+3t(1-\frac14t)+(3-\frac12t)(1+\frac12t)t^{\frac12}(1-\frac14t)^{\frac12}}{(1-\frac12t)^3}~.
\eea

\section{Conclusion and discussion}
\label{sec:discussion}
In this article, we have computed topological correlation functions of $\CN=2$ supersymmetric QCD with $N_f=1,2,3$ massive hypermultiplets. We expressed the instanton component to the correlation functions in terms of universal functions, which we demonstrated to match with those conjectured in the mathematical literature on virtual Segre numbers for projective surfaces~\cite{Gottsche:2020ass} (and previous literature).
In this final section, we discuss several interesting open problems for future work. 

\sss{Higher rank of $K$-theory class.}
We find perfect agreement between topological QCD correlators and the universal series as determined in Conjecture~\ref{conj_GK}~\cite[Conjecture 2.8]{Gottsche:2020ass}. This conjecture expresses the virtual Segre numbers for projective surfaces $S$ for any $K$-theory class $\alpha\in K(S)$ with arbitrary integer rank $\rk \, \alpha=s$. In the $\SU(2)$ case discussed in this paper, this proposes universal functions going beyond the range of theories originating from topological twists of asymptotically free supersymmetric gauge theories: supersymmetric QCD with gauge group $\SU(2)$ has a beta function proportional to $N_f-4$, and so is asymptotically free for $0\leq N_f \leq 3$, superconformal for $N_f=4$, and strongly coupled in the UV for $N_f>4$. 

Since $s=N_f$ counts the number of massive hypermultiplets for $\SU(2)$, the series suggested in~\cite{Gottsche:2020ass} for $s=-1$ and $s=5$ do not have an a priori well-defined physical origin. For instance, the $\SU(2)$ $N_f=5$ theory is strongly coupled in the UV and has a free non-Abelian Coulomb phase at $u=0$~\cite{Argyres:1995xn,Witten:1997sc}. Such theories however typically appear in geometric engineering constructions, thus allowing a UV completion in that broader framework.

\sss{The $\boldsymbol{N_f=4}$ superconformal theory.}
The remaining case for $\SU(2)$, which admits a gauge theoretic description and we have not studied in this article, is the $N_f=4$ theory, which is a massive deformation of a 4d $\CN=2$ SCFT. The SW curve is much more involved, since all the Coulomb branch couplings carry a dependence on the UV coupling $\tn$ in addition to the masses.
We could again study the equal mass configuration $\bfm = (m, m, m, m)$ in the decoupling limit $m \to \infty$, where the scale $\Lambda_0$ emerges from the double scaling limit $\Lambda_0^4 = 64 q_0^{1/2} m^4$, with $q_0=e^{2\pi i\tn}$~\cite{Seiberg:1994aj}.

The dictionary of Proposition~\ref{prop:Segre=ZSW} does not apply directly, since it is derived using the topological twist, decoupling factor and the SW analysis for $N_f=1,2,3$ equal masses. Other obstructions involve the gravitational couplings $A^\chi B^\sigma$ becoming dependent on $\tn$~\cite{Manschot:2019pog}, and the CB geometry receiving a non-trivial shift upon decoupling~\cite{Seiberg:1994aj}. 
Preliminary calculations of the contributions $Z_\SW^{\rm i}$ from the instanton component to the $N_f=4$ partition function give $q_0$-series that are modular forms. This is expected from the fact that the equal mass configuration exhibits a modular symmetry for $\tn$ under the congruence subgroup $\Gamma^0(4)$~\cite{Aspman:2021evt}. Matching all the universal functions $T_4$, $X_4$, etc. for $N_f=4$ is however much more difficult and requires a careful analysis, which we leave for future work.

\sss{Generic mass partition function.}
In this work, we have chosen the equal mass configuration $\bfm=(m,\dots, m)$ to calculate the instanton contribution to topological correlators, which we determine to all orders as a large $m$ series. The case where the masses $m_1,\dots, m_{N_f}$ are generic, or large but all distinct, is much more involved for several reasons.
First, the equal masses organise the $u$-plane singularities into two dedicated $I_1$ singularities, where an $I_{N_f}$ singularity moves to $\infty$ in the large mass limit. For generic masses, we have $2+N_f$ separate $I_1$ singularities. Moreover, for $N_f\geq 3$ it is in general not possible to express them in terms of radicals. 

For generic masses, there are two interesting limits in which topological partition functions can exhibit interesting singular behaviour. The first is the limit to Argyres--Douglas theories, where the $\CN=2$ theory becomes superconformal~\cite{Argyres:1995jj,Argyres:1995xn,Marino:1998uy,Moore:2017cmm}. Another origin of singular behaviour is the equal mass limit, where Higgs branches appear~\cite{Seiberg:1994aj,Moore:1997dj,Dedushenko:2017tdw,LoNeSha}. As a consequence of the moduli space being non-compact, the full topological partition function is in general singular as $m_i\to m_j$. A complete discussion of these non-trivial limits goes much beyond the scope of this paper, and we hope to report on it elsewhere.

\sss{Higher rank gauge groups.}
It is an interesting question how the results generalise to higher rank.  Reference~\cite{Gottsche:2020ass} proposes universal functions for virtual $\rm{SU}(\rho)$ Segre numbers, which generalise the $\SU(\rho)$ Donaldson invariants~\cite{Kronheimer2005}, and give explicit results for some $\rm{SU}(3)$ and $\rm{SU}(4)$ examples. Physically, the higher rank analogue of the $u$-plane integral has been studied  in~\cite{LoNeSha,Marino:1998bm}.  The Moore--Witten formula~\eqref{Z_contr} generalises, where an integral over the $N-1$-dimensional Coulomb branch remains, together with a discrete sum over the singularity locus $\mathcal D$. For gauge group ${\rm SU}(N)$, the divisor locus $\mathcal D$ is a stratified space, consisting of codimension $\ell$ strata $\mathcal D_i^\ell$. If $X$ is of simple type with $b_2^+(X)>1$, only the top codimension stratum $\mathcal D_i^{N-1}$ contributes. These are a collection of $h^\vee(\SU(N))=N$ points, which are precisely the supersymmetric vacua of the $\CN=1$ ${\rm SU}(N)$ theory. The contribution is given in~\cite[(9.17)]{Marino:1998bm}, which is again determined from the local expansions near the singularities. The calculation of those expansions is much more involved for higher rank, where some tools have been worked out in~\cite{DHoker:2020qlp,DHoker:2024vii,Aspman:2020lmf,DHoker:2022loi,Furrer:2022nio}.
 
We can then consider adding $N_f$ hypermultiplets of equal mass $m$ for ${\rm SU}(N)$, where $N_f\leq 2N$ ensures a well-defined UV theory.
The SW curves have been determined long time ago~\cite{Hanany:1995na,EGUCHI1996430}.
When $m$ is large, the hypermultiplets decouple and the pure ${\rm SU}(N)$ theory is recovered. Thus, in analogy with the $\SU(2)$ case, it is suggestive to consider that the Segre numbers match with the contribution from the  $\CN=1$ vacua in the large $m$ limit. While challenging, it would be interesting to test this conjecture, and to extend our techniques, dictionary and results to arbitrary rank $N$.

\sss{5d/6d uplift and Segre--Verlinde correspondence.}
Topological $\CN=2$ QCD can be considered a (rather elaborate) toy model for the formulation of an arbitrary 4d $\CN=2$ theory on a four-manifold, such as those of class $\CS$~\cite{Gaiotto:2009hg,Manschot:2021qqe,Moore:2024vsd}. It is feasible that the methods demonstrated in this article will be useful to study partition functions of 
other theories, including those originating from torus compactifications of 5d~\cite{Nakajima:2005fg,Gottsche:2006,Kim:2025fpz} and 6d~\cite{Gottsche:2018epz,Gukov:2018iiq,Pei:2025cud} theories followed by a partial topological twist. Such analyses will hopefully provide the physical explanation for various correspondences between invariants of four-manifolds, such as the Segre--Verlinde correspondence and its relation to strange duality~\cite{lepotier,johnson2020universal,Gttsche2018,Gttsche2019}. The Segre--Verlinde correspondence involves a conjectural relation between universal functions for virtual Segre numbers and $K$-theoretic invariants, and is under current investigation in algebraic geometry~\cite{johnson2020universal,Gottsche:2020ass,Marian2021,Gttsche2024,Marian2024}.

\appendix

\section{Modular and elliptic functions}
\label{app:modularforms}

In this Appendix, we set the notation for various modular and elliptic functions used throughout the paper. See for instance~\cite{erdelyiII,Zagier92} for comprehensive reviews. 

\sss{Modular forms.}
For the computation of Coulomb branch couplings, we require the elliptic Jacobi theta functions $\vartheta_j:\mathbb{H}\times
\mathbb{C}\to \mathbb{C}$, $j=1,\dots,4$. With $q=e^{2\pi i\tau}$, these are defined as
\be\label{jt_elliptic}
\begin{split}
&\vartheta_1(\tau,z)=i \sum_{r\in
  \mathbb{Z}+\frac12}(-1)^{r-\frac12}q^{r^2/2}e^{2\pi i
  rz}~, \\
&\vartheta_2(\tau,z)= \sum_{r\in
  \mathbb{Z}+\frac12}q^{r^2/2}e^{2\pi i
  rz}~,\\
&\vartheta_3(\tau,z)= \sum_{n\in
  \mathbb{Z}}q^{n^2/2}e^{2\pi i
  n z}~,\\
&\vartheta_4(\tau,z)= \sum_{n\in 
  \mathbb{Z}} (-1)^nq^{n^2/2}e^{2\pi i
  n z}~.
\end{split}
\ee
We also use notation $\jt_j=\jt_j(\tau)=\jt_j(\tau,0)$ or $\jt_j(z)=\jt_j(\tau,z)$. 
The $S$-transformation reads
\bea\label{jt_S-transformation}
\jt_1(-1/\tau,z/\tau)&=-i\sqrt{-i \tau}e^{\pi i z/\tau^2}\jt_1(\tau,z)~, \\
\jt_2(-1/\tau,z/\tau)&=\sqrt{-i \tau}e^{\pi i z/\tau^2}\jt_4(\tau,z)~,\\
\jt_3(-1/\tau,z/\tau)&=\sqrt{-i \tau}e^{\pi i z/\tau^2}\jt_3(\tau,z)~,\\
\jt_4(-1/\tau,z/\tau)&=\sqrt{-i \tau}e^{\pi i z/\tau^2}\jt_2(\tau,z)~,
\eea
and we have
\bea
\jt_1(\tau,0)&=0~,  \\
\jt_1'(\tau,0)&=-2\pi \eta(\tau)^3~. 
\eea
We furthermore require the Eisenstein series 
\be\label{eisenstein}
E_{k}(\tau)=1-\frac{2k}{B_k}\sum_{n=1}^\infty \sigma_{k-1}(n)\,q^n~,
\ee
and in particular $E_2$, which transforms under $\slz$ as
\be 
\label{E2trafo}
E_2\!\left(\frac{a\tau+b}{c\tau+d}\right) =(c\tau+d)^2E_2(\tau)-\frac{6\im}{\pi}c(c\tau+d)~.
\ee
The Dedekind eta function is defined as the infinite product
\bea\label{dedekind eta}
\eta(\tau)=q^{\frac{1}{24}}\prod_{n=1}^\infty(1-q^n)~.
\eea

\sss{The \ws{} elliptic functions.}
Let us also review some important properties of the \ws{} elliptic functions. In order to motivate it, consider the relation
\bea\label{elliptic integral}
z=\int_\infty^w \frac{dx}{y}
\eea
between $z$ and $w$, where $x$ and $y$ satisfy the \ws{} equation for an elliptic curve, $y^2=4x^3-g_2x-g_3$. Eq. (\ref{elliptic integral}) may be inverted to yield the \ws{} $\wp$ function.

In order to define it properly, consider a lattice $\Lambda=\mathbb Z\omega \oplus\mathbb Z\mathbb \omega'$, with $\tau=\omega'/\omega \in \mathbb H$.\footnote{In some standard literature~\cite{erdelyiII}, it is common to use as periods $2\omega$ and $2\omega'$. We use  $\omega$ and $\omega'$ instead.}
The \ws{} elliptic function is then a map from $\mathbb C/\Lambda$ to $\mathbb C$, 
\bea
\label{eq:WPv1}
\wp_\Lambda(z)\coloneqq \wp_\Lambda(z,\omega,\omega')\coloneqq \frac{1}{z^2}+\sum_{\lambda\in \Lambda\setminus \{0\}}\left(\frac{1}{(z-\lambda)^2}-\frac{1}{\lambda^2}\right)~,
\eea
where we use the subscript $\Lambda$ to stress that it is a lattice sum. The \ws{} $\wp_\Lambda$ function is homogeneous of degree $-2$, that is, it satisfies
\bea\label{wp homogeneity}
\wp_\Lambda(tz,t\omega,t\omega')=t^{-2}\wp_\Lambda(z,\omega,\omega')
\eea
for all $t\neq 0$. Thus, we can express it as a function of $z$ and $\tau=\omega'/\omega$ only, which we denote by 
\bea\label{eq:WPv2}
\wp(z,\tau)\coloneqq\wp_\Lambda(z,1,\tau)=\omega^2\wp_\Lambda(\tfrac{z}{\omega},\omega,\omega')~,
\eea
where we remove the subscript. Moreover, we sometimes drop the dependence on $\tau$, and denote $\wp(z)=\wp(z,\tau)$. 
The function $\wp$ is meromorphic in $\mathbb C$ with double poles in $\Lambda_\tau=\mathbb Z\oplus\tau\mathbb Z$. It is even, $\wp(-z)=\wp(z)$, and doubly periodic, $\wp(z+\lambda)=\wp(z)$ for all $\lambda\in\Lambda$. The Laurent expansion at $z=0$ reads
\bea
\wp(z)=\frac{1}{z^2}+\sum_{k=1}^\infty (2k+1)G_{2k+2}z^{2k}~, \qquad G_{k}(\Lambda)=\sum_{\lambda\in \Lambda\setminus \{0\}}\lambda^{-k}~,
\eea
where $G_k$ are the Eisenstein series, which relate as $G_k=2\zeta(k)E_k$ to the normalisations $E_k$ of before~\eqref{eisenstein}. We can express $\wp(z)$ in terms of the elliptic Jacobi functions,
\bea\label{wp jtf}
\wp(z)=\left(\frac{\pi\jt_2\jt_3\jt_4(z)}{\jt_1(z)}\right)^2-\frac{\pi^2}{3}(\jt_2^4+\jt_3^4)~,
\eea
where we suppress the $\tau$ dependence.
We require furthermore the identity 
\bea\label{wp z- tau}
\wp(z-\tfrac\tau2,\tau)=-\jt_4^{(2)}(\tau,z)-\tfrac{\pi^2}{3}E_2(\tau)~,
\eea
in the notation of~\eqref{3 equations generic ki}.

The zeros of $\wp'$ are the half-periods
\bea
e_\alpha=\wp(\tfrac{\omega_\alpha}{2})~,
\eea
with the notation $\omega_\alpha$ for $\alpha=1,2,3$, with $\omega_1=\omega$, $\omega_2=-\omega-\omega'$ and $\omega_3=\omega'$, such that $\sum_\alpha \omega_\alpha=0$. That is, 
\bea
\frac{\omega_\alpha}{2}=\int_\infty^{e_\alpha} \frac{dx}{y}~.
\eea
The half-periods obey $e_1+e_2+e_3=0$, and the \ws{} function satisfies the differential equation
\bea
\label{eq:Weierstrass}
\wp'(z)^2&=4(\wp(z)-e_1)(\wp(z)-e_2)(\wp(z)-e_3) \\
&=4\wp(z)^3-g_2 \wp(z)-g_3~,
\eea
which gives the uniformisation $(x,y)=(\wp(z),\wp'(z))$ of the elliptic curve $y^2=4x^3-g_2x-g_3$.

Finally, we introduce the \ws{} $\zeta$ and $\sigma$ functions, which are the (logarithmic) primitives of the $\wp$ function,
\bea
\wp(z)&=-\zeta'(z)~, \\
\zeta(z)&=\frac{d}{dz}\log \sigma(z)~.
\eea

\section{SW geometry and couplings}
\label{app:largemassSW}
In this Appendix, we introduce the SW curves for $\SU(2)$ SQCD, the explicit calculation of couplings from the prepotential, and an independent derivation of couplings from the SW curve.

\subsection{SW curves}
\label{app:curves}
We consider the $\SU(2)$ SQCD SW curves~\cite{Seiberg:1994aj}
\begin{equation}\label{eq:curves}
	\begin{aligned}  
		N_f=0:\quad y^2&=x^3-ux^2+\frac{1}{4}\Lambda_0^4x~, \\
		N_f=1:\quad y^2&=x^2(x-u)+\frac{1}{4}m\Lambda_1^3x-\frac{1}{64}\Lambda_1^6~, \\
		N_f=2:\quad y^2&=(x^2-\frac{1}{64}\Lambda_2^4)(x-u)+\frac{1}{4}m_1m_2\Lambda_2^2x-\frac{1}{64}(m_1^2+m_2^2)\Lambda_2^4~, \\
		N_f=3:\quad y^2&=
		x^2(x-u)-\frac{1}{64}\Lambda_3^2(x-u)^2-\frac{1}{64}(m_1^2+m_2^2+m_3^2)\Lambda_3^2(x-u)\\&\quad
		+\frac{1}{4}m_1m_2m_3\Lambda_3x-\frac{1}{64}(m_1^2m_2^2+m_2^2m_3^2+m_1^2m_3^2)\Lambda_3^2~.
	\end{aligned}
\end{equation} 
Bringing the curves in \ws{} form $y^2=x^3-g_2 x-g_3$,
we determine the \ws{} invariants $g_2$ and $g_3$. We work in the normalisation such that $g_2\sim \frac43u^2$ and $g_3\sim \frac{8}{27}u^3$. We furthermore introduce the monic polynomials $\PM_{N_f}$ of degree $N_f$. We collect the masses of the fundamental hypermultiplets in the mass vector $\bfm=(m_1,\dots,m_{N_f})$. 

For equal masses $\bfm=(m,\dots, m)$, we have for $N_f=1$:
\bea\label{g2g3 Nf=1}
g_2&=\frac{4 u^2}{3}-\Lambda_1^3 m~, \\
g_3&=\frac{\Lambda_1^6}{16}-\frac{1}{3} \Lambda_1^3 m u+\frac{8 u^3}{27}~, \\
\PM_1&=u-\frac43m^2~,
\eea
for $N_f=2$, 
\bea\label{g2g3 Nf=2}
g_2&=\frac{\Lambda_2^4}{16}-\Lambda_2^2 m^2+\frac{4 u^2}{3}~, \\
g_3&=\frac{\Lambda_2^4 m^2}{8}-\frac{1}{24} u \left(\Lambda_2^4+8 \Lambda_2^2 m^2\right)+\frac{8 u^3}{27}~, \\
\PM_2&=-\frac{\Lambda_2^4}{64}+2 m^4+m^2 \left(\frac{\Lambda_2^2}{8}-3 u\right)+u^2~,
\eea
and for $N_f=3$, 
\bea\label{g2g3 Nf=3}
g_2&=\frac{\Lambda_3^4}{3072}-\Lambda_3 m^3+\frac{1}{48} \Lambda_3^2 \left(9 m^2-4 u\right)+\frac{4 u^2}{3}~, \\
g_3&=\frac{8 u^3}{27}+\frac{5 \Lambda_3^2 u^2}{144}-\frac{\Lambda_3 u \left(\Lambda_3^3+768 m^3+288 \Lambda_3 m^2\right)}{2304} \\
   &\quad+\frac{\Lambda_3^2 \left(\Lambda_3^4+165888 m^4-4608 \Lambda_3
   m^3+864 \Lambda_3^2 m^2\right)}{884736}~, \\
\PM_3&=-\frac{1}{256} \left(16 m^2-\Lambda_3 m-4 u\right) (m (\Lambda_3+8 m)-8 u)^2~.
\eea
From these, we form the physical discriminant $\Delta=g_2^3-27g_3^2$, whose zeros give us the singularities. 
In the equal mass case $\bfm=(m,\dots, m)$, the singular configuration is $(I_{4-N_f}^{\infty,*},I_1,I_1,I_{N_f})$, in Kodaira's notation of singular fibres. We can express the two $I_1$ singularities for all $N_f$, including the superconformal case $N_f=4$ for completeness:
\bea\label{upm_exact}
\begin{tabular}{|c|| l|c|c|c|} 
\hline
$N_f$ & $\tu_{N_f}^\pm$ & $\mad$ & AD type \\
\hline\hline
0& $\pm \Lambda_0^2$ & & \\
1&$\frac{m^2}{3}+\frac{1}{24}d^{\frac13}\omega^{\mp1}+\frac{8m^4+27m\Lambda_1^3}{3d^{\frac13}}\,\omega^{\pm1}$ &  $\frac34\Lambda_1$&$II$\\
2&$\pm m\Lambda_2-\frac18\Lambda_2^2$&$\frac12\Lambda_2$ &$III$\\
3&$\frac{\Lambda_3^2-96m\Lambda_3\pm\sqrt{\Lambda_3(64m+\Lambda_3)^3}}{512}$ &$\frac18\Lambda_3$&$IV$\\
4&$-\frac{m^2}{3}\jt_2(\tn)^2\jt_3(\tn)^2\left(2\tfrac{\jt_2(\tn)^4+\jt_3(\tn)^4}{\jt_2(\tn)^2\jt_3(\tn)^2}\mp 6\right)$ & &
\\ \hline
\end{tabular}
\eea
where $\omega=e^{2\pi i/3}$,  and 
\begin{equation}
d=512m^6-4320m^3\Lambda_1^3-729\Lambda_1^6+\sqrt{27}(27\Lambda_1^4-64m^3\Lambda_1)^{\frac32}~.
\end{equation}
The second-to-last column lists the value of the equal mass $m=\mad$ such that the $u$-plane contains an Argyres--Douglas point~\cite{Argyres:1995xn}, and the last column gives its Kodaira type. 
We omit the SW curve for $N_f=4$, which we discuss briefly in Section~\ref{sec:discussion}, and hope to include it in future work.

In particular cases, SW curves become modular~\cite{Aspman:2021vhs,Aspman:2021evt,Closset:2021lhd,Magureanu:2022qym}. It shall be noted that the configurations of SQCD with $N_f=1,2,3$ generic masses are not modular. However, by tuning the masses to special values, the curves become modular, and the Coulomb branch is described by modular functions. For the equal mass configurations relevant to this paper, for every $N_f$, there is one branch point  
$u_{\rm bp}=\frac{4}{4-N_f}m^2$ on the Coulomb branch~\cite{Aspman:2021vhs}, and the fundamental domain is as in Fig. 11 of~\cite{Aspman:2023ate}.

\subsection{Prepotential}
The non-perturbative effective action of $\CN=2$ SQCD is characterised by the prepotential $F(a,\bfm)$. It can famously be derived by instanton counting techniques in the $\Omega$-background~\cite{Ne,NekOk,Nekrasov:2012xe}. 
The semi-classical part of $F$ reads~\cite{Ohta:1996hq, Ohta_1997,DHoker_1997, Ne}
\be\label{prepotential}
\begin{split}
&F_{\rm sc}(a,\bfm)=\frac{2\im}{\pi}  a^2 \left(\log(4a/\Lambda_{N_f})-\tfrac{3}{2}\right)+\frac{1}{2}\sum_{j=1}^{N_f}\left( \tfrac{m_j}{\sqrt{2}}\,a- c\left(a^2+\tfrac{m_j^2}{2}\right)\right)\\
&\quad -\frac{\im }{4\pi} \sum_{j=1}^{N_f} \sum_{\pm}\left(a\pm\tfrac{m_j}{\sqrt 2}\right)^2\log\left((a\pm\tfrac{m_j}{\sqrt 2})/\Lambda_{N_f}\right)~,
\end{split} 
\ee 
where $c=\frac{1}{2}-\frac{3i}{2\pi}+\frac{i}{2\pi}\log2$. 
It receives non-perturbative corrections~\cite{Seiberg:1994rs,Seiberg:1994aj}
\begin{equation}\label{F_np_def}
    F_{\rm np}(a,\bfm)=\sum_{k=1}^\infty F_k(\bfm) a^{-2k}~,
\end{equation}
which arise from contributions of $k$-instantons. Their sum 
\begin{equation}\label{prepotential_total}
    F(a,\bfm)=F_{\rm cs}(a,\bfm)+F_{\rm np}(a,\bfm)
\end{equation}
describes the exact low-energy dynamics, and encodes the Seiberg--Witten curves~\eqref{eq:curves} through $\tau=\frac{\partial^2F(a,\bfm)}{\partial a^2}$.

To determine the background couplings introduced in Section~\ref{sec:CB geometry}, we first need to compute the non-perturbative corrections $F_{\rm np}$. This can be done by direct integration of the SW periods, which admit a hypergeometric representation~\cite{Masuda:1996xj,Brandhuber:1996ng}
\begin{equation}\label{periods_hyperg}
	\begin{aligned}
		\frac{\partial a}{\partial u} =\,& \frac{1}{(48g_2)^{1/4}}\,{}_2F_1\left[\tfrac{1}{12},\tfrac{5}{12},1;\tfrac{12^3}{\CJ}\right], \\
		\frac{\partial a_D}{\partial u}=\,&\frac{i}{(48g_2)^{1/4}}\left(3\log(12)\,{}_2F_1\left[\tfrac{1}{12},\tfrac{5}{12},1;\tfrac{12^3}{\CJ}\right]-F^*\left[\tfrac{1}{12},\tfrac{5}{12},1;\tfrac{12^3}{\CJ}\right]\right).
	\end{aligned}
\end{equation}
Here, ${}_2F_1$ is the ordinary hypergeometric function, while $F^*$ is defined as 
\begin{equation}
	F^*\left[\alpha,\beta,1;z\right]={}_2F_1\left[\alpha,\beta,1;z\right]\log z+\sum_{n=0}^\infty\frac{(\alpha)_n(\beta)_n}{(n!)^2}z^n\sum_{r=0}^{n-1}\frac{1}{\alpha+r}+\frac{1}{\beta+r}-\frac{2}{r+1}.
\end{equation}
Moreover, $\CJ=12^3g_2^3/(g_2^3-27g_3^2)$ denotes the $\CJ$-invariant of the curve~\eqref{eq:curves}, where we use the \ws{} invariants~\eqref{g2g3 Nf=1}--\eqref{g2g3 Nf=3}.

In order to find $F_{\rm np}$, we compute the large $u$ series of $\frac{\partial a_D}{\partial u}$~\eqref{periods_hyperg} and integrate it order by order w.r.t. $u$ to find $a_D$ as a function of $u$. To express this as a series for large $a$ rather than $u$, we compute the large $a$ series of $\frac{\partial a}{\partial u}$ and integrate w.r.t. $u$ to obtain $a$ as a series at $u=\infty$. Inverting this series order by order gives $u$ as a function of large $a$, which we can insert into $a_D(u)$ to obtain $a_D$ as a function of $a$. Integration of this new series w.r.t. $a$ gives the full prepotential $F=F_{\rm cs}+F_{\rm np}$, up to integration constants. We fix the constants by adding terms proportional to $a^n$ with $n=0,1,2$ to the integration result.
In this way, we obtain arbitrarily many non-perturbative corrections $F_k(\bfm)$ to the semi-classical prepotential.\footnote{For \eg~$N_f=1$, we have computed $F_k$ up to $k=23$.} 
A faster method to determine the $F_k$ is to compute them directly by comparison with the SW curve.\footnote{An important consistency check is that the prepotential $F$ indeed derives the SW solution. From $F(a,\bfm)$ we compute $\tau=\frac{\partial^2 F}{\partial a^2}$, insert the series $a=a(u)$ into $q=e^{2\pi i\tau}$ and invert the resulting series. This gives $u=u(\tau)$, which can be compared directly with the curves~\eqref{eq:curves}.}

\sss{Homogeneity.}
The Seiberg--Witten prepotential enjoys a homogeneity property, which originates from special geometry and its relation to supergravity (see~\eg~\cite{deWit:1984wbb}). Viewing $F$ as a function of $a$, $\bfm$ and $\Lambda_{N_f}$, it is a homogeneous function of degree 2, which means that 
\bea
F(\lambda a,\lambda \bfm, \lambda \Lambda_{N_f})=\lambda^2 F(a,\bfm, \Lambda_{N_f})~,
\eea
for all $\lambda\neq 0$ and all $a$, $\bfm$ and $\Lambda_{N_f}$. Euler's homogeneous function theorem then states that 
\bea\label{homogeneity}
2F=a\frac{\partial F}{\partial a}+\bfm \frac{\partial F}{\partial \bfm}+\Lambda_{N_f} \frac{\partial F}{\partial \Lambda_{N_f}}~,
\eea
where the sum over the masses $\bfm$ is understood. 

\sss{Correction to prepotential.}
In order to find a precise match with the mathematical literature in Section \ref{sec:deriving_alg_functions}, we add to the above prepotential $F$ the terms
\bea \label{F shift}
\mathcal{F} = F-\frac{3i}{8\pi N_f}(\bfm^2)+\frac{i}{4\pi}(\bfm^2)\log(\tfrac{(\bfm)}{\Lambda_{N_f}})-\frac{i}{32\pi}\Lambda_2^2\delta_{N_f,2}-\frac{i}{32\pi}\Lambda_3(\bfm)\delta_{N_f,3}~,
\eea
where $F$ is the prepotential with semi-classical part $F_{\rm sc}$ \eqref{prepotential}, $(\bfm^p)=\sum_{j=1}^{N_f}m_j^p$ for $p=1,2$.
Such shifts originate from extra terms in Nekrasov's partition function, which account for the decoupling of the contribution of ${\rm U}(1)\subset {\rm U}(2)$ gauge fields~\cite{Alday:2009aq,Manschot:2019pog},
and are also required in the $\SU(2)$ $\nstar$ theory~\cite{Manschot:2019pog,Manschot:2021qqe}. They do not affect the low-energy theory. 
It is straightforward to obtain the shifts in the CB functions: 
\bea\label{shift functions}
u^{\CF}&= u^{F}+\tfrac18\Lambda_2^2\delta_{N_f,2}+\tfrac18 (\bfm)\Lambda_3\delta_{N_f,3}~, \\
G_{N_f}^{\CF}&= G_{N_f}^{F}+\tfrac18\delta_{N_f,2}+\tfrac{1}{8\Lambda_3}(\bfm)\delta_{N_f,3}~, \\
H_{N_f}^{\CF}&= H_{N_f}^{F}-\tfrac{2\sqrt2 i}{3\Lambda_1}m\delta_{N_f,1}-\tfrac{i}{\sqrt2\Lambda_2}(\bfm)\delta_{N_f,2}-\left(\tfrac{i}{4\sqrt2}+\tfrac{2\sqrt2 i}{3\Lambda_3}(\bfm)\right)\delta_{N_f,3}~.
\eea

\subsection{Calculation of couplings}
\label{app:coupling bf}
In this section, we sketch the method to compute the series $\tv_s$ (and hence $\tD_s=e^{-\pi i \tv_s}$), $\tC_s$, $\hat \tC_s$ and $\tH_s$ at the monopole singularity, from their definitions
\bea\label{couplings appendix}
v_j&=\sqrt{2}\frac{\partial^2 \CF}{\partial a \partial m_j}~, \\
C_{jk}&=\exp\left(-2\pi i \frac{\partial^2 \CF}{\partial m_j\partial m_k}\right)~, \\
H_{j}&=\frac{4\sqrt2\pi}{4-N_f}\frac{\partial^2 \CF}{\partial \Lambda_{N_f}\partial m_j}~,
\eea
as Coulomb branch functions, by exploiting the duality and modularity for large equal masses. 

Our aim is to compute these couplings for large equal masses $m$.  The general strategy is as follows: In the $m\to\infty$ limit, the equal mass theory decouples to the pure ($N_f=0$) $\SU(2)$ SYM, whose Coulomb branch is a modular curve for the congruence subgroup $\Gamma^0(4)$~\cite{Seiberg:1994rs,Seiberg:1994aj,Moore:1997pc}. Consequently, the SQCD Coulomb branch functions become Taylor series in $\frac 1m$ (or some root thereof), with coefficients being (quasi)-modular forms for $\Gamma^0(4)$~\cite{Aspman:2023ate}. The local expansion around the strong coupling singularities can then be obtained by a duality transformation.

\sss{Large mass series.}
The prepotential~\eqref{prepotential_total} gives a series at large $a$, with coefficients in the masses $\bfm$ and the scale $\Lambda_{N_f}$. In order to determine the couplings~\eqref{couplings appendix} as a large mass series with coefficients in $\tau$, we need to express the period $a$ as a large $m$ series with (modular) coefficients in $\tau$. To do so, we integrate Matone's relation~\cite{Matone:1995rx, Aspman:2021vhs}
\begin{equation}\label{Matone}
	\frac{du}{d\tau}=-\frac{16\pi i}{4-N_f}\frac{\Delta_{N_f}}{P^{\text{M}}_{N_f}}\left(\frac{da}{du}\right)^2~.
\end{equation}
Here, $\PM_{N_f}$ are the polynomials listed in~\eqref{g2g3 Nf=1}--\eqref{g2g3 Nf=3}, which encode the branch points on the fundamental domains, while $\Delta_{N_f}$ are the discriminants of the curve. For this, we need the order parameters 
$u_{(N_f)}$ of the $N_f$-theory in terms of $u_{(0)}$, the one of the pure $N_f=0$ theory. 
As described in~\cite{Aspman:2021vhs}, can find this large $m$ series using the ansatz
\be\label{uNf_large_m}
u_{(N_f)}=\sum_{n=0}^\infty c_n^{N_f}(u_{(0)})m^{-2n}~,
\ee
and iteratively find the polynomials $c_n^{N_f}$ by satisfying the relation $\CJ_{N_f}(u_{(N_f)})=\CJ_{0}(u_{(0)})$ order by order in $m^{-2}$. Using~\eqref{Matone}, this gives the large $m$ series of $\frac{da}{d\tau}=\frac{da}{du}\frac{du}{d\tau}$:
\begin{equation}
    \frac{da}{d\tau}=\sum_{n=0}^\infty f_n^{N_f}(\tau)m^{-2n}~,
\end{equation}
with coefficients $f_n^{N_f}(\tau)$ being weight 3 modular forms for $\Gamma^0(4)$, obtained from $u_{(0)}=-\frac{\Lambda_0^2}{2} \frac{\jt_2^4+\jt_3^4}{\jt_2^2\jt_3^2}$. Integration of $\frac{da}{d\tau}$ w.r.t. $\tau$ gives a similar series
\begin{equation}\label{a(tau)}
    a(\tau)=\sum_{n=0}^\infty \tilde f_n^{N_f}(\tau)m^{-2n}~,
\end{equation}
with coefficients $\tilde f_n^{N_f}$ now being quasi-modular forms.

To obtain the couplings~\eqref{couplings appendix}, we differentiate the prepotential $\CF$ w.r.t. to $a$ and $m_j$, set the masses equal to $m$, and then insert the series $a(\tau)$~\eqref{a(tau)}. The coefficients are strongly constrained by modularity, and a small number of coefficients in the $q$-series is sufficient to determine them exactly. 
By direct inspection, for each $N_f=1,2,3$,  the four independent functions admit the following large $m$ series:
\bea\label{large m series modular forms}
v_1(\tau)&=\frac{1}{\sqrt2\pi}\sum_{n=0}^\infty \frac{c^{v_1}_n f^{v_1}_n(\tau)}{(\jt_2(\tau)\jt_3(\tau))^{2n+1}}\left(\frac{\Lambda_0}{m}\right)^{2n+1}~, \\
C_{11}(\tau)&=\sum_{n=0}^\infty \frac{c_n^{C_{11}} f^{C_{11}}_n(\tau)}{(\jt_2(\tau)\jt_3(\tau))^{2n}}\left(\frac{\Lambda_0}{m}\right)^{2n+\frac{4}{4-N_f}}~, \\
C_{12}(\tau)&=\sum_{n=0}^\infty \frac{c_n^{C_{12}} f^{C_{12}}_n(\tau)}{(\jt_2(\tau)\jt_3(\tau))^{2n}}\left(\frac{\Lambda_0}{m}\right)^{2n}~, \\
H(\tau)&=\sqrt 2i\sum_{n=0}^\infty \frac{c_n^{H} f^{H}_n(\tau)}{(\jt_2(\tau)\jt_3(\tau))^{2n}}\left(\frac{\Lambda_0}{m}\right)^{2n-\frac{4}{4-N_f}}~,
\eea
Here,  $c^{v_1}_n$, $c_n^{C_{11}}$, $c_n^{C_{12}}$ and $f_n^H\in\mathbb Q$ are rational coefficients, 
$f^{v_1}_n\in\mathbb Z[\jt_2^4,\jt_3^4]$ are modular forms  of weight $4n$ for $\Gamma^0(4)$ with integer coefficients, while $f^{C_{11}}_n,f^{C_{12}}_n,f_n^H\in \mathbb Z[E_2,\jt_2^4,\jt_3^4]$ are quasi-modular forms of weight $2n$ for $\Gamma^0(4)$ with integer coefficients. All four functions $f^{v_1}_n$, $f^{C_{11}}_n$,   $f^{C_{12}}_n$ and $f^H_n$ are invariant under the exchange of $\jt_2^4$ and $\jt_3^4$. 

\subsection{Large mass expansions}
\label{app: large mass expansions}

Let us confirm this prediction by an explicit calculation, as outlined above.

\sss{$\boldsymbol{N_f=1}$.}
We compute:
\beasmall \label{v1 Nf=1 large m}
v_1&=-\frac{1}{\sqrt{2}\pi}  \frac{1}{\vartheta_2\vartheta_3}\frac{\Lambda_0}{m}+\frac{5}{48\sqrt{2}\pi} \frac{\vartheta_2^4+\vartheta_3^4}{\vartheta_2^3\vartheta_3^3} \left(\frac{\Lambda_0}{m}\right)^3\\
&\quad -\frac{33}{2560\sqrt{2}\pi} \frac{\jt_2^8+4\jt_2^4\jt_3^4+\jt_3^8}{\vartheta_2^5\vartheta_3^5} \left(\frac{\Lambda_0}{m}\right)^5 \\
& \quad +\frac{17}{57344\sqrt2 \pi}\frac{(\jt_2^4+\jt_3^4)(3\jt_2^8+52\jt_2^4\jt_3^4+3\jt_3^8)}{\jt_2^7\jt_3^7}\left(\frac{\Lambda_0}{m}\right)^7~, \\
& \quad +\frac{23}{4718592\sqrt2\pi}\frac{35(\jt_2^{16}+\jt_3^{16})-736\jt_2^4\jt_3^4(\jt_2^8+\jt_3^8)-2034\jt_2^8\jt_3^8}{\jt_2^9\jt_3^9}\left(\frac{\Lambda_0}{m}\right)^9+\CO(m^{-{11}})~,
\eeasmall
while 
\beasmall
 C_{11}&=\left(\frac{\Lambda_0}{m}\right)^{\frac43}-\frac{E_2+2\jt_2^4+2\jt_3^4}{12\jt_2^2\jt_3^2}\left(\frac{\Lambda_0}{m}\right)^{\frac{10}{3}} \\
   &\quad +\frac{E_2^2+8(\jt_2^8+\jt_3^8)+28\jt_2^4\jt_3^4+9E_2(\jt_2^4+\jt_3^4)}{288\jt_2^4\jt_3^4}\left(\frac{\Lambda_0}{m}\right)^{\frac{16}{3}} \\
   &\quad -\frac{20E_2^3+420E_2^2(\jt_2^4+\jt_3^4)+22(\jt_2^4+\jt_3^4)(31(\jt_2^8+\jt_3^8)+305\jt_2^4\jt_3^4)+3E_2\left(571(\jt_2^8+\jt_3^8)+1679\jt_2^4\jt_3^4\right)}{207360\jt_2^6\jt_3^6}\left(\frac{\Lambda_0}{m}\right)^{\frac{22}{3}} \\
   &\quad +\CO(m^{-\frac{22}{3}})~,
\eeasmall
and 
\beasmall
H_1&=\frac{2\sqrt2 i}{3}\left(\frac{m}{\Lambda_0}\right)^{\frac43}\Bigg[
1+\frac{-E_2+\jt_2^4+\jt_3^4}{\jt_2^2\jt_3^2}\left(\frac{\Lambda_0}{m}\right)^2-\frac{-E_2(\jt_2^4+\jt_3^4)+\jt_2^8+\jt_3^8+5\jt_2^4\jt_3^4}{96\jt_2^4\jt_3^4}\left(\frac{\Lambda_0}{m}\right)^4 \\
&\quad \frac{-E_2(\jt_2^8+\jt_3^8+29\jt_2^4\jt_3^4)+(\jt_2^4+\jt_3^4)(\jt_2^8+\jt_3^8+95\jt_2^4\jt_3^4)}{7680\jt_2^6\jt_3^6}\left(\frac{\Lambda_0}{m}\right)^6 \\
&\quad -\frac{E_2(\jt_2^4+\jt_3^4)(16(\jt_2^8+\jt_3^8)-61\jt_2^4\jt_3^4)-16(\jt_2^{16}+\jt_3^{16})+158\jt_2^4\jt_3^4(\jt_2^8+\jt_3^8)+498\jt_2^8\jt_3^8)}{71680\jt_2^8\jt_3^8}\left(\frac{\Lambda_0}{m}\right)^8
\Bigg] \\
&\quad +\CO(m^{-10})~,
\eeasmall
where for $H_1$ we incorporate the shift~\eqref{shift functions}. We have computed $v_1$ up to $\CO(m^{-21})$, $C_{11}$ up to $\CO(m^{-\frac{34}{3}})$ and $H_1$ up to $\CO(m^{-\frac{38}{3}})$.

Using the modular $S$-transformation $\jt_2\to \sqrt{-i \tau}\jt_4$, $\jt_3\to \sqrt{-i \tau}\jt_3$ \eqref{jt_S-transformation}, $E_2\to \tau^2E_2 - 6i\tau/\pi$ (\ref{E2trafo}) followed by evaluating at $q=0$, we find the values at strong coupling $\tau=0$. Using $z=\frac{\Lambda_0^2}{2m^2}$, this gives
\bea\label{C1 D1 H1}
\tfrac{m}{\Lambda_1}\tC_1=\, \, &1-\tfrac{z}{2}+\tfrac{3 z^2}{8}-\tfrac{21 z^3}{64}+\tfrac{5 z^4}{16}-\tfrac{1287 z^5}{4096}+\CO(z^6)~, \\
\tD_1=\, \, &1-\sqrt{z}+\tfrac{z}{2}+\tfrac{z^{\frac{3}{2}}}{4}-\tfrac{3 z^2}{8}-\tfrac{7 z^{\frac{5}{2}}}{64}+\tfrac{21 z^3}{64}+\tfrac{13 z^{\frac{7}{2}}}{256}-\tfrac{5 z^4}{16}-\tfrac{133 z^{\frac{9}{2}}}{8192}+\tfrac{1287 z^5}{4096} \\
\, \, &-\tfrac{315
   z^{\frac{11}{2}}}{32768}-\tfrac{21 z^6}{64}+\tfrac{17143 z^{\frac{13}{2}}}{524288}+\tfrac{46189 z^7}{131072}-\tfrac{117549 z^{\frac{15}{2}}}{2097152}-\tfrac{99 z^8}{256}+\tfrac{10963581 z^{\frac{17}{2}}}{134217728} \\
\, \, &+\tfrac{7243275
   z^9}{16777216}-\tfrac{59743005 z^{\frac{19}{2}}}{536870912}-\tfrac{1001 z^{10}}{2048}+\tfrac{1258550623 z^{\frac{21}{2}}}{8589934592}+\tfrac{300540195 z^{11}}{536870912}
   z^{\frac{23}{2}} +\CO(z^{12})~, \\
-\tfrac{i}{\sqrt2}\tfrac{\Lambda_1}{m} 
   \tH_1=\, \, &z-\tfrac{z^2}{4}+\tfrac{5 z^3}{32}-\tfrac{z^4}{8}+\tfrac{231 z^5}{2048}-\tfrac{7 z^6}{64}+\tfrac{7293 z^7}{65536} +\CO(z^8)~.
\eea
Using these series, it is then easy to verify that the remaining equations in~\eqref{universal functions from SW} are satisfied. In particular, according to~\eqref{CB functions from universal}, the first line above is $\tfrac{m}{\Lambda_1}\tC_1=W_1$, while $\tD_1^2=Y_{1,1}$, and the third line is equal to $R_1$.

\sss{$\boldsymbol{N_f=2}$.}
Similarly, for $N_f=2$ we find:
\beasmall\label{v1 Nf=2}
v_1&=-\frac{1}{\sqrt{2}\pi}  \frac{1}{\vartheta_2\vartheta_3}\frac{\Lambda_0}{m}+\frac{1}{24\sqrt{2}\pi} \frac{\vartheta_2^4+\vartheta_3^4}{\vartheta_2^3\vartheta_3^3} \left(\frac{\Lambda_0}{m}\right)^3 +\frac{3}{640\sqrt{2}\pi} \frac{\jt_2^8-6\jt_2^4\jt_3^4+\jt_3^8}{\vartheta_2^5\vartheta_3^5} \left(\frac{\Lambda_0}{m}\right)^5 \\
& \quad -\frac{5}{7168\sqrt2 \pi}\frac{(\jt_2^4+\jt_3^4)(\jt_2^8-6\jt_2^4\jt_3^4+\jt_3^8)}{\jt_2^7\jt_3^7}\left(\frac{\Lambda_0}{m}\right)^7+\CO(m^{-9})~,
\eeasmall
while
\beasmall \label{Cij Nf=2}
         C_{11}&=\left(\frac{\Lambda_0}{m}\right)^{2}-\frac{E_2+2(\jt_2^4+\jt_3^4)}{12\jt_2^2\jt_3^2}\left(\frac{\Lambda_0}{m}\right)^{4} \\
   &+\frac{E_2^2+2(\jt_2^8+\jt_3^8)+34\jt_2^4\jt_3^4+6E_2(\jt_2^4+\jt_3^4)}{288\jt_2^4\jt_3^4}\left(\frac{\Lambda_0}{m}\right)^{6} \\
   &+\frac{5E_2^3+60E_2^2(\jt_2^4+\jt_3^4)+E_2(57(\jt_2^8+\jt_3^8)+888\jt_2^4\jt_3^4)-2(\jt_2^4+\jt_3^4)(61(\jt_2^8+\jt_3^8)-850\jt_2^4\jt_3^4)}{51840\jt_2^6\jt_3^6}\left(\frac{\Lambda_0}{m}\right)^{8}\\
   &+\CO(m^{-10})~,
\eeasmall
\beasmall 
C_{12}&=1+\frac{-E_2+\jt_2^4+\jt_3^4}{12\jt_2^2\jt_3^2}\left(\frac{\Lambda_0}{m}\right)^{2}-\frac{-E_2^2+\jt_2^8+\jt_3^8+8\jt_2^4\jt_3^4}{288\jt_2^4\jt_3^4}\left(\frac{\Lambda_0}{m}\right)^{4}
   \\
&+\frac{5E_2^3+15E_2^2(\jt_2^4+\jt_3^4)-78E_2(\jt_2^8+\jt_3^8-\jt_2^4\jt_3^4)+2(\jt_2^4+\jt_3^4)(29(\jt_2^8+\jt_3^8)-185\jt_2^4\jt_3^4)}{51840\jt_2^6\jt_3^6}\left(\frac{\Lambda_0}{m}\right)^{6} \\
&+\CO(m^{-8})~,
\eeasmall
and 
\beasmall
H_2&=2\sqrt2 i\Bigg[
\frac{-E_2+\jt_2^4+\jt_3^4}{12\jt_2^2\jt_3^2}
+\frac{-E_2(\jt_2^4+\jt_3^4)+\jt_2^8+\jt_3^8-4\jt_2^4\jt_3^4}{144\jt_2^4\jt_3^4}\left(\frac{\Lambda_0}{m}\right)^2 \\
&+\frac{E_2(\jt_2^8+\jt_3^8+14\jt_2^4\jt_3^4)-(\jt_2^4+\jt_3^4)(\jt_2^8+\jt_3^8-10\jt_2^4\jt_3^4)}{5760\jt_2^6\jt_3^6}\left(\frac{\Lambda_0}{m}\right)^4 \\
&+\frac{2E_2(\jt_2^{12}+\jt_3^{12}-5\jt_2^4\jt_3^4(\jt_2^4+\jt_3^4))-2(\jt_2^{16}+\jt_3^{16})+11\jt_2^4\jt_3^4(\jt_2^8+\jt_3^8)-34\jt_2^8\jt_3^8}{20160\jt_2^8\jt_3^8}\left(\frac{\Lambda_0}{m}\right)^6 +\CO(m^{-8})\Bigg]~,
\eeasmall
From these, we again compute:
\bea\label{C2 D2 H2}
\tfrac{m}{\Lambda_2}\tC_2=\, \, &1-\tfrac{z}{2}+\tfrac{3 z^2}{8}-\tfrac{5 z^3}{16}+\CO(z^4)~,\\
\hat \tC_2=\, \, &1+\tfrac{z}{2}-\tfrac{z^2}{8}+\tfrac{z^3}{16}+\CO(z^4)~,\\
\tD_2=\, \, &1-\sqrt{z}+\tfrac{z}{2}-\tfrac{z^2}{8}+\tfrac{z^3}{16}-\tfrac{5
   z^4}{128}+\CO(z^5)~, \\
   -\tfrac{i}{\sqrt2}\tfrac{\Lambda_2}{m} 
   \tH_2=\, \, &z+\CO(z^4)~.
\eea
It is again straightforward to verify the rest of the equations in~\eqref{universal functions from SW}.

\sss{$\boldsymbol{N_f=3}$.}
Finally, for $N_f=3$ we find:
\beasmall
v_1&=-\frac{1}{\sqrt{2}\pi}  \frac{1}{\jt_2\jt_3}\frac{\Lambda_0}{m}
-\frac{1}{48\sqrt{2}\pi} \frac{\jt_2^4+\jt_3^4}{\jt_2^3\jt_3^3} \left(\frac{\Lambda_0}{m}\right)^3\\
&\quad 
+\frac{1}{2560\sqrt{2}\pi} \frac{7(\jt_2^8+\jt_3^8)-12\jt_2^4\jt_3^4}{\jt_2^5\jt_3^5} \left(\frac{\Lambda_0}{m}\right)^5 \\
& \quad +\frac{1}{57344\sqrt2 \pi}\frac{(\jt_2^4+\jt_3^4)(65(\jt_2^8+\jt_3^8)-124\jt_2^4\jt_3^4)}{\jt_2^7\jt_3^7}\left(\frac{\Lambda_0}{m}\right)^7 \\
&\quad +\frac{5}{4718592\sqrt2 \pi}\frac{209(\jt_2^{16}+\jt_3^{16})-112\jt_2^4\jt_3^4(\jt_2^8+\jt_3^8)-198\jt_2^8\jt_3^8}{\jt_2^9\jt_3^9}\left(\frac{\Lambda_0}{m}\right)^9 \\
&\quad +\frac{21}{92274688\sqrt2 \pi}\frac{(\jt_2^4+\jt_3^4)\left(85(\jt_2^{16}+\jt_3^{16})+16\jt_2^4\jt_3^4(\jt_2^8+\jt_3^8)-206\jt_2^8\jt_3^8\right)}{\jt_2^9\jt_3^9}\left(\frac{\Lambda_0}{m}\right)^{11} \\
&\quad+\CO(m^{-13})~,
\eeasmall
while
\beasmall \label{Cij Nf=3}
         C_{11}&=\left(\frac{\Lambda_0}{m}\right)^{4}-\frac{E_2+2(\jt_2^4+\jt_3^4)}{12\jt_2^2\jt_3^2}\left(\frac{\Lambda_0}{m}\right)^{6} \\
   &+\frac{E_2^2+3E_2(\jt_2^4+\jt_3^4)-4(\jt_2^8+\jt_3^8)+40\jt_2^4\jt_3^4}{288\jt_2^4\jt_3^4}\left(\frac{\Lambda_0}{m}\right)^{8} \\
   &+\frac{20E_2^3+60E_2^2(\jt_2^4+\jt_3^4)-3E_2(149(\jt_2^8+\jt_3^8)-779\jt_2^4\jt_3^4)-2(\jt_2^4+\jt_3^4)(19(\jt_2^8+\jt_3^8)-2095\jt_2^4\jt_3^4)}{207360\jt_2^6\jt_3^6}\left(\frac{\Lambda_0}{m}\right)^{10} \\
   &\quad+\CO(m^{-12})~,
\eeasmall
\beasmall 
C_{12}&=1+\frac{-E_2+\jt_2^4+\jt_3^4}{12\jt_2^2\jt_3^2}\left(\frac{\Lambda_0}{m}\right)^{2}+\frac{E_2^2-3E_2(\jt_2^4+\jt_3^4)+2(\jt_2^8+\jt_3^8)-11\jt_2^4\jt_3^4}{288\jt_2^4\jt_3^4}\left(\frac{\Lambda_0}{m}\right)^{4}
\\ &+\frac{-20E_2^3+120E_2^2(\jt_2^4+\jt_3^4)+E_2(-93(\jt_2^8+\jt_3^8)+363\jt_2^4\jt_3^4)-7(\jt_2^4+\jt_3^4)(\jt_2^8+\jt_3^8-25\jt_2^4\jt_3^4)}{207360\jt_2^6\jt_3^6}\left(\frac{\Lambda_0}{m}\right)^{6} \\
 &\quad+\CO(m^{-8})~,
\eeasmall
and 
\beasmall
H_3&=4\sqrt2 i\Bigg[
\frac{-E_2+\jt_2^4+\jt_3^4}{24\jt_2^2\jt_3^2}\left(\frac{m}{\Lambda_0^2}\right)^2
+\frac{5E_2(\jt_2^4+\jt_3^4)-5(\jt_2^8+\jt_3^8)+29\jt_2^4\jt_3^4}{576\jt_2^4\jt_3^4}\\
&+\frac{E_2(27(\jt_2^8+\jt_3^8)-37\jt_2^4\jt_3^4)-(\jt_2^4+\jt_3^4)(27(\jt_2^8+\jt_3^8)-55\jt_2^4\jt_3^4)}{15360\jt_2^6\jt_3^6}\left(\frac{\Lambda_0}{m}\right)^2 \\
&+\frac{E_2(\jt_2^4+\jt_3^4)(128(\jt_2^{12}+\jt_3^{12})-243\jt_2^4\jt_3^4)-128(\jt_2^{16}+\jt_3^{16})+74\jt_2^4\jt_3^4(\jt_2^8+\jt_3^8)+134\jt_2^8\jt_3^8}{430080\jt_2^8\jt_3^8}\left(\frac{\Lambda_0}{m}\right)^4 \\
&+\CO(m^{-6})\Bigg]~.
\eeasmall
From these, we compute again
\bea\label{C3 D3 H3}
\tfrac{m}{\Lambda_3}\tC_3=\, \, &1-\tfrac{z}{2}+\tfrac{3 z^2}{8}-\tfrac{17 z^3}{64}+\CO(z^4) \\
\hat \tC_3=\, \, &1+\tfrac{z}{2}+\tfrac{z^3}{64}+\CO(z^4)~,\\
\tD_3=\, \, &1-\sqrt{z}+\tfrac{z}{2}-\tfrac{z^{3/2}}{4}+\tfrac{z^2}{8}-\tfrac{3
   z^{5/2}}{64}+\tfrac{z^3}{64}-\tfrac{z^{7/2}}{256}+\tfrac{3
   z^{9/2}}{8192}-\tfrac{z^5}{4096}+\tfrac{3 z^{11/2}}{32768}+\CO(z^6)~, \\
   -\tfrac{i}{\sqrt2}\tfrac{\Lambda_3}{m} 
   \tH_3=\, \, &z-\tfrac{z^2}{4}+\tfrac{z^3}{32}-\tfrac{z^5}{2048}+\CO(z^6)~,
\eea
and check that they satisfy~\eqref{universal functions from SW}.

\subsection{Couplings as period integrals}
\label{app:v as period integral}

In this final section, we derive an expression of the coupling $v$~\eqref{couplings appendix} in terms of the SW curve.\footnote{See also~\cite{Sonnenschein_1996,DHoker:1996yyu, Gottsche:2006, Gottsche:2010ig} for similar approaches.} For simplicity, we restrict to $N_f=1$, with the SW curve given in~\eqref{eq:curves}. We express the curve as a product over the three roots $X_j$, $y^2=(x-X_1)(x-X_2)(x-X_3)$.
In the weak coupling region, $0\ll u/\Lambda_0^2 \in \mathbb{R}^+$, and decoupling limit $m\to \infty$, the leading terms of the roots are
\be
\begin{split}
X_1&= \frac{\Lambda_0^4}{16m^2}+\dots~,\\
X_2&= \frac{\Lambda_0^4}{4u}-\frac{\Lambda^4_0}{16m^2}+\dots~,\\
X_3&= u-\frac{\Lambda_0^4}{4u}+\frac{\Lambda_0^8}{64u^2m^2}+\dots~.\\
\end{split}
\ee
We choose the two branch cuts to run between $X_1$ and $X_2$, and from $X_3$ to $\infty$. The curve can be represented as a two-sheeted cover with $y=\pm \sqrt{\dots}$~. We denote the branch points as $X_{j\pm}$ for $j=1,2,3$ on each sheet. We let the domain for the phase $\phi$ of an exponent $e^{i\phi}$ be $\phi\in [0,2\pi)$. We take the branch of the square root along the positive real axis.

The goal is to express the coupling
\be \label{v def app}
v=\sqrt2\frac{d a_D}{d m}
\ee 
in terms of the SW curve and differential. The dual period $a_D$ is defined as the contour integral over the $B$-cycle of the SW differential~\cite{Seiberg:1994aj},
\be 
\label{eq:aaDAB}
a=\oint_A \lambda~,\qquad a_D=\oint_B \lambda~,\qquad \lambda=-\frac{\sqrt{2}}{4\pi} \frac{y}{x^2}\,dx~,
\ee 
where we choose the $A$-cycle as the circular clockwise contour with center $0$ and surrounding the roots $X_{1+}$ and $X_{2+}$, while we choose the $B$-cycle as a curve from $X_{2+}$ to $X_{3+}$ and then continuing on the second sheet from $X_{3-}$ to $X_{2-}$ and back to $X_{2+}$. The choice of $A$-cycle is such that in the decoupling limit $m\to\infty$ the singularity of $\lambda$ merges with $X_1$ without coinciding with the $A$-cycle.

The differential $\lambda$ has a singularity at $z_\pm=(x=0, y=\pm i \Lambda_1^3/8)$, with residue
\be 
\mathop{\mathrm{Res}}_{z_\pm} \lambda=\pm \frac{1}{2\pi i} \frac{m}{\sqrt{2}}~.
\ee 
For the differential $\partial \lambda/\partial m$, we find
\be
\frac{\partial \lambda}{\partial m}=-\frac{\sqrt{2}}{32\pi}\,\Lambda_1^3\,\frac{dx}{x\,y}~,
\ee 
with residue
\be \label{dldm residue}
\mathop{\mathrm{Res}}_{z_\pm} \frac{\partial \lambda}{\partial m}=\pm \frac{1}{2\pi i} \frac{1}{\sqrt{2}}~.
\ee 
We consider the period $a$ and $m$ as independent variables. This gives the relation
\be 
\begin{split} 
0&=\frac{da}{dm}=\oint_A \frac{d\lambda}{dm}=\oint_A \frac{\partial \lambda}{\partial m}+\frac{\partial u}{\partial m}\frac{\partial \lambda}{\partial u}\\
&=\frac{\partial u}{\partial m}\,\frac{da}{du}+\oint_A \frac{\partial \lambda}{\partial m}~.
\end{split}
\ee 
We can apply this to the coupling~\eqref{v def app},
\be
\begin{split} 
\frac{v}{\sqrt2}&=\oint_B \frac{\partial \lambda}{\partial m}+\frac{\partial u}{\partial m} \frac{da_D}{da}\frac{da}{du}\\
&=\oint_B \frac{\partial \lambda}{\partial m}-\tau \oint_A \frac{\partial \lambda}{\partial m}~.
\end{split}
\ee 
In order to write this more symmetrically, we introduce the holomorphic 1-form $\varphi$ on the curve, with periods 
\be 
\oint_A \varphi=1~,\qquad \oint_B\varphi =\tau~,
\ee 
such that $\varphi=\frac{du}{da}\frac{\partial\lambda}{\partial u}$, where
$\frac{\partial\lambda}{\partial u}=\frac{\sqrt2}{8\pi}\frac{dx}{y}$. 
Then $v$ can be expressed as
\be \label{v as period integrals}
\frac{da}{du}\frac{v}{\sqrt2}=\oint_A \frac{\partial\lambda}{\partial u} \oint_B \frac{\partial \lambda}{\partial m}-  \oint_B \frac{\partial\lambda}{\partial u}\oint_A \frac{\partial \lambda}{\partial m}~.
\ee 
Such a difference of products of periods for holomorphic and meromorphic differentials may be simplified using reciprocity laws, see \eg~\cite[Chapter 2.2]{Griffiths1994}. It allows us to express~\eqref{v as period integrals} as
\be
2\pi i \sum_\pm \mathrm{ Res}_{z_\pm}\left[\frac{\partial \lambda}{\partial m} \right] \int_{z_b}^{z_\pm} \frac{\partial \lambda}{\partial u}~,
\ee
where $z_b$ is a base point located away from the cycles $A$ and $B$, and the integration contours from $z_b$ to $z_\pm$ do not cross the $A$- or $B$-contour. The expression is then independent of the choice of base point.
Inserting $\partial \lambda/\partial u$ from above and using $z_b=X_{1}$ gives
\bea\label{dadu v}
\frac{da}{du}v= \frac{1}{2\sqrt 2\pi}\int_{X_1}^0\frac{dx}{y}~.
\eea
In order to make use of results on the \ws{} function, we aim to relate this to an integral from $\infty$ to 0. To this end, recall that the $A$-contour can be deformed for this integrand to a curve from $X_{1+}$ to $X_{2+}$ above the branch cut, and then from $X_{2+}$ to $X_{1+}$ below the branch cut. Moreover, the $A$-cycle is homologically equivalent to a contour from $X_{3+}$ to $\infty$ and back to $X_{3+}$ on the other side of the branch cut. Therefore  
\bea
\frac{da}{du}= \frac{1}{2\sqrt 2\pi}\int_{X_3}^{\infty}\frac{dx}{y}~,
\eea
where the contour lies just above the real axis in the $x$-plane to avoid the branch cut. Moreover, we have from Eq. (\ref{eq:aaDAB})
\bea
 \frac{da_D}{du}= \frac{1}{2\sqrt 2\pi}\int_{X_2}^{X_3}\frac{dx}{y}~.
\eea
We find that a linear combination of these periods gives the desired integral from $\infty$ to $0$,
\bea
\frac{da}{du}v-\frac{da}{du}-\frac{da_D}{du}-\frac{da}{du}=\frac{1}{2\sqrt 2\pi}\left(\int_{X_1}^0+\int_{X_2}^{X_1}+\int_{X_3}^{X_2}+\int^{X_3}_{\infty}\right)\frac{dx}{y}~.
\eea
Finally, to use the technology of \ws{} functions, we bring the $N_f=1$ curve to \ws{} form \eqref{eq:Weierstrass} by the shift $x\mapsto x+u/3$ and $y\mapsto y/2$, such that 
\bea
\frac{da}{du}v-\frac{da_D}{du} - 2\frac{da}{du}=\frac{1}{\sqrt 2\pi}\int_{\infty}^{-\frac u3}\frac{dx}{y}~.
\eea
We can now use Eq. (\ref{elliptic integral}) for an elliptic curve in \ws{} form,
\bea
-\tfrac u3=\wp_\Lambda\left(\sqrt2\pi(\tfrac{da}{du}(v-2)-\tfrac{da_D}{du}),\omega,\omega'\right)=\wp_\Lambda\left(2\sqrt2\pi\tfrac{da}{du}(\tfrac{v}{2}-\tfrac\tau2-1)\right)~,
\eea
with periods $\omega=2\sqrt{2}\pi\, da/du$ and $\omega'=2\sqrt{2}\pi\, da_D/du$. Rewriting $\frac{da_D}{du}=\tau\frac{da}{du}$, using the homogeneity relation of $\wp$~\eqref{wp homogeneity} and its periodicity, we express it in terms of the elliptic function $\wp$~\eqref{eq:WPv2}, which gives 
\bea
-\tfrac{8\pi^2}{3}u\left(\tfrac{da}{du}\right)^2 =\wp(\tfrac v2-\tfrac \tau2,\tau)~.
\eea
Writing this more explicitly in terms of theta functions, we substitute the identity~\eqref{wp z- tau}, such that
\bea
\tfrac{\pi^2}{3}\left(8\left(\tfrac{da}{du}\right)^2u- E_2\right)&=\jt_4^{(2)}(\tau,\tfrac v2)~.
\eea
The lhs is essentially the contact term~\eqref{contactterm explicit}, from which the result~\eqref{3 equations c} for $N_f=1$ follows:
\bea
8 \left(\pi\Lambda_{1}\tfrac{da}{du}\right)^2G_{1}&=\jt_4^{(2)}(\tau,\tfrac{v}{2})~.
\eea
We expect that a similar analysis can be carried out to reproduce the contact terms for $N_f=2,3$ in Eq. \eqref{3 equations c}.

\newpage
\section{Symbol list} 
\label{app:symbol list}
\begin{longtable}{|p{2.3cm}|p{10cm}|p{1cm}|}
\hline
Symbol & Meaning & Ref. \\
\hline
$a$ & Low-energy scalar field & \ref{sec:CB geometry}  \\
$a_D$ & Scalar dual to $a$ & \ref{sec:CB geometry} \\
$\alpha$ & $K$-theory class in $K(S)$ &\ref{sec: segre numbers} \\
$A$ & Effective gravitational coupling & \eqref{A B couplings} \\
$B$ & Effective gravitational coupling & \eqref{A B couplings} \\
$c$ & Seiberg--Witten basic class on $X$ & \ref{sec: contr singularities} \\
$c_1$, $c_2$ & Chern classes of sheaves on $S$ & \ref{sec: segre numbers} \\
$c_1(\alpha)$, $c_2(\alpha)$ & Chern classes of $K$-theory class $\alpha\in K(S)$ & \eqref{c1 c2 definition} \\
$\chi$ & Topological Euler characteristic of $X$ & \ref{top twist} \\ 
$\chih$ & Holomorphic Euler characteristic, or $\frac14(\chi+\sigma)$ &  \ref{top twist}  \\
$C_{ij}$ , $\tC_{N_f}$, $\hat \tC_{N_f}$ & Exponentiated quadratic coupling to background fluxes  & \eqref{C D couplings} \\
$C'$ & Exceptional divisor of the blowup & \ref{sec:blowup} \\
$D_j$, $\tD_{N_f}$ & Exponentiated linear coupling to background fluxes & \eqref{C D couplings} \\
$\decoup$ & Decoupling factor & \eqref{decoup equal mass}\\
$\Delta$ & Discriminant of SW curve & \eqref{sec:CB geometry}\\
$E$ & Principal $\SO(3)$ bundle over $X$ & \ref{top twist} \\
$E_k$ & Eisenstein series & \eqref{eisenstein} \\
$\eta$ & Dedekind $\eta$-function & \eqref{dedekind eta} \\
$F,\CF$ & Prepotential & \ref{sec:CB geometry} \\
$g_2, g_3$ & \ws{} invariants & \ref{sec:CB geometry}\\
$G_{N_f}$, $\tG_{N_f}$ & Contact term for surface observables & \eqref{contactterm} \\
$\segregen_{\segre}$ & Generating function of virtual Segre numbers & \eqref{segre gen} \\
$\segregen_{\universal}$ & Restriction of $\universal_\alpha$ & \eqref{segre generating} \\
$H_{N_f}$,  $\tH_{N_f}$ & Coupling of background fluxes to surfaces & \eqref{contactterm}\\ 
$I(\bfx)$ & Observable for surface class $\bfx$ & \eqref{UV corr} \\
$\CI_\mu^X$ & $u$-plane integrand on $X$ & \eqref{blowup u-plane integrand} \\
$J$ & Period point & \ref{sec: contr singularities} \\
$k$ & Instanton number of principal bundle $E\to X$ & \ref{top twist} \\
$\bfk_j$ & Background fluxes & \ref{top twist} \\
$k_j'$ & Background fluxes along exceptional divisor & \eqref{components exceptional divisor} \\
$(\bfk)$ & Combinations of $\bfk_j$ & \ref{sec: contr singularities} \\
$\kappa$ & The combination $3\chih+4\bfmu^2$ & \ref{sec:contribution instanton} \\
$K$ & Canonical class of $X$ & \ref{top twist}  \\
$K_X$ & Canonical bundle of $X$ & \ref{top twist}  \\
$K(S)$ & Grothendieck group of coherent sheaves on $S$ & \ref{sec: segre numbers} \\
$L$ & Middle cohomology lattice $H^2(X,\mathbb Z)$ modulo torsion & \ref{top twist} \\
$L_j$ & Line bundles comprising $K$-theory class $\alpha$ & \ref{sec:comparison} \\
$\CL_j$ & Line bundle coupled to $j$'th hypermultiplet & \ref{top twist} \\ 
$\Lambda_{N_f}$ & Scale of SQCD with $N_f$ flavours & \ref{sec:CB geometry} \\
$\lambda$ & Seiberg--Witten differential & \eqref{eq:aaDAB} \\
$m$ & Bare mass of the hypermultiplets& \ref{sec:CB geometry} \\
$\bfm $ & Mass vector $(m,\dots, m)$& \ref{sec:CB geometry} \\
$\tm_{N_f}$ & Coefficient in local expansion near singularity  & \eqref{u_exp} \\
$\bfmu$ & 't Hooft flux for gauge bundle & \ref{top twist}  \\
$\mud$ & Donaldson map & \ref{sec: segre numbers}\\
$M$ & Moduli space of semistable sheaves on $S$ with fixed $(\rho, c_1,c_2)$ & \ref{sec: segre numbers} \\
$\CM_{k}^{\CQ,N_f}$ & Moduli space of $\CQ$-fixed equations & \ref{top twist}\\
$\Mi$ & Instanton component of above moduli space & \ref{top twist}\\ 
$\CM_k^{\rm m}$ & Monopole component of above moduli space & \ref{top twist}\\ 
$N_f$ & Number of hypermultiplets& \ref{sec:CB geometry} \\
$\nu$ & $u$-plane measure factor & \eqref{nu measure} \\ 
$\CO$ & Observable in TQFT & \ref{top twist} \\
$\langle \CO\rangle$ & Correlation function of $\CO$ on $X$& \ref{top twist}\\
$\langle \CO\rangle^{\rm i}$ & Contribution of instanton component to $\langle\CO\rangle$ & \eqref{O instanton}\\
$p,p'$ & Fugacity for the point class & \eqref{observables} \\
$\PM_{N_f}$ & Matone's polynomial & \ref{sec:CB geometry} \\
$\wp$ & \ws{} $\wp$-function & \ref{app:modularforms}\\
$\Phi$ & $u$-plane integral & \eqref{u plane integral} \\
$\Psi^J_\bfmu$ & Siegel--Narain theta function & \eqref{u plane integral} \\
$Q_s$ & Universal function & \eqref{Vs Ws Xs Qs Rs Ts}\\
$\rho$ & Rank of semistable sheaves on $S$ & \ref{sec: segre numbers}\\
$R_s$ & Universal function& \eqref{Vs Ws Xs Qs Rs Ts} \\
$R[[z]]$ & Formal power series in $z$ over a ring $R$ & \ref{sec: segre numbers} \\
$\CR_{\mu'}$ & Blowup factor & \eqref{blowup R} \\
$s$ & Rank of $\alpha\in K(S)$ & \ref{sec: segre numbers} \\
$\sigma$ & Signature $b_2^+-b_2^-$ of $X$ & \ref{top twist} \\
$S$ & Smooth projective surface & \ref{sec: segre numbers}\\
$\segre_\alpha$ & Virtual Segre numbers of $S$ & \eqref{virtual Segre def} \\
$S_s$, $S_{1,s}$ & Universal function& \eqref{Ys Zs Ss} \\
$\SW$ & Seiberg--Witten invariant, differential geometry convention & \eqref{SWcont}\\
$\mathcal{SW}$ & Seiberg--Witten invariant, algebraic geometry convention & \eqref{SW shift}\\
$t$ & Change of variables& \eqref{changeofvar}\\
$\tau$ & Low-energy effective coupling on the CB & \ref{sec:CB geometry}\\
$\jt_j$ & Jacobi theta function & \eqref{jt_elliptic} \\ 
$T_s$ & Universal function & \eqref{Vs Ws Xs Qs Rs Ts}\\
$u$ & Coulomb branch coordinate & \ref{sec:CB geometry} \\
$\tu_{N_f}^{(\pm)}$ & CB singularities for $N_f=0,\dots, 3$ equal masses & \ref{sec:CB geometry} \\
$\universal_\alpha$ & Product of universal functions & \eqref{universal series} \\
$v_j$ & Linear coupling to background fluxes & \eqref{vj wij} \\ 
$V_s$ & Universal function & \eqref{Vs Ws Xs Qs Rs Ts}\\
$w_{ij}$ & Quadratic coupling to background fluxes & \eqref{vj wij} \\
$W_s$ & Universal function& \eqref{Vs Ws Xs Qs Rs Ts} \\
$W_k^j$ & Index bundle for Dirac equation of $j$'th multiplet & \ref{top twist}  \\
$x,x'$ & Surface class & \eqref{observables} \\
$\tx_{N_f}$  & Period $\frac{da}{du}$ at singularity & \eqref{dadu def} \\ 
$X$ & Smooth, oriented, compact, simply-connected four-manifold & \ref{top twist} \\
$\widehat X$ & Blowup of $X$ at a point & \ref{sec:blowup} \\
$X_s$ & Universal function& \eqref{Vs Ws Xs Qs Rs Ts} \\
$Y_s$, $Y_{1,s}$ & Universal function & \eqref{Ys Zs Ss}\\
$z$ & Fugacity for Segre numbers & \eqref{virtual Segre def}  \\ 
$\bfz$ & Elliptic variable in $u$-plane integral & \eqref{elliptic variable u-plane} \\
$Z$ & Topological partition function of $\CN=2$ $\SU(2)$ SQCD &  \eqref{Z_contr} \\
$Z^{\rm i}$ & Contribution of instanton component to partition function & \eqref{O instanton}\\ 
$Z^{\rm m}$ & Contribution of monopole component to partition function & \ref{top twist} \\
$Z_s$, $Z_{11,s}$ & Universal function& \eqref{Ys Zs Ss} \\
$Z_\SW$ & Seiberg--Witten contribution to partition function & \eqref{Z_contr} \\
$Z_{\SW,\pm}$ & Contribution from $\tu_{N_f}^\pm$ to SW partition function & \eqref{ZSW pm} \\
$Z_u$ & Contribution from $u$-plane to partition function & \eqref{Z_contr}  \\
\hline
\end{longtable}

\bibliography{uplane} 
\bibliographystyle{JHEP} 
\end{document}